\documentclass[twocolumn]{aastex631}

\usepackage{amsmath}
\usepackage{booktabs}
\usepackage{multirow}
\usepackage{threeparttable}
\usepackage{ulem}
\usepackage{soul}
\usepackage{comment}
\usepackage{natbib}
\usepackage{scalerel}

\begin{document}

\title{A Unified Origin of Faraday Rotation toward 3C~84: The Circumnuclear Ambient Medium within the Parsec-Scale Bondi Radius of the Host Galaxy NGC~1275}

\author[0000-0001-9799-765X]{Minchul Kam}
\affiliation{\normalfont Institute of Astronomy and Astrophysics, Academia Sinica, P.O. Box 23-141, Taipei 10617, Taiwan, R.O.C.}
\affiliation{\normalfont Department of Physics and Astronomy, Seoul National University, Gwanak-gu, Seoul 08826, Republic of Korea}

\author[0000-0001-6558-9053]{Jongho Park}
\affiliation{\normalfont School of Space Research, Kyung Hee University, 1732, Deogyeong-daero, Giheung-gu, Yongin-si, Gyeonggi-do 17104, Republic of Korea}
\affiliation{\normalfont Institute of Astronomy and Astrophysics, Academia Sinica, P.O. Box 23-141, Taipei 10617, Taiwan, R.O.C.}
\affiliation{\normalfont G-LAMP NEXUS Institute, Kyung Hee University, Yongin, 17104, Republic of Korea}

\author[0000-0003-0465-1559]{Sascha Trippe}
\affiliation{\normalfont Department of Physics and Astronomy, Seoul National University, Gwanak-gu, Seoul 08826, Republic of Korea}
\affiliation{\normalfont SNU Astronomy Research Center, Seoul National University, Gwanak-gu, Seoul 08826, Republic of Korea}

\author[0000-0003-1157-4109]{Do-Young Byun}
\affiliation{\normalfont Korea Astronomy and Space Science Institute, Daedeok-daero 776, Yuseong-gu, Daejeon 34055, Republic of Korea}

\author[0000-0001-6988-8763]{Keiichi Asada}
\affiliation{\normalfont Institute of Astronomy and Astrophysics, Academia Sinica, P.O. Box 23-141, Taipei 10617, Taiwan, R.O.C.}

\author[0000-0003-0292-3645]{Hiroshi Nagai}
\affiliation{\normalfont National Astronomical Observatory of Japan, 2-21-1 Osawa, Mitaka, Tokyo 181-8588, Japan}
\affiliation{\normalfont Department of Astronomical Science, The Graduate University for Advanced Studies, SOKENDAI, 2-21-1 Osawa, Mitaka, Tokyo 181-8588, Japan}

\author[0000-0002-2709-7338]{Motoki Kino}
\affiliation{\normalfont National Astronomical Observatory of Japan, 2-21-1 Osawa, Mitaka, Tokyo 181-8588, Japan}
\affiliation{\normalfont Kogakuin University of Technology $\&$ Engineering, Academic Support Center, 2665-1 Nakano, Hachioji, Tokyo 192-0015, Japan}

\author[0000-0001-6947-5846]{Luis C. Ho}
\affiliation{\normalfont Kavli Institute for Astronomy and Astrophysics, Peking University, Beijing 100871, China}
\affiliation{\normalfont Department of Astronomy, School of Physics, Peking University, Beijing 100871, China}

\author[0000-0002-9378-4072]{Andrew C. Fabian}
\affiliation{\normalfont Institute of Astronomy, University of Cambridge, Madingley Road, Cambridge CB3 0HA, UK}

\author[0000-0001-6094-9291]{Jeffrey Hodgson}
\affiliation{\normalfont Department of Physics and Astronomy, Sejong University, 209 Neungdong-ro, Gwangjin-gu, Seoul, Republic of Korea}

\author[0009-0007-8554-4507]{Kunwoo Yi}
\affiliation{\normalfont School of Space Research, Kyung Hee University, 1732, Deogyeong-daero, Giheung-gu, Yongin-si, Gyeonggi-do 17104, Republic of Korea}
\affiliation{\normalfont G-LAMP NEXUS Institute, Kyung Hee University, Yongin, 17104, Republic of Korea}
\affiliation{\normalfont Department of Astronomy and Space Science, Kyung Hee University, 1732, Deogyeong-daero, Giheung-gu, Yongin-si, Gyeonggi-do 17104, Republic of Korea}

\author[0000-0002-7114-6010]{Kenji Toma}
\affiliation{\normalfont Frontier Research Institute for Interdisciplinary Sciences, Tohoku University, Sendai 980-8578, Japan}
\affiliation{\normalfont Astronomical Institute, Graduate School of Science, Tohoku University, Sendai 980-8578, Japan}

\author[0000-0002-4991-9638]{Junghwan Oh}
\affiliation{\normalfont Joint Institute for VLBI ERIC (JIVE), Oude Hoogeveensedijk 4, 7991 PD Dwingeloo, The Netherlands }

\author[0000-0001-7003-8643]{Taehyun Jung}
\affiliation{\normalfont Korea Astronomy and Space Science Institute, Daedeok-daero 776, Yuseong-gu, Daejeon 34055, Republic of Korea}

\begin{abstract}
We present multi-frequency polarimetric observations of 3C~84 obtained with the Korean VLBI Network at 43–141~GHz, the Very Long Baseline Array at 43~GHz, and the High Sensitivity Array at 8~GHz from 2015 to 2024. We find that the Faraday rotation measure (RM) decreases systematically with distance from the black hole over 1–8~pc, following a single power-law trend of $\mathrm{RM} \propto r^{-2.7\pm0.2}$. Notably, RM measurements from earlier studies across the same distance range follow the same relation. This consistency across epochs, frequencies, and independent datasets indicates a common and stable external Faraday screen. These results naturally identify the circumnuclear ambient medium within the parsec-scale Bondi radius of the host galaxy NGC~1275 as the origin of the Faraday rotation, thereby resolving a long-standing question about its physical origin.
From the RM profile, we derive radial distributions of the electron density and magnetic-field strength in the circumnuclear ambient medium that are consistent with independent constraints. The derived density lies below that of the free-free absorption disk and, when extrapolated inward, remains below the density of the broad-line region. The magnetic-field strength gradually increases from $0.1–1.5~\mu$G at the Bondi radius to milligauss-to-gauss levels toward the black hole, providing the first spatially resolved constraint on the magnetic-field strength at parsec-scale distances in an elliptical galaxy. Together, these results present a spatially resolved and physically consistent picture of the circumnuclear environment in NGC~1275.
\end{abstract}

\keywords{galaxies: active --- galaxies: jets --- radio continuum: galaxies}

\section{Introduction} \label{sec:intro}

A wide range of observations over the past decades have revealed that the dominant emission mechanism of active galactic nuclei (AGN) jets is synchrotron radiation \citep{bradt08, beckmann12}. This naturally explains the widespread detection of linearly polarized emission in AGN jets \citep[e.g.,][]{lister05, jorstad07, hovatta12, park18, lister18}. However, some nearby AGN jets exhibit remarkably weak polarization, with fractional polarization below 1\% at centimeter wavelengths \citep[e.g.,][]{homan06}. A representative case is 3C~84, a nearby radio AGN located at the center of the elliptical galaxy NGC~1275 \citep[z = 0.0176;][]{strauss92, hitomi16}. 

NGC~1275 is the brightest cluster galaxy (BCG) located at the center of the Perseus cluster \citep[e.g.,][]{forman72, fabian74, fabian81}. It hosts a radio-loud active nucleus associated with the radio source 3C~84. Owing to its proximity, high brightness, and clear signatures of ongoing AGN feedback, the 3C~84 jet has been extensively monitored over the past decades across a wide range of frequencies \citep[e.g.,][]{dutson14, agudo18_3, agudo18_1, paraschos22, sinitsyna25}. A distinctive morphological feature of this source is the presence of multiple radio lobes aligned roughly along the north–south direction, which has been interpreted as evidence for intermittent jet activity \citep{pedlar83, pedlar90, walker94, silver98, dhawan98, nagai10, fujita17}. 

On parsec scales, two distinct episodes of jet activity are evident. The outer lobes, located at de-project distances of 7--8~pc from the core, are thought to have been launched in the 1950s \citep{asada06}. More recently, a new jet (hereafter the \textit{inner jet}) emerged in 2003 \citep{nagai10, suzuki12}. Its termination region, commonly referred to as component C3, has since propagated at a nearly constant apparent speed of $0.26c$ \citep{kam24}. 

Despite its brightness and well-resolved jet, 3C~84 has long been known to exhibit exceptionally weak or undetectable linear polarization at radio wavelengths, from the 1960s until recently \citep[e.g.,][]{hobbs68, allen68, mccullough69, wardle71, hobbs78, homan99, homan04, homan06, trippe12}. As a result, it has been widely used as a primary instrumental polarization calibrator in radio observations \citep[e.g.,][]{roberts94}. However, the physical origin of its low polarization remained uncertain for decades. A widely considered explanation for the unusually low polarization is strong Faraday depolarization. In this scenario, intrinsically polarized jet emission undergoes substantial Faraday rotation as it propagates through magnetized plasma along the line of sight \citep[e.g.,][]{wardle71, trippe12, plambeck14, bower17}. The resulting EVPA rotation is given by $\Delta\phi$ = RM$\lambda^2$, where $\lambda$ is the observing wavelength and RM is the rotation measure, which depends on the electron density and the line-of-sight magnetic-field strength in the foreground plasma. If the plasma is inhomogeneous within a telescope beam, radiation passing through different regions will experience different magnitudes of Faraday rotation, and the beam-averaged emission becomes depolarized \citep{burn66, sokoloff98, pasetto18}. Consequently, if strong Faraday depolarization indeed occurs in 3C~84, we expect (i) higher fractional polarization at shorter wavelengths, where Faraday rotation is weaker, and (ii) a large RM dispersion across the observing beam, comparable to or proportional to the RM itself, sufficient to depolarize the emission within the observing beam. 

To investigate this, polarimetric observations have been conducted from centimeter to millimeter and submillimeter wavelengths. At centimeter wavelengths, the Very Long Baseline Array (VLBA) observations at 5-22~GHz detected linear polarization with fractional polarization levels of 0.8-7.5\% in the southern lobe, located $\sim$14~mas from the core \citep{taylor06}. The RM measured at this location is $\sim$7000 $\rm rad\ m^{-2}$, which was interpreted as arising from filamentary ionized gas on kiloparsec scales. In contrast, polarization in the inner jet is below 0.2\% over the same frequency range, implying the RM exceeding 10000 $\rm rad\ m^{-2}$ to account for the Faraday depolarization \citep{taylor06}.

At higher frequencies, observations with the IRAM Plateau de Bure Interferometer (PdBI) at 81-348 GHz found no significant polarization \citep{trippe10, trippe12}, placing upper limits of 0.5\% at 239~GHz and 1.9\% at 348~GHz. At these frequencies, the emission is expected to be dominated by the inner jet, the same region in which \citet{taylor06} reported non-detections of polarization. Accordingly, \citealt{trippe12} suggested that strong Faraday depolarization by dense ionized gas could explain these non-detections, requiring the RM dispersion of at least $\rm 1\times10^{6}\ rad\ m^{-2}$ to suppress polarization even at 354~GHz. 

Later, \citealt{plambeck14} detected linear polarization at the 1-3\% level during 2011-2013 in the 1.3 mm band using the Submillimeter Array (SMA) and the Combined Array for Research in Millimeter Wavelength Astronomy (CARMA). They measured an average RM of $\rm 8.7\times10^{5}\ rad\ m^{-2}$ and suggested that the Faraday rotation originates within 0.4~pc from the central black hole. This RM is approximately two orders of magnitude larger than the RM of $\rm 7000\ rad\ m^{-2}$ measured in the southern lobe at 7--8~pc from the black hole \citep{taylor06}. To account for the large RM near the core, \citealt{plambeck14} considered two possible Faraday screens: (i) a boundary layer of the inner jet, and (ii) a radiatively inefficient accretion flow (RIAF), which could intersect the line of sight toward the jet given its geometrically thick and optically thin structure \citetext{see \citealp{narayan05, ho08, yn14} for reviews}. However, they found that RIAF models predict RM values several orders of magnitude higher than observed. To reconcile this discrepancy, they suggested that if the RIAF is responsible for the Faraday rotation, it must have a flattened, disk-like geometry that significantly reduces the effective path length through the plasma contributing to the RM.

Subsequent studies explored alternative interpretations. \citet{li16} argued that the discrepancy between the observed RM and the RM predicted by RIAF models could be alleviated if the RM is weighted by the jet luminosity distribution along the line of sight. In contrast, \citet{kim19} proposed that a jet boundary layer more naturally explains the RM measured with the SMA and CARMA. 

Meanwhile, \citet{nagai17} investigated the polarization and RM in the termination region of the inner jet. They analyzed archival VLBA 43~GHz data (hereafter BU data) obtained between 2015 December and 2016 April as part of the VLBA-BU-BLAZAR program \citep{jorstad16, jorstad17, weaver22}. They detected enhanced polarization at the 1-2\% level in the termination region of the inner jet (component C3) located $\sim$3~mas from the core. The measured RM over a 256~MHz bandwidth was $\rm (3-6)\times 10^{5}\ rad\ m^{-2}$. This value is comparable to the $\rm 8.7\times10^{5}\ rad\ m^{-2}$ measured with the CARMA and SMA, which was attributed to emission originating from the core \citep{plambeck14}, where the higher electron density and magnetic-field strength are expected. \citet{nagai17} suggested that the comparable RM values could be explained if a dense clump of ionized gas is present near the jet head.

However, despite the various scenarios proposed to explain the diverse RM measurements across different frequencies and spatial scales, it has remained challenging to reconcile these results due to the lack of spatially resolved RM measurements. In this paper, we show for the first time that the observed RM values in 3C~84 can be consistently explained by a single physical origin: the circumnuclear ambient medium of the host galaxy NGC~1275. Furthermore, by deriving the radial profiles of the electron density and magnetic-field strength from the RM measurements, we provide a spatially resolved and physically consistent picture of the circumnuclear environment in NGC~1275. 

The paper is organized as follows. Section~\ref{sec:Data} describes the datasets used in our analyses. Section~\ref{sec:analysis} presents the polarization images and the measured RM. Section~\ref{sec:Discussions} interprets the observational results, discusses the physical origin of the RM and derives the physical properties of the Faraday screen. Section~\ref{sec:Conclusions} summarizes our main results and presents the conclusions. 

The Hubble constant $H_{\rm 0}$ are measured to be different between the early and local universe, a discrepancy often referred to as the Hubble tension (see \citealp{valentino21} for a review). Given that 3C~84 resides in the local universe, we adopt the cosmological parameters $H_{\rm 0}= \rm 73.30\ km\ s^{-1}\ Mpc^{-1}$ and $q_{0}=-0.51\pm0.024$ obtained from the Type Ia supernovae observations \citep{riess22}. These correspond to $\Omega_{\rm m}=\rm 0.327$ and $\Omega_{\rm \Lambda}=\rm 0.673$ in a flat universe, yielding a linear scale of 1 mas = 0.343 pc at a distance of 73.2 Mpc. We adopt a jet viewing angle of 45$^\circ$ when calculating the de-projected distances from the core \citep{walker94, asada06, fujita17}. 


\begin{table*}[ht]
    \centering 
    \caption{Summary of KVN observations \label{tab:kvn_summary}}
    \begin{tabular}{ccc ccc c}
        \hline
        Telescopes & Exp. code & Date & Rec. rate & Observed Frequency & Bandwidth & Used Frequency  \\
        & & & [Gbps] & [GHz]& [MHz] & [GHz]  \\
        & (1) & (2) & (3) & (4) & (5) & (6)  \\
        \hline
        \multirow{23}{*}{YS, US, TN} & p18st01b & 2018 Feb. 14 & 1 & 43/139 & 64 & 139  \\
        & n18jp01a & 2018 Feb. 20 & 8 & 86/129 & 512 & 86/129 \\
        & n18jp01b & 2018 Feb. 21 & 8 & 94/141 & 512 & 94/141 \\
        \cline{2-7}
        & p18st01o & 2019 Jan. 02 & 1 & 22/86 & 64 & 86  \\
        & p18st01p & 2019 Jan. 03 & 1 & 43/94 & 64 & 94  \\
        \cline{2-7}
        & p19st01g & 2019 May 02 & 1 & 22/86 & 64 &86  \\
        & p19st01h & 2019 May 03 & 1 & 43/94 & 64 & 94  \\
        \cline{2-7}
        & \multirow{3}{*}{p21st02e} & \multirow{3}{*}{2021 Dec. 01} & \multirow{3}{*}{16} & 43.1/43.2/46.0/46.1 & \multirow{2}{*}{128} & \multirow{3}{*}{86.2/86.3/92.0/92.1}  \\
        & & & & 86.2/86.3/92.0/92.1 & & \\
        \cline{5-6}
        & & & & 21.55/129.3 & 512 & 129.3 \\ 
        \cline{2-7}
        & \multirow{3}{*}{p22st02b} & \multirow{3}{*}{2022 Sep. 15} & \multirow{3}{*}{16} & 43.1/43.2/46.0/46.1 & \multirow{2}{*}{128} & \multirow{3}{*}{43.1/43.2/46.0/46.1}  \\
        & & & & 86.2/86.3/92.0/92.1 & & \\
        \cline{5-6}
        & & & & 21.55/129.3 & 512 & \\ 
        \cline{2-7}
        & \multirow{3}{*}{p22stf02} & \multirow{3}{*}{2022 Oct. 20} & \multirow{3}{*}{16} & 43.1/43.2/46.0/46.1 & \multirow{2}{*}{128} & 43.1/43.2/46.0/46.1  \\
        & & & & 86.2/86.3/92.0/92.1 & & 86.2/86.3/92.0/92.1 \\
        \cline{5-6}
        & & & & 21.55/129.3 & 512 & \\ 
        \cline{2-7}
        & \multirow{3}{*}{p22st02d} & \multirow{3}{*}{2022 Nov. 05} & \multirow{3}{*}{16} & 43.1/44.0/45.5/46.5 & \multirow{2}{*}{128} & 43.1/44.0/45.5/46.5  \\
        & & & & 86.2/88.0/91.0/93.0 & & 86.2/88.0/91.0/93.0  \\
        \cline{5-6}
        & & & & 21.55/129.3 & 512 & \\ 
        \cline{2-7}
        & \multirow{3}{*}{p22st02e} & \multirow{3}{*}{2022 Dec. 27} & \multirow{3}{*}{16} & 43.1/44.0/45.5/46.5 & \multirow{2}{*}{128} & 43.1/44.0/45.5/46.5  \\
        & & & & 86.2/88.0/91.0/93.0 & & 86.2/88.0/91.0/93.0 \\
        \cline{5-6}
        & & & & 21.55/129.3 & 512 & \\ 
        \cline{2-7}
        & \multirow{2}{*}{t23jp01a} & \multirow{2}{*}{2023 Feb. 05} & \multirow{2}{*}{8} & 86.2/88.0/90.0/92.0 & \multirow{2}{*}{128} & \multirow{2}{*}{86.2/88.0/90.0/92.0} \\
        & & & & 129.1/129.2/129.4/129.5 & & \\ 
        \hline
        \multirow{8}{*}{YS, US, TN, PC} & \multirow{2}{*}{n24jp01a} & \multirow{2}{*}{2024 Nov. 08} & \multirow{2}{*}{8} & 43.1/44.0/45.5/46.5 & \multirow{2}{*}{128} & 43.1/44.0/45.5/46.5 \\ 
        & & & & 86.2/88.0/91.0/93.0 & & 86.2/88.0/91.0/93.0  \\
        \cline{2-7}
         & \multirow{2}{*}{n24jp01c} & \multirow{2}{*}{2024 Nov. 19} & \multirow{2}{*}{8} & 43.1/44.0/45.5/46.5 & \multirow{2}{*}{128} &  \\ 
        & & & & 86.2/88.0/91.0/93.0 & & 86.2/88.0/91.0/93.0  \\
        \cline{4-6}
         & \multirow{2}{*}{n24jp01d} & \multirow{2}{*}{2024 Nov. 20} & \multirow{2}{*}{8} & 35.6/36.1/37.1/38.1 & \multirow{2}{*}{128} & 106.8/108.3/111.3/114  \\ 
        & & & & 106.8/108.3/111.3/114.0 & &  \\
        \cline{2-7}
         & \multirow{2}{*}{n24jp01e} & \multirow{2}{*}{2024 Dec. 27} & \multirow{2}{*}{8} & 43.1/44.0/45.5/46.5 & \multirow{2}{*}{128} & 43.1/44.0/45.5/46.5 \\ 
        & & & & 86.2/88.0/91.0/93.0 & & 86.2/88.0/91.0/93.0  \\
        \hline
    \end{tabular}
\tablecomments{Columns indicate (1) the experiment code, (2) observing date, (3) recording rate, (4) all frequencies used for the observations, (5) bandwidth for each frequency, and (6) frequencies used for analysis. Data were excluded if polarization was not reliably detected, if no reference EVPA was available for absolute EVPA calibration, or if any other instrumental issues were present.}
\end{table*}

\section{Observations and data reduction} \label{sec:Data}

\subsection{KVN 43--141 GHz}
To reliably measure the RM in 3C~84, it is essential to conduct quasi-simultaneous multi-frequency polarimetric observations at millimeter wavelengths (e.g., above 43~GHz). The Korean VLBI Network (KVN) is uniquely suited for this purpose, as it offers simultaneous multi-frequency capability at high frequencies. It consisted of three telescopes, KVN Yonsei (YS), KVN Ulsan (US), and KVN Tamna (TN), until 2024, when a fourth telescope, KVN Pyeongchang (PC), joined the operation. 

We monitored 3C~84 with the KVN from 2018 February to 2024 December. Most of the data were obtained as part of the Plasma-physics of Active Galactic Nuclei (PAGaN) blazar monitoring project, a KVN Key Science Program started in 2017 \citep{park18}. The instrumental and frequency setups were updated over time as improved observational capabilities became available, as described below (see Table~\ref{tab:kvn_summary} for a summary).

-- Setup~1 (3 telescopes with 1 and 8~Gbps): Five of the seven epochs obtained in 2018 and 2019 were recorded with Mark5B at a recording rate of 1~Gbps. The received signals were digitized into 2-bit data streams and divided into four sub-bands (IFs) of 16~MHz bandwidth for both right circular polarization (RCP) and left circular polarization (LCP) receivers and for two different frequency bands ($\rm 16~MHz\times4~IFs\times2~pol.\times2~bands.$). The four 16~MHz IFs are contiguous in frequency, with no gaps between them, and are therefore quoted as a total bandwidth of 64~MHz per frequency band in Table~\ref{tab:kvn_summary}.

The other two epochs obtained on 2018 February 20 and 21 were recorded with Mark6 at a recording rate of 8~Gbps. The Fila10G interface was used to convert the 2-bit digitized data into the VLBI Data Interchange Format (VDIF) and send them to the Mark6. As there is no digital filtering function in Fila10G, the data obtained during this period consists of a single IF with a bandwidth of 512~MHz for both circular polarization receivers and for two different frequency bands ($\rm512~MHz\times1~IFs\times2~pol.\times2~freq$.). 

As these setups allow simultaneous observations in two frequency bands in dual-polarization mode, we conducted 2--3 consecutive 24-hour observing runs to secure adequate frequency coverage and spacing (e.g., 86, 94, 129, 139, and 141~GHz). This frequency coverage is essential for reliably measuring the RM in 3C~84, which is expected to exceed $10^5$ rad m$^{-2}$ \citep{trippe12, plambeck14, nagai17}. 

-- Setup~2 (3 telescopes with 16~Gbps): A new wide-band VLBI sampler, OCTAD, was installed at KVN in 2020. It consists of four analog-to-digital converters (ADC), four digital signal processing (DSP) modules, and a VDIF formatter. The ADCs digitize the incoming intermediate-frequency (IF) signals, which are then digitally down-converted and filtered in the DSP modules. The processed data are converted in VDIF and output as four 10GbE streams. 
By using both Fila10G and OCTAD, we obtained polarization data in all four frequency bands simultaneously, with two frequency bands allocated to each of Fila10G and OCTAD. For the OCTAD-equipped bands, we used the $\rm 128~MHz\times16~IF$ mode, which assigns four IFs to both RCP and LCP receivers in each of the two frequency bands. These four IFs can be spaced up to $\sim$7~GHz within each band. For the other two bands, Fila10G was used with $\rm 512~MHz\times1~IF$ per polarization. As a result, we simultaneously obtained EVPAs at 10 different frequencies: 21.55, 43.1, 44, 45.5, 46.5, 86.2, 88, 91, 93, and 129~GHz (OCTAD: $\rm 128~MHz\times4~IFs\times2~pol.\times2~bands.$, Fila10G: $\rm 512~MHz\times1~IF\times2~pol.\times2~bands$.). This represents a significant improvement over earlier setups, which only allowed polarization measurements at two frequencies simultaneously. This capability is particularly powerful for measuring the large RM in 3C~84, as the short frequency intervals enable us to reliably solve the n$\pi$ ambiguity between EVPAs at adjacent frequency pairs. To take advantage of this capability, we began recording 16~Gbps data with Mark6 in 2022 September.

\begin{figure}[t!]
\centering 
\includegraphics[width=0.47\textwidth]{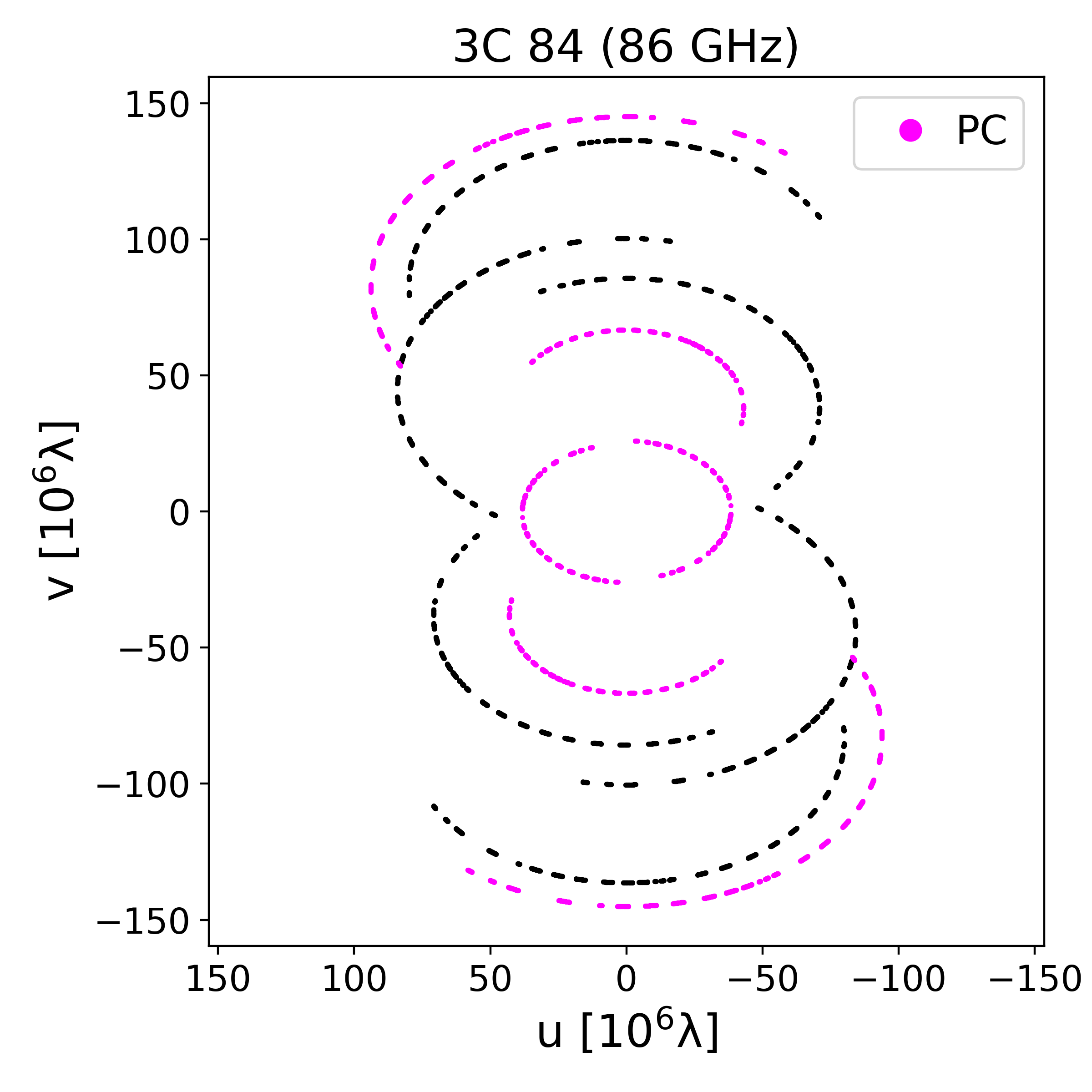}
\caption{The $(u,v)$ coverage of the KVN for 3C~84 at 86~GHz on 2024 November 8–9. Baselines involving the Pyeongchang (PC) are highlighted in magenta.
\label{fig:uv_ekvn}}
\end{figure}

-- Setup~3 (4 telescopes with 8~Gbps): The fourth telescope of the KVN, Pyeongchang (PC), began participating in KVN VLBI observations in the 2024B semester (Figure~\ref{fig:uv_ekvn}). This not only doubles the number of baselines but also enables amplitude self-calibration, thereby significantly improving the sensitivity and accuracy of the total intensity and linear polarization maps. 
We conducted three observing sessions with the four telescopes in 2024 November and December. All three sessions included a one-day observation in the Q and W bands at the following frequencies: 43.1, 44, 45.5, 46.5, 86.2, 88, 91, and 93~GHz (OCTAD: $\rm 128~MHz\times4~IFs\times2~pol.\times2~bands.$). In the second session, we carried out an additional one-day observation with the following frequencies: 35.8, 36, 37, 38, 106.8, 108, 111, and 114~GHz. 

\begin{table*}[ht]
    \centering 
    \caption{KVN polarization calibration details \label{tab:kvn_cal}}
    \begin{tabular}{ccccc} 
        \hline
        Exp. code & Used Freq. [GHz] & D-term Cal. & EVPA Cal. & $\sigma_{\chi}$ [$^\circ$] \\
        (1) & (2) & (3) & (4) & (5)  \\
        \hline
        p18st01b & 139 & OJ~287, 3C~84 & OJ~287, 3C~279 & 2.20  \\
        n18jp01a & 86/129 & OJ~287, 3C~84 & OJ~287, 3C~279 & 0.93/3.01 \\
        n18jp01b & 94/141 & OJ~287, 3C~84 & OJ~287, 3C~279 & 0.71/2.95 \\
        \hline
        p18st01o & 86 & OJ~287, 3C~84 & OJ~287 & $\langle \sigma_{\chi,{\rm W}} \rangle$ \\
        p18st01p & 94 & OJ~287, 3C~84 & OJ~287 & $\langle \sigma_{\chi,{\rm W}} \rangle$  \\
        \hline
        p19st01g & 86 & OJ~287, 3C~84 & OJ~287 & $\langle \sigma_{\chi,{\rm W}} \rangle$ \\
        p19st01h & 94 & OJ~287, 3C~84 & OJ~287 & $\langle \sigma_{\chi,{\rm W}} \rangle$ \\
        \hline
        \multirow{2}{*}{p21st02e} & 86.2/86.3/92.0/92.1 & OJ~287, 3C~84, 0003-066 & OJ~287, 3C~279 & 0.73/0.66/1.60/1.21 \\
        & 129.3 & OJ~287, 3C~84 & OJ~287 & $\langle \sigma_{\chi,{\rm D}} \rangle$ \\ 
        \hline
        p22st02b & 43.1/43.2/46.0/46.1 & OJ~287, 3C~84 & OJ~287, 3C~279 & 0.95/0.42/2.05/3.25 \\
        \hline
        \multirow{2}{*}{p22stf02} & 43.1/43.2/46.0/46.1 & OJ~287, 3C~84 & OJ~287, 4C~39.25 & 1.04/1.64/2.47/2.34 \\
        & 86.2/86.3/92.0/92.1 & OJ~287, 3C~84, BL Lac & OJ~287, BL Lac & 0.28/0.39/0.57/0.19 \\
        \hline
        \multirow{2}{*}{p22st02d} & 43.1/44.0/45.5/46.5 & OJ~287, 3C~84 & OJ~287, 3C~279 & 0.48/0.21/0.08/0.83 \\
        & 86.2/88.0/91.0/93.0 & OJ~287, 3C~84, BL Lac & OJ~287, 3C~279 & 0.59/0.44/0.44/0.66 \\
        \hline
        \multirow{2}{*}{p22st02e} & 43.1/44.0/45.5/46.5 & OJ~287, 3C~84 & OJ~287, 3C~345 & 0.18/0.87/0.55/0.03 \\
        & 86.2/88.0/91.0/93.0 & OJ~287, 3C~345, BL Lac & OJ~287, 3C~345 & 0.56/0.61/0.63/0.97 \\
        \hline
        t23jp01a & 86.2/88.0/90.0/92.0 &  OJ~287, 3C~84 & OJ~287 & $\langle \sigma_{\chi,{\rm W}} \rangle$ \\
        \hline
        \multirow{2}{*}{n24jp01a} & 43.1/44.0/45.5/46.5 & OJ~287, NRAO~150 & OJ~287, 3C~279, BL Lac & 0.35/1.00/0.35/0.43 \\ 
        & 86.2/88.0/91.0/93.0 & OJ~287, 3C~84 & OJ~287 & $\langle \sigma_{\chi,{\rm W}} \rangle$ \\
        \hline
        n24jp01c & 86.2/88.0/91.0/93.0 & 3C~84, NRAO~150, BL~Lac & OJ~287, 3C~345 & 0.56/0.91/1.00/1.28 \\
        n24jp01d & 106.8/108.3/111.3/114.0 & 3C~84, OJ~287, NRAO~150, BL~Lac  & OJ~287, 3C~345 & 1.09/2.17/2.16/1.78 \\
        \hline
        \multirow{2}{*}{n24jp01e} & 43.1/44.0/45.5/46.5 & OJ~287, 3C~84 & OJ~287, 3C~279 & 3.16/2.75/2.11/1.95 \\ 
        & 86.2/88.0/91.0/93.0 & OJ~287, 3C~84 & OJ~287, 3C~279 & 0.89/2.47/0.52/2.54 \\
        \hline
    \end{tabular}
\tablecomments{Columns indicate (1) the experiment code, (2) observing frequencies, (3) D-term calibrators, (4) absolute EVPA calibrators, and (5) uncertainties in the absolute EVPA calibration at each frequency. When only a single EVPA calibrator is available, we adopted the mean EVPA calibration uncertainty in the corresponding frequency band: $\langle \sigma_{\chi,{\rm W}} \rangle = 0.86^\circ$ in the W band and $\langle \sigma_{\chi,{\rm D}} \rangle = 2.98^\circ$ in the D band.}
\end{table*}

We performed data reduction using the NRAO’s
Astronomical Image Processing System \citep[AIPS;][]{greisen03}. 
We applied the Frequency Phase Transfer (FPT) technique to calibrate rapid tropospheric water-vapor fluctuations whenever applicable \citep{rioja11, algaba15}. Phase solutions were first obtained at lower frequencies (e.g., 22, 43~GHz) and then transferred to calibrate tropospheric phase errors at higher frequencies (e.g., 86, 129~GHz). For the 8/16~Gbps observations, phase solutions derived in the Q band (e.g., 43.1, 44, 45.5, 46.5~GHz) were transferred to the W band (e.g., 86.2, 88, 91, 93~GHz). Specifically, we performed global fringe fitting in the K/Q bands using short solution intervals of 12--15s and transferred these fringe solutions to the W/D bands. This procedure effectively corrects rapid tropospheric fluctuations at higher frequencies, leaving only slower residual errors, such as those introduced by the ionosphere. As these residual errors have much longer coherence times, we performed global fringe fitting at higher frequencies with longer solution intervals of 3 minutes, which increases the fringe detection rate. The CLEAN imaging, including phase and amplitude self-calibration, was performed using the Difmap package \citep{shepherd97}. 

The instrumental polarization (D-term) calibration was performed using a calibration pipeline, \texttt{GPCAL} \citep{park21_gpcal, park23_freq, park23_time}. This allows us to significantly improve the accuracy of D-term calibration compared to the AIPS task \texttt{LPCAL} by using multiple calibrators simultaneously and not being limited by the similarity approximation \citetext{e.g., \citealp{park21_m87}}. We primarily used OJ~287 and 3C~84 as D-term calibrators. OJ~287 is compact and strongly polarized, while 3C 84 can be reasonably assumed to be an effectively unpolarized source \citep[e.g.,][]{kim19}, consistent with its integrated fractional polarization being below 1\% across the KVN frequency bands. The D-terms of the KVN are summarized in Appendix~\ref{app:dterm}.

For the absolute EVPA calibration, we performed the KVN single-dish observations close in time (typically within 1--2 days) to the VLBI observations \citetext{see \citealp{kam23} for details}. Each KVN telescopes can perform observations in two different frequency bands simultaneously in dual-polarization mode \citep{lee11}. YS and TN were primarily used for observations in the W/D and K/Q bands, respectively, while US was used in either K/Q bands or W/D bands given the weather condition of the other two station. We mainly observed OJ~287 and 3C~279 as the calibrators, considering that they are typically compact to KVN across all frequency bands and exhibit strong polarization. We reduced the single-dish data as described in \citealt{kang15} and \citealt{kam23}. 

When KVN single-dish data in the W/D band were unavailable, we utilized the Atacama Large Millimeter/submillimeter Array (ALMA) polarimetric monitoring data reduced with the Analytic Matrix for ALMA POLArimetry (AMAPOLA) pipeline \citep{kameno23}. The reliability of the AMAPOLA measurements has been validated through comparison with independent observations from other instruments \citep{kam25}. The uncertainties arising from the absolute EVPA calibration $\sigma_{\chi}$ are summarized in Table~\ref{tab:kvn_cal}. 

Measuring the uncertainty in polarization is not straightforward. The EVPA uncertainty arises from several sources: random noise, D-term error, CLEAN error, and the absolute EVPA calibration error. To estimate the random, D-term, and CLEAN contributions together, we used the rms noise measured at the center of the Stokes $Q$ and $U$ residual maps, where the source is located. Since the on-source rms is typically 2–7 times larger than the off-source rms, this central rms appears to reflect those uncertainties. Using these rms values, the EVPA uncertainty derived from the maps was calculated following \citealt{hovatta12} as
\begin{equation}
\sigma_{\chi,{\rm map}} = \frac{1}{2}
\frac{\sqrt{\sigma_U^2 Q^2 + \sigma_Q^2 U^2}}{Q^2 + U^2},
\end{equation}

\noindent where $\sigma_Q$ and $\sigma_U$ are the rms noises measured at the center of the residual $Q$ and $U$ images, respectively. The final EVPA uncertainty $\sigma_{\rm EVPA}$ was then obtained by combining the uncertainty from the maps $\sigma_{\rm \chi,map}$ and the absolute EVPA calibration uncertainty $\sigma_{\chi}$ in quadrature: 
\begin{equation}
    \sigma_{\rm EVPA}=\sqrt{\sigma_{\rm \chi,map}^2+\sigma_{\chi}^2}.
\end{equation}

To verify whether the estimated EVPA uncertainties are reasonable, we compared the EVPA values measured in adjacent frequencies with sufficiently small frequency separations, where Faraday rotation is negligible (Appendix~\ref{app:evpa}). The observed EVPA differences between these frequency pairs are consistent within the estimated uncertainties, confirming that our uncertainty estimates reliably reflect the true measurement accuracy.

The uncertainty in fractional polarization described in \citealt{hovatta12} formally applies to the ideal case when the flux densities are accurately known. In practice, however, the absolute uncertainty in the total and polarized flux densities in VLBI observations is difficult to quantify precisely and are commonly assumed to be 5--10\% \citep[e.g.,][]{lister05, lu12}. These systematic uncertainties typically exceed the formal errors derived from the map noise. We therefore adopt a conservative uncertainty of 10\% of the measured fractional polarization as the polarization measurement uncertainty. 

We briefly summarize why some KVN observations listed in Table~\ref{tab:kvn_summary} were not included in the RM analysis. The 22~GHz data were not used because polarization is rarely detected at this frequency, most likely due to strong Faraday depolarization. Even when weak polarization is present, only a single frequency is available within the K band, and the large frequency gap to higher frequency bands prevents solving the $n\pi$ ambiguity and obtaining a reliable RM measurement. Similarly, the 35.6–38.1~GHz data obtained on 2024 November 19 were excluded because polarization was not reliably detected, likely because Faraday depolarization remains significant at these frequencies. These low-frequency data were therefore primarily used for FPT.

The 43~GHz observations obtained between 2018 February and 2019 May were also excluded because only a single frequency was available within the Q band. The large frequency separation from higher frequency bands prevents solving the $n\pi$ ambiguity and deriving a reliable RM.

Several 129~GHz datasets obtained between 2022 September and 2023 February were not used because either no reliable single-dish EVPA reference was available for absolute EVPA calibration or the data quality was insufficient to obtain reliable RM measurements.

\subsection{VLBA at 43 GHz}
We analyzed 18 epochs of the BU data obtained between 2015 December and 2017 May, during which the polarized emission is consistently detected at C3 in all four IFs. 3C~84 was added to the source list of the BU program in 2010 November and has been monitored nearly every month along with other sources. The on-source time for 3C~84 was 20--30 minutes per integration (7--10 scans of approximately 2--3 minutes each), and the scans were distributed over a 7--8 hour time span to cover a sufficient (u,v) coverage. During this period, the data were recorded with four 64~MHz IFs, yielding a total bandwidth of 256~MHz. This enabled in-band RM measurements across the full 256~MHz bandwidth \citep[e.g.,][]{nagai17}. The frequency of the four IFs are 43.008, 43.088, 43.152, and 43.216~GHz from 2015 December to 2017 April, and 43.693, 43.761, 43.821, and 43.883~GHz in 2017 May. We applied uniform weighting, leading to the average beam size being $\rm 0.30\times0.14~mas$, with a position angle (P.A.) of -3.8$^\circ$. The EVPA uncertainty was estimated following the formula suggested for the VLBA, adopting the D-term uncertainty as 5\% \citep{hovatta12, kim19}. The uncertainties in the fractional polarization measurements with the VLBA were also assumed to be 10\% of the measured fractional polarization.  

\begin{figure*}[t!]
\centering 
\includegraphics[width=\textwidth]{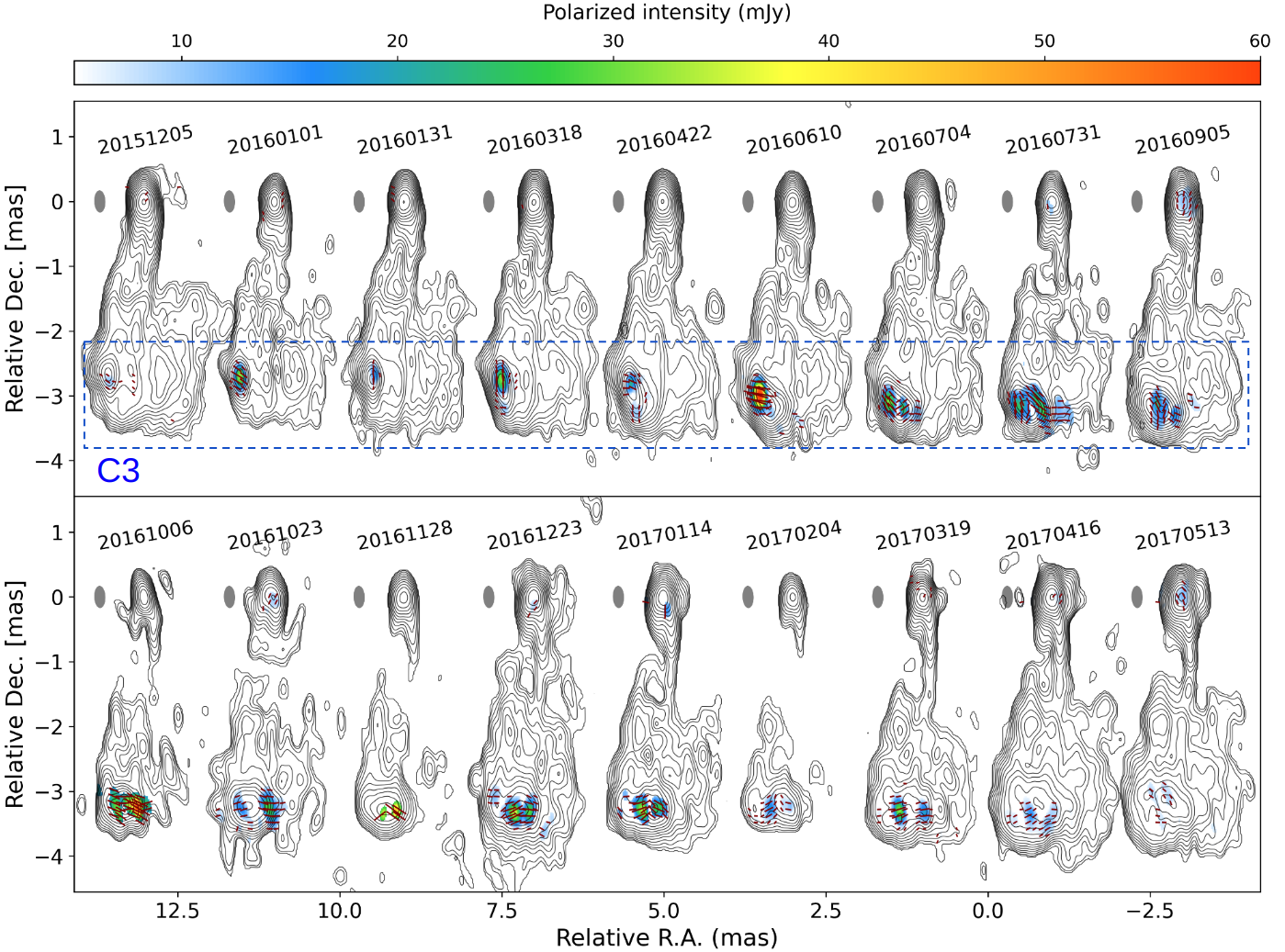}
\caption{BU images of 3C~84 from 2015 December to 2017 May. Contours show the total intensity, and colors represent the polarized intensity in units of mJy. The observing epochs, displayed above each image, are given in yyyymmdd format (e.g., 20151205). 
\label{fig:vlba_img1}}
\end{figure*}

\subsection{High Sensitivity Array at 8~GHz}

We observed a set of low-luminosity AGNs with the high sensitivity array (HSA) at multiple frequencies to study their jet collimation and acceleration processes \citep[e.g.,][]{park21_ngc315, park24_ngc315, yi24}. 3C~84 was observed as a calibrator. In this study, we focus on the 8~GHz band data of 3C~84 to investigate its linear polarization and associated RM in its southern lobe.

The data were recorded in both right- and left-hand circular polarizations with two-bit quantization across four IFs at starting frequencies of 8.112, 8.240, 8.368, and 8.496~GHz. We used the digital downconverter (DDC) system at a recording rate of 4~Gbps, yielding a total bandwidth of 512~MHz per polarization. The total observation time was $\approx 11$~hours; 3C~84 was observed in 12 scans with an on-source time of 2~minutes per scan.

All ten VLBA stations, the phased VLA, and the Effelsberg 100\,m telescope participated in the observations. Weather conditions were good, and no major technical issues occurred at any station. Notably, Effelsberg used a wideband C/X-band receiver with linear feeds. We therefore converted the mixed-polarization visibilities involving Effelsberg to a circular basis using \texttt{PolConvert} \citep{MartiVidal2016}. Gain ratio solutions between the two polarizations were derived from a short segment of the bright calibrator 3C~279 and applied to all Effelsberg visibilities.

Standard data reduction was carried out using the NRAO AIPS package \citep{greisen03}, following the methods of our previous works \citep[e.g.,][]{park26}. We first performed an iterative imaging and self-calibration process using Difmap \citep{shepherd97} to construct the initial total intensity models. Subsequently, we applied an additional amplitude and phase self-calibration using the AIPS task \texttt{CALIB} with a 10\,s solution interval. This step was necessary to correct for residual gain ratio offsets between the two polarizations that are not fully addressed by the Difmap imaging routine. Following this external calibration, we resumed the iterative imaging process in Difmap until the model $\chi^2$ minimized and stabilized.

\begin{figure*}[t!]
\centering 
\includegraphics[width=\textwidth]{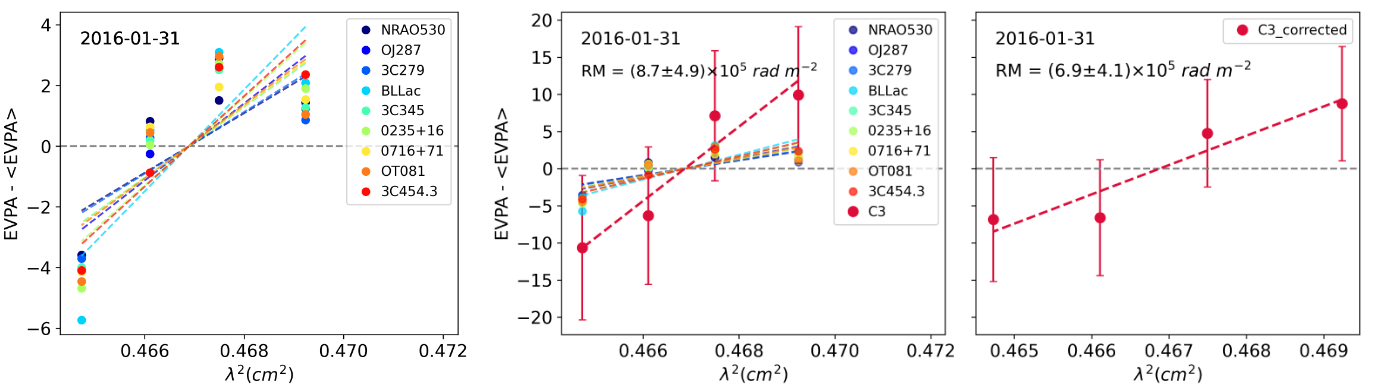}
\caption{Calibration procedure for the relative EVPA (defined as the EVPA in each IF relative to the average EVPA across the four IFs) of C3 obtained with the BU data. \textit{Left}: Relative EVPA values of the comparison sources in each IF, $\chi_{\rm cal,i}$, whose intrinsic RM are expected to be significantly smaller than $\rm 10^5\ rad\ m^{-2}$ at 43~GHz. 
\textit{Middle}: The relative EVPA values of C3 in each IF, $\chi_{\rm C3,i}$, shown together with those of the comparison sources. \textit{Right}: Relative EVPA values of C3 after subtracting the mean relative EVPA values from the comparison sources in each IF, $\chi'_{\rm C3,i} = \chi_{\rm C3,i} - \langle\chi_{\rm cal,i}\rangle$.
\label{fig:evpa_correction}}
\end{figure*}

The D-term was determined and corrected using the \texttt{GPCAL} pipeline \citep{park21_gpcal}. We used 3C~84, assumed to be unpolarized, and the compact polarized source J0205+3212 as D-term calibrators. 

For the EVPA calibration---specifically to correct the gain phase offset between the two polarizations at the reference antenna---we employed VLA data acquired concurrently with the HSA observations. We calibrated the absolute orientation by assuming that the integrated EVPA of J0205+3212 in our VLBI data is equivalent to the value derived from the VLA-only data. The uncertainties in the EVPA and fractional polarization are derived in the same manner as for the VLBA data.

\section{Results} \label{sec:analysis}

\subsection{VLBA in 2015--2017}

Polarization in 3C~84 was rarely detected with the VLBA at 43~GHz before 2015 \citep{nagai17}. After the emergence of a compact, bright hotspot in the eastern limb of C3 in late 2015 \citep{nagai17, kino18}, polarized emission was consistently detected around the hotspot from 2015 December to 2017 May (Figure~\ref{fig:vlba_img1}). 

In early 2016, the hotspot was compact, but it gradually became elongated, split into multiple structures, and shifted from the eastern limb to the southern part of C3 during mid-2016. It then settled in the southern region of C3 and faded after 2017 May (see also Figure~1 in \citealp{kam24}). During this evolution, twin polarized regions appeared at the eastern and western edges of the hotspot. These twin polarized regions disappeared in 2017 June, followed by the disruption of the hotspot \citep{kino21}. 

This polarization behavior at C3 can be explained by magnetic-field compression at shocks formed where the jet head interacts with the surrounding medium \citep{laing80, meisenheimer89}. The location of polarized emission within C3 depends on the direction and angle at which the jet impacts the surrounding medium \citep{kam24}. As the dominant jet-ambient interaction site at the jet head evolves with time, the structure and position of the polarized regions are therefore expected to vary accordingly.  
  


The 256~MHz bandwidth of the VLBA consists of four IFs, each with 64~MHz. This enables us to measure RM at C3 during this period, if it exceeds a few $10^5$ rad m$^{-2}$ \citep[e.g.,][]{nagai17}. However, a typical EVPA calibration uncertainty of $5^\circ\sim6^\circ$ across the 256~MHz bandwidth corresponds to an apparent RM of $\rm \sim2\times10^{5}\ rad\ m^{-2}$. This implies that even a small EVPA calibration uncertainty can produce a significant artificial RM of order $\rm 10^5\ rad\ m^{-2}$, comparable to or larger than the expected RM \citep[e.g.,][]{kim19}. 

The uncertainty in the EVPA introduced by the calibration process are systematically the same for all the sources monitored within the BU program, since the calibration process is identical for all sources (private communication with Svetlana Jorstad). Then the systematic uncertainty can be estimated using sources known to have the RM much smaller than $\rm \sim2\times10^{5}\ rad\ m^{-2}$. 

We selected 6$\sim$9 sources in each epoch that satisfy the following conditions: (i) RM over $\rm 1\times10^{5}\ rad\ m^{-2}$ has never been observed at 43~GHz or below, and (ii) polarization is detected at all four IFs. For each source, we obtained the EVPA at the polarized intensity peak and calculated the relative EVPA in each IF (relative to the average EVPA taken over four IFs). We found that the relative EVPA values in these sources are indeed correlated, suggesting that they are introduced by the common calibration process (Figure~\ref{fig:evpa_correction}). 

To correct this effect, we (i) calculate the average of the relative EVPA values in each IF obtained from the comparison sources, and (ii) subtract this average value from the relative EVPA of C3 in each IF. Although the absolute EVPA value is lost during this process, we can more accurately determine the intrinsic EVPA rotation as a function of frequency. This method is valid because our interest lies in understanding how much the EVPA rotates with frequency, rather than its absolute value. 

After correcting the systematic uncertainties in the EVPA, we adopted two methods to determine the RM of C3. First, we obtained the RM at the polarized intensity peak within C3, based on the assumption that this is the pixel where we expect the highest signal-to-noise ratio (SNR) of the polarized intensity, and thus the most reliable RM measurement. As can be seen in Figure~\ref{fig:vlba_img1}, however, the structure of the polarized region in C3 is complex and sometimes even exhibits two separate polarized features. To determine the RM value that well represents this entire region, we also computed the weighted mean RM over this region, following \citealt{park19_rm}. We first calculated the RM value at each pixel, then obtained the weighted average (RM$_{\rm weighted}$) and its standard error. We then multiplied this standard error by $\sqrt{n\Sigma_{\rm FWHM}/\Sigma_{\rm RM}}$, where $n$ is the number of pixels where the RM is detected, $\Sigma_{\rm RM}$ is the size of the region where the RM is detected, and $\Sigma_{\rm FWHM}$ is the full width at half maximum (FWHM) of the synthesized beam. All the RM values using these two methods are summarized in Table~\ref{tab:rm_bu}. The average and standard deviation of the RM from these two approaches are $\rm \langle RM_{\rm peak}\rangle = (3.9\pm0.6)\times10^5\ rad\ m^{-2}$ and $\rm \langle RM_{\rm weighted}\rangle = (3.4\pm0.4)\times10^5\ rad\ m^{-2}$, respectively, which are consistent within the uncertainty. 

\begin{table}[t!]
\begin{center}
    \caption{RM at C3 from the BU data. \label{tab:rm_bu}}
    \begin{tabular}{ccc}
        \hline 
        Epoch & RM$_{\rm peak}$ & RM$_{\rm weighted}$  \\
        (yyyy-mm-dd) & ($\rm 10^5\ rad\ m^{-2}$) &  ($\rm 10^5\ rad\ m^{-2}$) \\
        \hline 
        2015-12-05 & 2.5$\pm$5.5 & 1.8$\pm$6.4 \\ 
        2016-01-01 & 2.5$\pm$2.3 & 2.4$\pm$1.9 \\ 
        2016-01-31 & 6.9$\pm$4.1 & 6.8$\pm$4.2 \\ 
        2016-03-18 & 2.4$\pm$4.2 & 3.1$\pm$2.5 \\ 
        2016-04-22 & 0.7$\pm$4.6 & 2.3$\pm$2.5 \\ 
        2016-06-10 & 1.9$\pm$1.7 & 1.3$\pm$1.2 \\ 
        2016-07-04 & 1.9$\pm$2.7 & 2.3$\pm$1.6 \\ 
        2016-07-31 & 2.8$\pm$2.3 & 2.3$\pm$1.0 \\ 
        2016-09-05 & 3.4$\pm$4.1 & 4.8$\pm$2.2 \\ 
        2016-10-06 & 7.7$\pm$2.4 & 6.1$\pm$1.9 \\ 
        2016-10-23 & 6.6$\pm$2.2 & 5.7$\pm$1.4 \\ 
        2016-11-28 & 5.5$\pm$1.8 & 5.5$\pm$1.6 \\ 
        2016-12-23 & 2.5$\pm$1.7 & 2.9$\pm$1.0 \\ 
        2017-01-14 & 4.6$\pm$2.2 & 4.6$\pm$1.6 \\ 
        2017-02-04 & 2.9$\pm$2.4 & 2.7$\pm$2.0 \\ 
        2017-03-19 & 4.5$\pm$1.5 & 3.9$\pm$1.2 \\ 
        2017-04-16 & 4.9$\pm$3.5 & 3.4$\pm$1.9 \\ 
        2017-05-13 & 6.9$\pm$5.6 & 4.4$\pm$5.4 \\ 
        \hline 
        Average & 3.9$\pm$0.6 & 3.4$\pm$0.4 \\ 
        \hline
    \end{tabular}
\end{center}
\end{table}

\begin{figure*}[t!]
\centering 
    \includegraphics[width=0.95\textwidth]{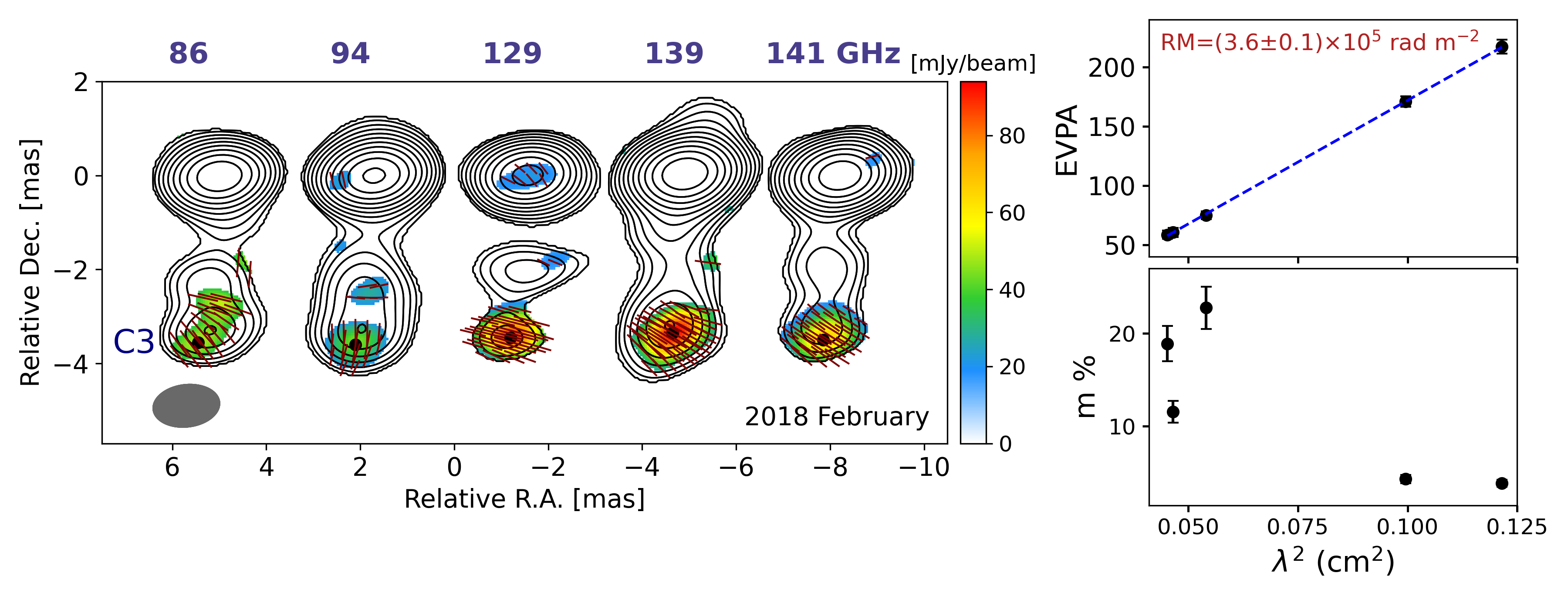}
    \caption{\textit{Left}: KVN images of 3C~84 at 86, 95, 129, 139, and 141~GHz obtained on 2018 February 13--21. Contours and colors represent total and polarized intensity, and bars show the EVPA. All images are convolved with the beam size at the lowest frequency (86.2~GHz) \textit{Right}: The EVPA values at the polarized intensity peak within C3 are shown on the top. Blue dashed line shows the best fit to the EVPA values. The fractional polarization values at the same position are shown at the bottom. 
\label{fig:18feb}}
\end{figure*}

\begin{figure*}[t!]
\centering 
    \includegraphics[width=0.95\textwidth]{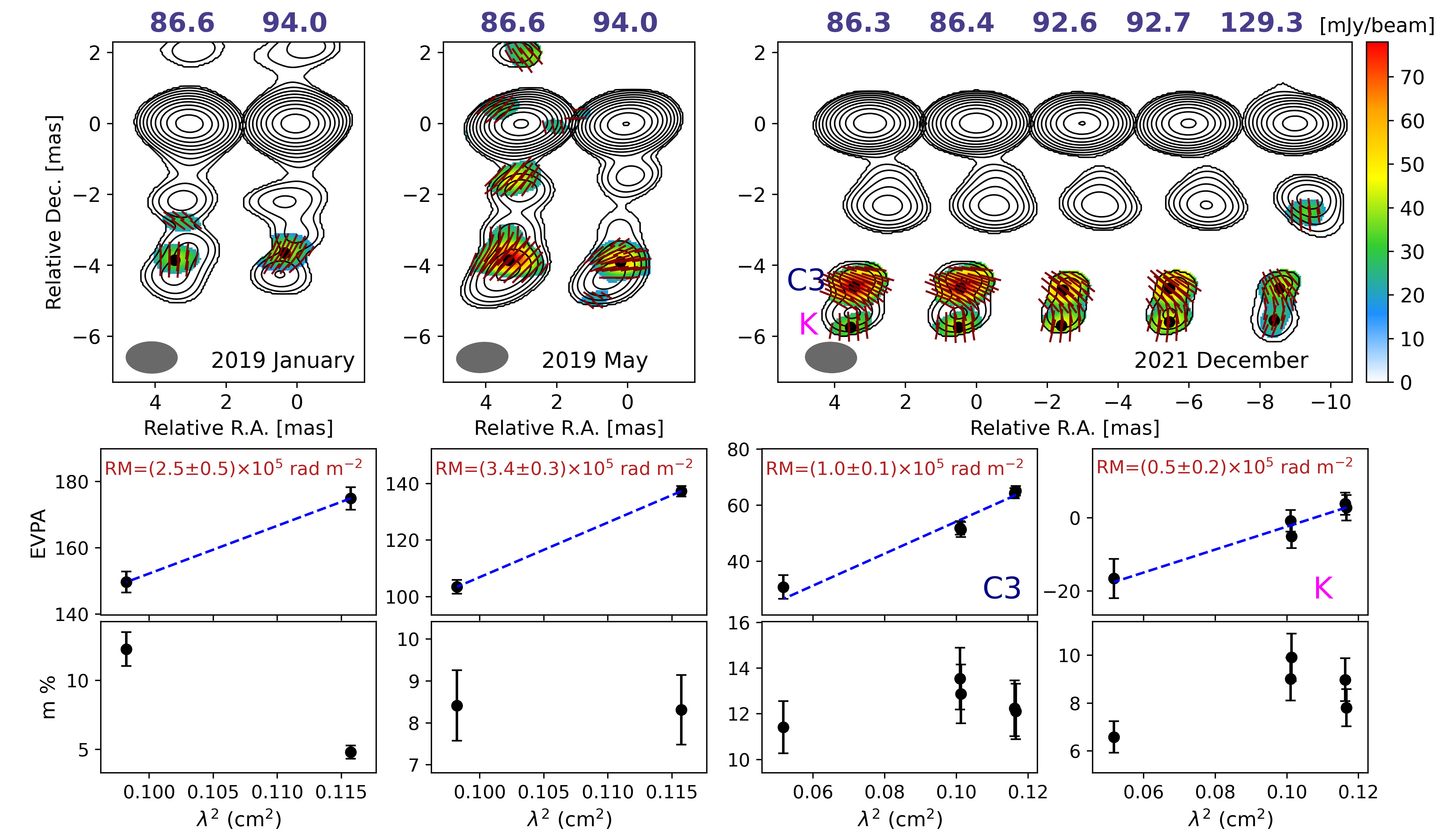}
    \caption{Same as Figure~\ref{fig:18feb}. \textit{Left and middle}: Observations at 86.6 and 94.0~GHz obtained on 2019 January 2--3 and May 2--3, respectively. \textit{Right}:  Observations at 86.3, 86.4, 92.6, 92.7, and 129.3~GHz obtained on 2021 December 1. 
\label{fig:19jan21dec}}
\end{figure*}

\subsection{KVN in 2018--2024}
After 2017 May, polarization was no longer reliably detected at C3 in the BU 43 GHz data for the next few years. In contrast, polarized emission from C3 was consistently detected with the KVN at 43--141~GHz between 2018 February to 2024 December, because the KVN provides access to higher observing frequencies, where Faraday depolarization is smaller than at 43 GHz (see Appendix~\ref{app:comparison} for comparison of the KVN and BU data). 

Figure~\ref{fig:18feb} shows a clear detection of polarization at C3 across all frequencies from 86 to 141~GHz in 2018 February. To determine the RM at C3, we first convolved all images with the beam at the lowest frequency (e.g., 86~GHz in 2018 February) and then aligned the images by shifting the brightest pixel in the core to the map center at (0, 0). We did not apply additional core-shift corrections because the expected core shift is smaller than the position uncertainties in the KVN images. The position uncertainty in VLBI images is typically taken as one-fifth of the synthesized beam \citep[e.g.,][]{lister09, lister16}. For 3C~84 observed with the KVN, the typical beam size is 1.6$\times$0.9 mas at 86~GHz, implying uncertainties of at least 0.2~mas. In contrast, the expected core-shift between 86 and 141~GHz is at least 2--3 times smaller \citep{giovannini18, paraschos21, savolainen23, park24_3c84}, and therefore negligible compared to the uncertainty. Consequently, we aligned the images by centering the brightest pixel in the core, and measured the EVPA at the polarized intensity peak closest to the total intensity peak within C3 at each frequency. We measured the separation between the polarized-intensity peak in C3 and the total-intensity peak in the core at each frequency, and adopted the mean of these separations as the representative distance at which the RM was measured. The uncertainty in this distance was estimated as the quadrature sum of two contributions: (i) 20\% of the synthesized beam minor axis, which is oriented nearly parallel to the jet direction in most cases, and (ii) the standard deviation of the measured peak separations across frequency.

\begin{figure*}[t!]
\centering 
    \includegraphics[width=\textwidth]{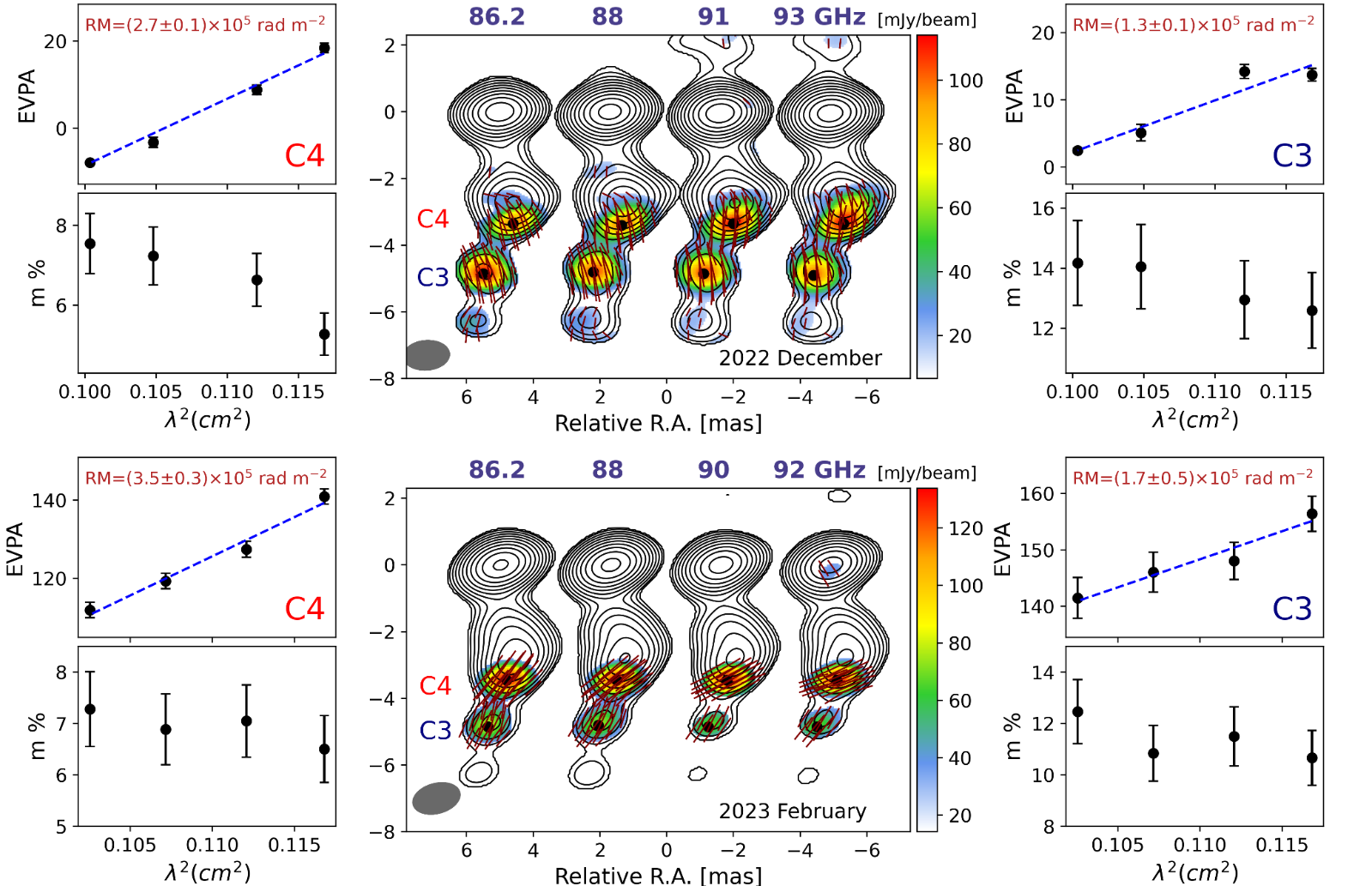}
    \caption{Same as Figure~\ref{fig:18feb}. Observations at 86.2, 88.0, 91.0, and 93.0~GHz on 2022 December 27 (top) and 2023 February 5 (bottom). The EVPA, RM, and fractional polarization measured at the polarized intensity peaks of C4 are shown in the left panels, and those of C3 are shown in the right panels. 
\label{fig:22dec23feb_w}}
\end{figure*}

The right panel of Figure~\ref{fig:18feb} shows the EVPA and fractional polarization at C3 in 2018 February. The EVPA rotates by more than 150$^\circ$ between 86 and 141~GHz while showing a good $\lambda^2$ fit. This suggests that the Faraday rotation is external to the jet, since internal Faraday rotation is not expected to produce rotations larger than $\sim45^\circ$ with such a good $\lambda^2$ fit \citep{burn66, cioffi80}. The EVPA rotation corresponds to the RM of $\rm (3.6\pm0.1)\times10^5\ rad\ m^{-2}$. The fractional polarization at the same position is about 3\% at 86 and 94~GHz and exceeds 10\% at higher frequencies. Such an increasing trend in fractional polarization with frequency is consistent with Faraday depolarization. 

Polarized emission from C3 is also clearly detected in 2019 January and May (Figure~\ref{fig:19jan21dec}). The RM at C3 measured between 86.6 and 94~GHz is $\rm (2.5\pm0.5)\times10^5\ rad\ m^{-2}$ in 2019 January and $\rm (3.4\pm0.3)\times10^5\ rad\ m^{-2}$ in 2019 May. These values are consistent with the RM previously measured at C3 within 2--3 times the measurement uncertainty. 

In 2021 December, however, the RM at C3 shows a clear decrease (Figure~\ref{fig:19jan21dec}). In this epoch, the RM measured between 86.3, 86.4, 92.6, 92.7, and 129.3~GHz in 2021 December is $\rm (1.0\pm0.1)\times10^5\ rad\ m^{-2}$. In addition, a new polarized feature appears downstream of C3 at all observed frequencies. We refer to this downstream feature as component K. In higher-resolution VLBA images, this region contains multiple faster-moving subcomponents that split from C3 \citep{kam24}, and the KVN detection likely represents a blend of these subcomponents. As with C3, the polarization observed at component K can be explained by magnetic-field compression, as component K also propagates through the ambient medium. Notably, the RM at component K is $\rm (0.5\pm0.2)\times10^5\ rad\ m^{-2}$, smaller than the value at C3. This relation between C3 and K persists in subsequent epochs, even if polarization from component K gradually becomes fainter and becomes detectable only in the Q band (Appendix~\ref{app:kvn_images_2022}). 

\begin{figure*}[ht!]
\centering 
    \includegraphics[width=0.95\textwidth]{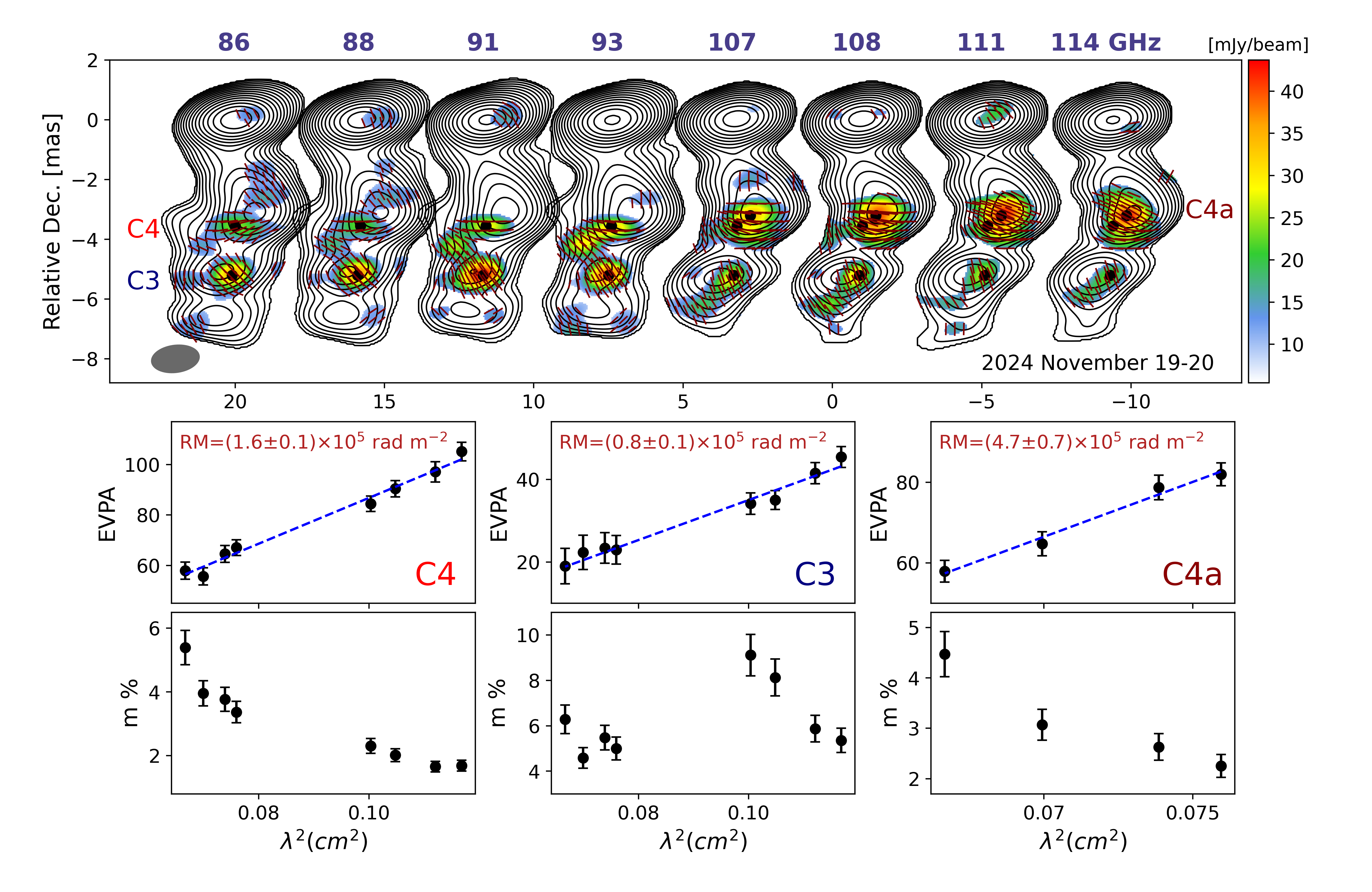}
    \caption{Same as Figure~\ref{fig:18feb}. Observations at 86.2, 88.0, 91.0, 93.0, 106.8, 108.3, 111.3, and 114.0~GHz on 2024 November 19--20. The EVPA, RM, and fractional polarization measured at the polarized intensity peaks of C3, C4, and C4a are shown in the bottom panels. 
\label{fig:nov2024_01cd}}
\end{figure*}

\begin{figure*}[ht!]
\centering 
    \includegraphics[width=0.75\textwidth]{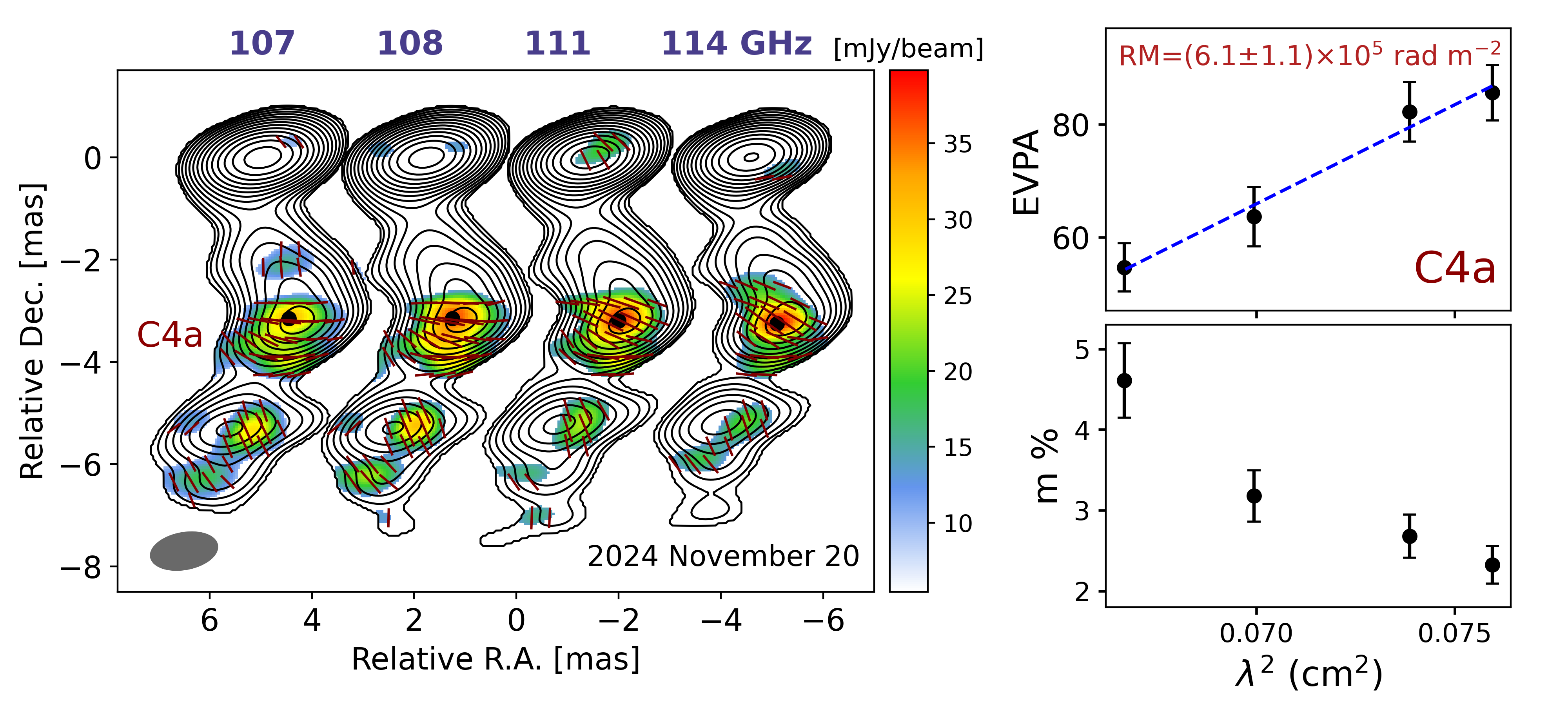}
    \caption{Same as Figure~\ref{fig:nov2024_01cd}, but for images at 106.8, 108.3, 111.3, and 114.0~GHz, all convolved with the 106.8~GHz beam.
\label{fig:nov2024_01d_w}}
\end{figure*}

One notable development observed since 2022 December is the emergence of a new polarized feature upstream of C3. This feature lies within the upstream component which we designate as C4, a localized bright feature located at $\sim$3~mas from the core in 2022 December (Figure~\ref{fig:22dec23feb_w}). C4 was first clearly identified in the KVN images in 2021 December at a distance of $\sim$2~mas from the core (Figure~\ref{fig:19jan21dec}). At that epoch, polarization at C4 was detected only at the highest frequency (129~GHz), while no polarization was detected at lower frequencies (86--93~GHz). As C4 moved outward to $\sim$3~mas in 2022 December and 2023 February, polarization became consistently detected, particularly in its southern part, across all observing frequencies from 86~GHz upward.

Similar to C3 and K, the polarization observed at C4 can be explained by magnetic-field compression caused by the interaction with the ambient medium. This interpretation is supported by the location of C4 at a region where the jet undergoes a pronounced change in direction. In VLBA 43~GHz images obtained in late 2021 (Figure~1 in \citealt{kam24}), the jet upstream of this location is oriented toward the southwest, while downstream of C4 it turns toward the southeast. This transition creates a convex bend in the jet structure at the position where C4 later emerges. Such bending is expected to enhance the interaction between the jet and its surroundings, providing favorable conditions for magnetic-field compression and the emergence of polarized emission.

The bending at the location of C4 may be part of a larger instability affecting the inner jet as a whole. Near the core, the jet was initially oriented toward the southeast in the early 2010s but gradually shifted toward the southwest by the early 2020s \citep{kam24}. At the same time, a downstream bend developed in which the jet turns from southwest toward southeast at the location where C4 later appears. These changes suggest that the inner jet is undergoing a progressive distortion, possibly driven by an instability in the jet flow \citep{hiura18, paraschos25}. As this instability-driven bending became more pronounced after 2021, the curvature of the jet at the C4 location increased, naturally leading to stronger interaction between the jet and the surrounding ambient medium and the enhanced polarized emission observed there\footnote{We note that $\gamma$-ray flares between 2022 November and 2023 January \citep{godambe24, cao25} coincide with the emergence of polarization at C4 in 2022 December and 2023 February.}. 

This scenario provides a natural explanation for the detection of polarization at C4. Notably, polarization becomes consistently detectable at C4 only from 2022 December onward, even though the component was already evident in 2021 December. This can be naturally explained by reduced Faraday depolarization at larger distances downstream (see Section~\ref{sec:Discussions_origin}).

The RM at C4 is $\rm (2.7\pm0.1)\times10^5\ rad\ m^{-2}$ in 2022 December and $\rm (3.5\pm0.3)\times10^5\ rad\ m^{-2}$ in 2023 February, both larger than that at C3 in the same epochs. The fractional polarization at C4 is 6--8\%, lower than 10--14\% observed at C3 during these observations.

In 2022 December, polarization was only faintly visible at component K, and in 2023 February even the total intensity was only marginally detected in the W band images. Although the polarized features associated with component K formally exceed the SNR threshold of 7, their positions are not consistent across frequencies and therefore cannot be regarded as reliable detections. We thus conclude that polarization is not robustly detected at component K in the W-band during these epochs (see also Appendix~\ref{app:kvn_images_2022} and Appendix~\ref{app:kvn_images_2024} for the detection of polarization at component K in the Q band).

In 2024 November and December, we conducted three observing epochs using four telescopes, including Pyeongchang. Each epoch, carried out on November~8, November~19, and December~27, consisted of simultaneous observations at eight frequencies, with four frequencies in the Q band (43.1, 44.0, 45.5, and 46.5~GHz) and four in the W band (86.2, 88.0, 91.0, and 93.0~GHz).

For the epoch on November 19, an additional one-day observation was conducted on November 20 to extend the frequency coverage. This epoch also employed eight simultaneous observing frequencies, with four frequencies in the Q band (35.5, 36, 37, and 38~GHz) and four in the W band (106, 108, 111, and 114~GHz). As a result, the combined dataset covers 35.5--46.5~GHz in the Q band and 86.2--114~GHz in the W band. While polarization was not reliably detected in the Q band during this epoch, it was clearly detected across all W-band frequencies. 

\begin{table}[t!]
\begin{center}
    \caption{RM measured with the KVN. \label{tab:rm_kvn}}
    \begin{tabular}{llccc}
        \hline 
        Obs. Date & Comp. & RM & Distance  \\
         & & [rad m$^{-2}$] & [mas] \\
        \hline 
        2018 Feb. 14--21 & C3 & $(3.6\pm0.1)\times10^5$ & $3.5\pm0.4$ \\ 
        2019 Jan. 2--3 & C3 & $(2.5\pm0.5)\times10^5$ & $3.8\pm0.4$ \\ 
        2019 May 2--3 & C3 & $(3.4\pm0.3)\times10^5$ & $3.9\pm0.4$ \\ 
        2021 Dec. 1 & C3 & $(1.0\pm0.1)\times10^5$ & $4.7\pm0.4$ \\ 
        & K & $(0.5\pm0.2)\times10^5$ & $5.7\pm0.4$ \\ 
        2022 Sep. 15 & C3 & $(1.9\pm0.2)\times10^5$ & $3.9\pm0.8$ \\
        & K & $(0.9\pm0.1)\times10^5$ & $6.0\pm0.8$ \\ 
        2022 Oct. 21 & C3 & $(1.0\pm0.2)\times10^5$ & $5.0\pm0.4$ \\ 
        & K & $(0.9\pm0.2)\times10^5$ & $6.0\pm0.8$ \\ 
        2022 Nov. 5 & C3 & $(1.3\pm0.2)\times10^5$ & $4.9\pm0.3$ \\ 
        & K & $(0.7\pm0.1)\times10^5$ & $6.2\pm0.7$ \\ 
        2022 Dec. 28 & C4 & $(2.7\pm0.1)\times10^5$ & $3.4\pm0.4$ \\ 
        & C3 & $(1.3\pm0.1)\times10^5$ & $4.9\pm0.4$ \\ 
        & K & $(0.5\pm0.1)\times10^5$ & $5.6\pm0.8$ \\ 
        2023 Feb. 5 & C4 & $(3.5\pm0.3)\times10^5$ & $3.5\pm0.4$ \\ 
        & C3 & $(1.7\pm0.5)\times10^5$ & $4.9\pm0.4$ \\ 
        2024 Nov. 8 & C4 & $(2.6\pm0.4)\times10^5$ & $3.8\pm0.4$ \\ 
        & C3 (W) & $(1.1\pm0.4)\times10^5$ & $5.2\pm0.4$ \\ 
        & C3 (Q) & $(1.5\pm0.1)\times10^5$ & $4.9\pm0.8$ \\ 
        & K & $(0.8\pm0.1)\times10^5$ & $7.4\pm0.9$ \\ 
        2024 Nov. 19--20 & C4a & $(6.1\pm1.1)\times10^5$ & $3.2\pm0.3$ \\
        & C4 & $(1.6\pm0.1)\times10^5$ & $3.4\pm0.4$ \\ 
        & C3 & $(0.8\pm0.1)\times10^5$ & $5.2\pm0.4$ \\ 
        2024 Dec. 27 & C4 & $(2.6\pm0.5)\times10^5$ & $3.8\pm0.4$ \\ 
        & C3 (W) & $(1.6\pm0.3)\times10^5$ & $5.3\pm0.4$ \\ 
        & C3 (Q) & $(1.4\pm0.1)\times10^5$ & $4.9\pm0.9$ \\ 
        & K & $(0.5\pm0.1)\times10^5$ & $6.8\pm0.9$ \\ 
        \hline 
    \end{tabular}
\end{center}
\end{table}

Figure~\ref{fig:nov2024_01cd} shows images at eight frequencies spanning 86.2--114~GHz, obtained on 2024 November 19 and 20. Polarized emission is consistently detected at components C3 and C4 across all frequencies, in agreement with previous epochs. A new feature revealed in these observations is the emergence of additional polarized emission within C4 at higher frequencies. At lower frequencies (86–93~GHz), the polarized-intensity peak is located in a downstream region of C4, and polarization from this region is detectable at all frequencies up to 114~GHz. At higher frequencies (106--114~GHz), however, the polarized-intensity peak shifts systematically upstream relative to its position at lower frequencies (86--93~GHz), due to the presence of this additional upstream polarized component. In addition, the EVPA distribution in this upstream region, as well as its frequency-dependent rotation, differs from that of the downstream polarized region, indicating distinct polarization properties between the two regions.

Motivated by these differences, we derived the RM separately for the two polarized regions. For the downstream region, identified with C4, the EVPA was measured at all frequencies from 86 to 114~GHz at the polarized-intensity peak defined at 86~GHz. For the upstream region, designated as C4a, the EVPA was measured only at 106--114~GHz, at the polarized-intensity peak identified at 106~GHz. The EVPA at C4 shows clear Faraday rotation with a good $\lambda^2$ fit across 86--114~GHz, corresponding to the RM of $\rm (1.6\pm0.1)\times10^5\ rad\ m^{-2}$. In contrast, the RM at C4a is $\rm (4.7\pm0.7)\times10^5\ rad\ m^{-2}$, nearly three times larger than that at C4. 

This difference is naturally explained by Faraday depolarization, under the assumption that the RM dispersion scales with the RM magnitude (i.e., $\sigma_{\rm RM} \propto |{\rm RM}|$). The larger RM measured at C4a therefore implies a larger RM dispersion, leading to stronger Faraday depolarization that suppresses polarized emission at lower frequencies (86--93~GHz) and allows it to be detected only at higher frequencies, where Faraday depolarization is weaker. In contrast, the smaller RM at C4 corresponds to a smaller RM dispersion, enabling polarized emission to remain detectable at these lower frequencies. 

The RM at C4a becomes even larger when it is derived from images convolved with the smaller beam at 106.8~GHz, corresponding to the lowest frequency used in the C4a RM measurement. Figure~\ref{fig:nov2024_01d_w} shows images at 106.8, 108.3, 111.3, and 114.0~GHz, convolved with the beam size at 106.8 GHz, rather than with the larger beam at 86.2~GHz. Using this smaller beam, the derived RM at C4a is $\rm (6.1\pm1.1)\times10^5\ rad\ m^{-2}$, which is larger than the value obtained from images convolved with the 86~GHz beam. This increase may reflect reduced contamination from the nearby downstream component C4, which has a smaller RM, due to the improved spatial separation provided by the smaller beam.

In the observations obtained on 2024 November 8 and December 27, polarization and RM are also consistently detected at C3 and C4, again showing systematically larger RM values at C4 than at C3 (see also Appendix~\ref{app:kvn_images_2024}). In contrast, polarized emission is not reliably detected at component K across the W band, consistent with the 2022 December and 2023 February epochs discussed above. Although a few polarized features formally exceed an SNR of 7, their positions vary with frequency and lack spatial coherence across the band. As in the earlier epochs, we therefore do not regard these features at component K as reliable detections.

\begin{figure*}[t!]
\centering 
    \includegraphics[width=0.9\textwidth]{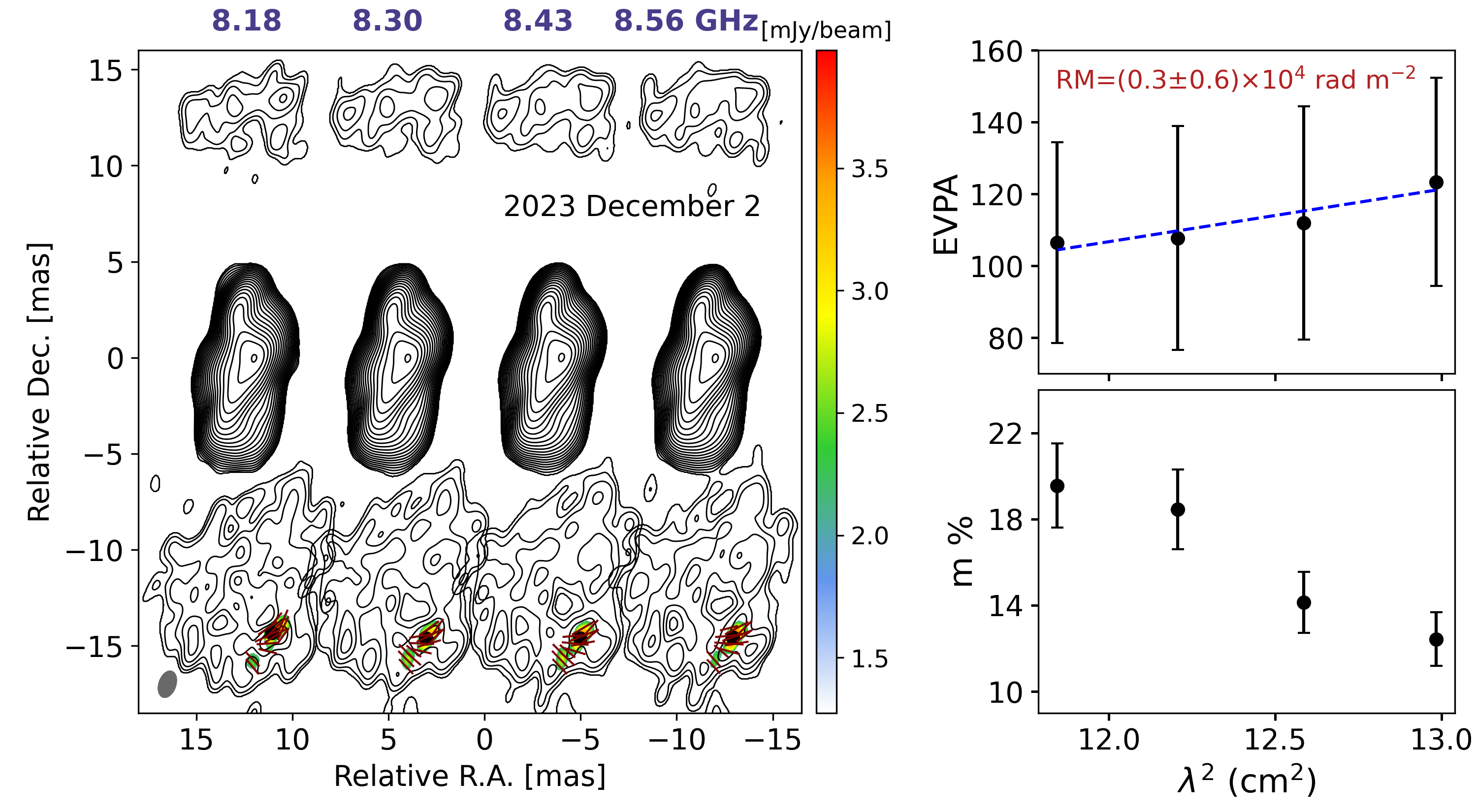}
    \caption{HSA images of 3C~84 obtained at 8.176, 8.304, 8.432, and 8.560~GHz on 2023 December 2. All images are convolved with the beam size at 8.176~GHz. The EVPA, RM, and fractional polarization at the polarized intensity peak in the southern lobe are shown in the right panels. 
\label{fig:hsa}}
\end{figure*}

\subsection{HSA in 2023}
Figure~\ref{fig:hsa} shows X-band images of 3C~84 obtained with the HSA on 2023 December 2. At these lower frequencies, the northern and southern lobes are clearly visible. Polarization is not detected in the inner jet, likely due to severe Faraday depolarization at these low frequencies \citep{taylor06}. In contrast, polarized emission is consistently detected across all frequencies at the tip of the southern lobe, which was ejected in the 1950s and is currently located 14--15~mas from the core \citep{asada06}. This region coincides with the location where the RM of $\rm 7000~rad~m^{-2}$ was previously reported \citep{taylor06}. The RM measured at this position in our data is $(0.4\pm0.5)\times10^4~\rm rad~m^{-2}$, which is significantly smaller than the values measured toward the inner jet with the KVN and VLBA, but consistent with the previously reported RM at a similar location.

The fractional polarization at the polarized-intensity peak is $\sim$20\%, which is higher than 0.8--7.5\% at 8--22~GHz measured in 2004 November \citep{taylor06}.  

\subsection{Absence of detectable core polarization}\label{sec:3.5}

A notable result is that, although polarized emission is robustly and repeatedly detected in the downstream jet components, polarization has never been reliably detected at the core up to 141~GHz. Weak polarized features occasionally appear at the core position in some BU epochs (see Figure~\ref{fig:vlba_img1}), and in some cases they are not extremely weak. While these features could in principle represent real and time-variable core polarization, their behavior differs from that of the polarized emission detected in the C3 region. The C3 polarization is detected continuously over many epochs between 2015 December and 2017 May, and although its exact position is not identical at every epoch, it remains relatively consistent or evolves smoothly. In contrast, the weak polarized features in the core do not appear persistently and, when detected, their positions vary irregularly from epoch to epoch.

A similar pattern is seen in the KVN data. Weak polarized features occasionally appear at the core in some epochs, but they are not consistently detected across all frequencies within the same observing session. This is particularly noteworthy because the KVN observations extend to higher frequencies (86–141~GHz), where Faraday depolarization is expected to be substantially weaker. If intrinsic core polarization were present at a detectable level, it would be more easily detectable across these higher frequencies. In practice, however, the KVN observations do not show such consistent multi-frequency core polarization. For example, in the KVN images obtained in 2018 February (Figure~\ref{fig:18feb}), weak polarized emission appears at some frequencies such as 129~GHz, but it is absent at other frequencies within the same epoch. Likewise, in the KVN images obtained on 2024 November 19--20 (Figure~\ref{fig:nov2024_01cd}), weak polarized features are present at certain frequencies but absent at others during the same session. Although the BU and KVN epochs are not simultaneous and intrinsic variability cannot be entirely excluded, the absence of coherent high-frequency core polarization further weakens the interpretation of the occasional core features as robust intrinsic core polarization. This suggests that the weak polarized features occasionally seen at the core may instead arise from residual instrumental polarization.

Such artifacts are generally negligible in AGN jets with strongly polarized cores, such as blazars, because they are masked by much brighter intrinsic polarized emission. However, they become more apparent in jets with weak core polarization such as M87 and 3C~84, and may therefore be misinterpreted as true core polarization \citep[e.g., see Figure 1 and 2 in][]{park21_m87}. Therefore, we conclude that the weak polarized features occasionally appearing at the core position in the VLBA and KVN images of 3C~84 do not constitute robust detections of intrinsic core polarization.

\subsection{Summary of polarization and RM detections}\label{sec:obs_sum}
From 2015 to 2024, we detected polarization and RM toward three regions of the inner jet (C4, including C4a; C3; and K) using the KVN and VLBA across 43--141~GHz. These observations provide the first detections of polarization from components C4 and K. In addition, polarization and RM are detected at the tip of the southern lobe with the HSA at 8~GHz, consistent with the location where polarization and RM were previously reported in 2004 with the VLBA observations at 8--22~GHz \citep{taylor06}. 

\begin{figure*}[t!]
\centering 
\includegraphics[width=\textwidth]{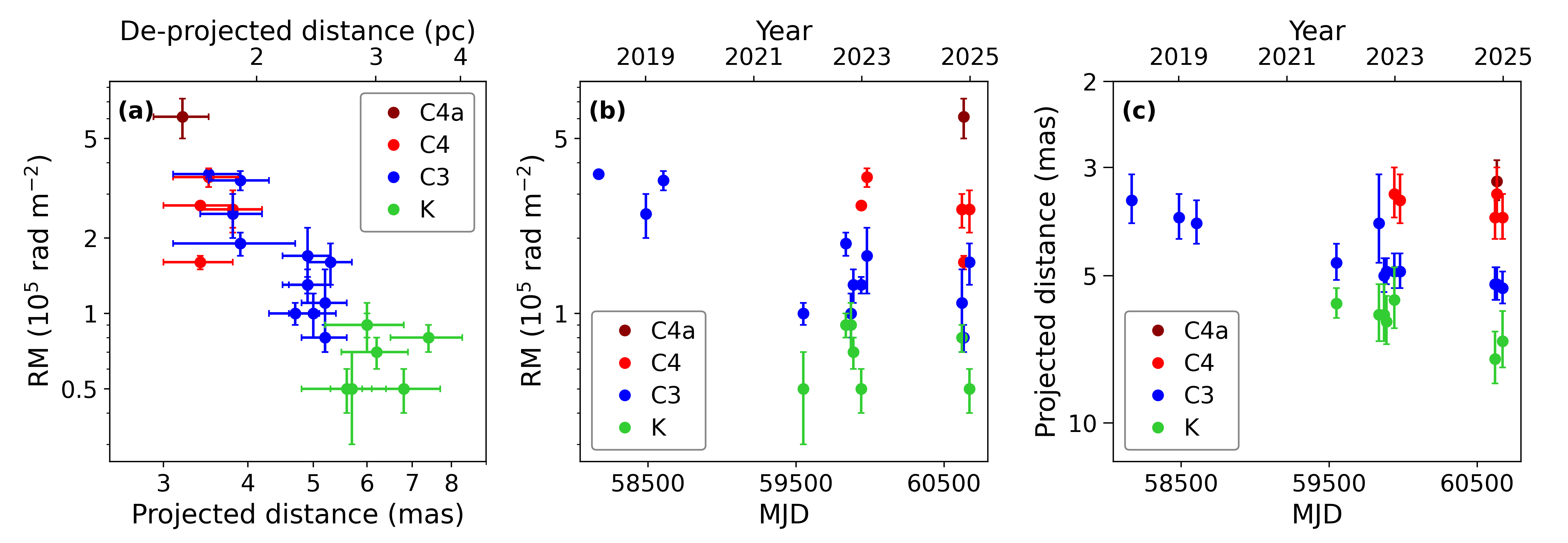}
\caption{Panel (a) shows the RM measured with the KVN as a function of distance from the core, with both axes plotted on a logarithmic scale. Panel (b) presents the temporal variation of the RM at each component (logarithmic scale on the y-axis). Panel (c) shows the projected distance of each component as a function of time (logarithmic scale on the y-axis). The distances increase from top to bottom, following the jet direction.
\label{fig:rm_time}}
\end{figure*}

\begin{table*}[t!]
\begin{center}
    \caption{RM measured with other instruments. \label{tab:rm_others}}
    \begin{tabular}{lllcccc}
        \hline 
        Instrument & Obs. Date & Component & RM & Distance & Reference  \\
         & & & [rad m$^{-2}$] & [mas] & \\
        \hline 
        VLBA & 2015 Dec.--2017 May & C3 & $(3.9\pm0.6)\times10^5$ & $3.1\pm0.2$ & This work \\ 
        HSA & 2023 Dec. 05 & southern lobe & $(3.0\pm6.0)\times10^3$ & $15.4\pm0.8$ & This work \\ 
        VLBA & 2004 Nov. 11 & southern lobe & $(7.0\pm1.0)\times10^3$ & $15.0\pm1.0$ & \citealt{taylor06} \\
        CARMA, SMA & 2011 May--2013 Aug. & unresolved & $(8.7\pm2.3)\times10^5$ & $2.1\pm0.2$ & \citealt{plambeck14} \\ 
        IRAM PdBI & 2011 Mar. 9, 11 & unresolved & $>1\times10^6$ & $1.7\pm0.2$ & \citealt{trippe12} \\ 
        \hline 
    \end{tabular}
\tablecomments{The positions for the unresolved RM measurements were inferred using the C3 positions in the BU data obtained during the observing periods \citep{jorstad16}. The average position and its standard deviation were adopted as the representative position and its uncertainty. For the VLBA measurements for the southern lobe, the RM uncertainty is taken to be $\rm 1\times10^3\ rad\ m^{-2}$, reflecting the range of detected RM values. The adopted position of ($15\pm1$)~mas accounts for the spatial extent of the RM-detected region together with a possible core-shift of 0.8~mas.}
\end{center}
\end{table*}

\section{Discussion}\label{sec:Discussions}

\subsection{External Faraday rotation}\label{sec:external}
Our wide-band, multi-frequency monitoring of 3C~84 provides clear evidence regarding the origin of the Faraday rotation. Whenever the RM is sufficiently high and frequency coverage is wide enough, we observe large EVPA rotations exceeding 45$^\circ$ that still follow a good $\lambda^2$ fit. For example, at C3 in 2018 February, the EVPA rotates by more than 150$^\circ$ between 86 and 141~GHz while maintaining a good $\lambda^2$ fit (Figure~\ref{fig:18feb}). Similar large rotations are also seen in 2024 November and December (Figure~\ref{fig:novdec2024_q}). This behavior suggests that the Faraday rotation originates external to the jet \citep{burn66, park19_rm}. In the case of internal Faraday rotation, emission from different depths within the jet experiences different amount of Faraday rotation, and the superposition of these contributions limits the observable $\lambda^2$-coherent rotation to below $\sim45^\circ$. When the rotation exceeds this level, the EVPA deviates from the $\lambda^2$ law, as theoretically expected \citep{burn66, cioffi80} and observed in practice \citep{pasetto21, wielgus24}. Thus, the large EVPA rotations showing good $\lambda^2$-fit seen in 3C~84 support an external Faraday screen.  



\begin{figure*}[t!]
\centering 
\includegraphics[width=0.55\textwidth]{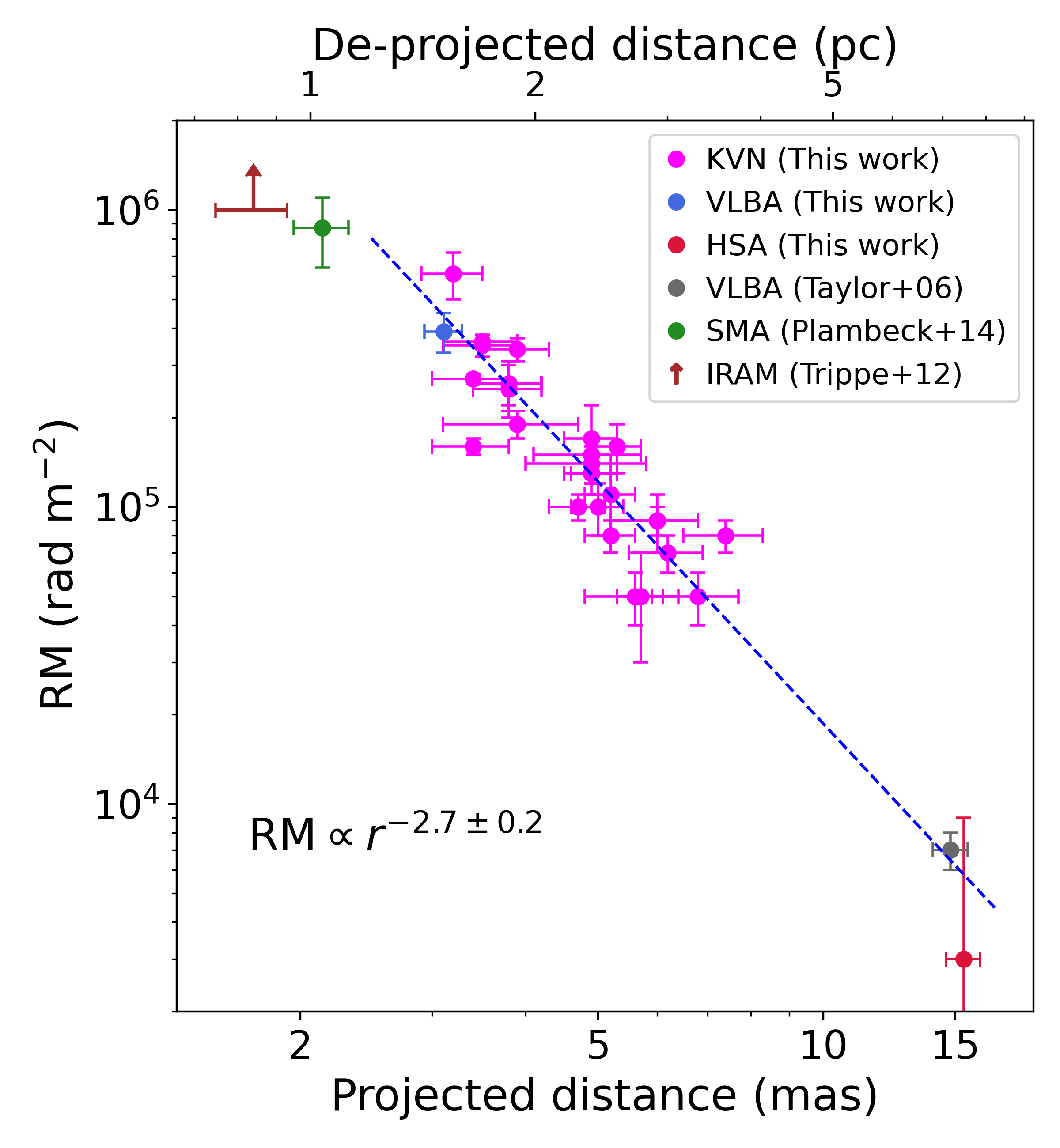}
\caption{The RM values measured at different distances from the central black hole are presented. The blue data point represents the RM at C3 obtained with the BU data. The black and red data points indicate the RM measured toward the southern lobe with the VLBA \citep{taylor06} and with the HSA, respectively. The magenta data points represent the RM values obtained with the KVN. The blue dashed line represent the best fit to the spatially resolved RM measurements. The green data point shows the unresolved RM measured with CARMA and SMA, plotted assuming that the polarized emission is dominated by C3 \citep{plambeck14}. The brown horizontal bar with an upward arrow indicates the lower limit on the RM derived from unresolved IRAM PdBI observations \citep{trippe12}, also plotted at the distance of C3. These unresolved measurements are shown for reference and are not used for the fitting.
\label{fig:rm_dist}}
\end{figure*}

\subsection{RM as a function of distance} \label{sec:Discussions_origin}

The most important result from our monitoring is the clear systematic decrease of the RM with increasing distance from the core: it is consistently largest at C4, followed by C3, and smallest at K whenever these components are observed in the same epoch. This relationship persists even as the components move outward and their individual RM values change over time.

To investigate this further, we analyzed how RM varies with distance from the core and with time. For each RM measurement, the distance from the core was determined as the mean position of the polarized intensity peaks across the frequencies used to measure that RM. As estimating positional uncertainty in VLBI images is not straightforward, it is typically assumed to be 20\% of the beam size, a value based on practical experience \citep[e.g.,][]{lister09, lister16, lister19, lister21}. Since our analysis focuses on the separation of jet components from the core and the minor axis of the beam is nearly aligned with the jet direction, we adopt the minor axis as the effective beam size. In addition, we account for the standard deviation of the polarized intensity peak positions across the frequencies. The final positional uncertainty is obtained by combining these two contributions in quadrature. All RM values measured with the KVN and their corresponding distances from the core are summarized in Table~\ref{tab:rm_kvn}.

Figure~\ref{fig:rm_time}a shows the RM measured with the KVN as a function of distance from the core. Notably, the RM decreases systematically with increasing distance, regardless of their origin (C4a, C4, C3, and K) or observing epoch. The Pearson correlation coefficient between RM and distances is $r=-0.604$ with $p=4.1\times10^{-4}$, indicating a negative correlation. Together with the evidence for the external Faraday rotation (EVPA rotations exceeding 45$^\circ$ and non-detection of negative RM), this suggests that a common external Faraday screen affects all these components, with RM steeply decreasing with distance from the core. 

In contrast, most RMs show no significant trend as a function of time (Figure~\ref{fig:rm_time}b). Only C3 shows a gradual decrease, but this component alone has been monitored long enough to clearly trace its outward motion from the core (Figure~\ref{fig:rm_time}c). The apparent temporal decrease in RM for C3 is therefore naturally explained by its increasing distance from the core, which causes it to probe regions of the Faraday screen with progressively smaller RM \citep[see also Figure~5 in][]{kam24}. This indicates that RM at a fixed location is largely stable over time, and that the observed changes in the RM at C3 mainly reflect spatial changes rather than temporal variations in the Faraday screen.

To test whether this relation between RM and distance extends to larger distances, we included the RM measured at the southern lobe, as well as the RM at C3 measured with the VLBA during epochs preceding the KVN observations (Table~\ref{tab:rm_bu}). Two RM values are available for the southern lobe. One is $\rm 7\times10^3\ rad\ m^{-2}$ between 5 and 22~GHz on 2004 November 11, and we measured $\rm (3\pm6)\times10^3\ rad\ m^{-2}$ with the HSA at 8~GHz on 2023 December~5. While a core-shift of $\sim$0.8~mas is suggested at 8~GHz by \citealt{paraschos21}, smaller values have been suggested by other studies \citep{giovannini18, savolainen23, park24_3c84}. We therefore adopt 0.8~mas as the core-shift at 8~GHz and take the same value as the positional uncertainty (Table~\ref{tab:rm_others}). The corresponding RM values and distances have been added to Figure~\ref{fig:rm_dist}.

Remarkably, the RM values toward the southern lobe follow the same monotonic decreasing trend with distance seen in the RM at C4a, C4, C3, and K, which probe the Faraday screen ahead of the inner jet (Figure~\ref{fig:rm_dist}). Despite the southern lobe and inner jet representing physically distinct structures separated by a factor of 2–5 in distance, their RM values fall on the same relation. Moreover, the two RM measurements at the southern lobe are separated by nearly 20 years, but they are consistent within the uncertainty and simply extend the same decreasing trend with distance from the core. This continuity across both spatial scale and epoch provides strong evidence for the long-term stability of the Faraday screen. 

The Pearson correlation coefficient between RM and distance is $r=-0.876$ with $p=2.1\times10^{-9}$, revealing an even stronger negative correlation than that seen in the KVN-only measurements toward the inner jet. Consequently, these results strongly indicate that a coherent external Faraday screen is responsible for all spatially resolved RM measurements in 3C~84. 

In addition, the RM toward both the inner jet and the southern lobe is consistently positive across all observing epochs and frequencies, indicating that the line-of-sight magnetic field is systematically directed toward us over scales of 1–8~pc from the black hole.

We fitted the RM measurements using the Python package LINMIX, which performs linear regression accounting for uncertainties in both the $x$ and $y$ directions within a hierarchical Bayesian framework \citep{kelly07}. The best-fit relation yields $\mathrm{RM} \propto d^{-k}$ with $k = 2.7 \pm 0.2$. This is much steeper than the relation found in M87, which shows roughly $k\sim1$ when fitted linearly \citep{park19_rm}.

Additional constraints at higher frequencies are also consistent with this picture under the assumption that the unresolved CARMA, SMA, and IRAM PdBI measurements in 2010--2014 are likewise dominated by emission from C3. These sub-arcsecond measurements obtained between 2010 and 2014 have frequently been assumed as representing the core. However, polarization has never been reliably detected at the core throughout our monitoring (see Section~\ref{sec:obs_sum}), suggesting that the polarized emission measured by these instruments could be dominated by C3, the only prominent polarized component until 2019. Adopting this interpretation, the average RM from CARMA and SMA between 2011 and 2013, $\rm (8.7\pm2.3)\times10^5\ rad\ m^{-2}$, and the lower limit of $\rm 10^6\ rad\ m^{-2}$ derived from the strong Faraday depolarization seen with IRAM PdBI, provide additional data points for the RM--distance relation. Since these sub-arcsecond arrays cannot resolve the inner jet, we estimated the contemporaneous positions of C3 using the monthly BU 43~GHz VLBA monitoring data \citep{jorstad16}, adopting the C3 positions from \citealt{kam24}  derived with MODELFIT in Difmap. For each instrument, the adopted position and its uncertainty correspond to the mean and standard deviation of the C3 locations over its observing period (Table~\ref{tab:rm_others}). When these RM values and corresponding distances are added to Figure~\ref{fig:rm_dist}, they fall on the same RM-distance relation observed in the spatially resolved VLBI measurements, extending the trend to even closer distances to the black hole and to larger RM values.  

Altogether, this RM-distance relation unifies measurements obtained over decades, with multiple instruments and across different spatial scales, suggesting that they all originate from a single, common external medium surrounding both the inner jet and southern lobe. This can be naturally explained if both inner jet and the southern lobe are embedded in the same Faraday screen, implying that the screen must extend beyond the southern lobe, to a distance of at least 7--8~pc from the black hole.

\subsection{The origin of Faraday rotation}

Our RM measurements and analyses point to a consistent picture: both inner jet and the southern lobe are surrounded by a common external Faraday screen. In this section, we explore the nature of the medium that responsible for the observed RM in 3C~84. We begin by carefully examining the black hole mass, Bondi radius, and Eddington ratio of NGC~1275, as these parameters are critical for the overall physical interpretation. We then discuss scenarios that are unlikely to represent the Faraday screen, and identify the medium that best matches the observations.

\subsubsection{Black hole mass, Bondi radius, and Eddington ratio}

We adopt a black hole mass of $M_{\rm BH}=1.1\times10^9\ M_{\odot}$, a Bondi radius of 11.8~pc, and an Eddington ratio of 0.04--0.09\% for NGC~1275/3C~84. In the following, we describe the justification for these adopted values. 

The adopted black hole mass is derived from modeling the rotation of the molecular gas disk traced by the CO(2-1) emission line, incorporating the contribution of the molecular gas mass \citep{nagai19}. This estimate is consistent, within the uncertainty, with the upper limit of $M_{\rm BH} = 8^{+7}_{-2} \times 10^{8}\ M_\odot$ obtained from modeling the rotation of the H$_2$ disk in the inner $\sim$50~pc of NGC~1275 \citep{scharwachter13}. This upper limit represents the total enclosed mass within 50~pc, including contributions from the black hole, stars, and molecular gas. The stellar mass in this region is negligible ($3.7\times10^6\ M_\odot$) \citep{wilman05}, while the molecular gas mass is $4\times10^8\ M_\odot$ with substantial uncertainty. Consequently, \citealt{scharwachter13} suggested their enclosed mass estimate $M_{\rm BH} = 8^{+7}_{-2} \times 10^{8}\ M_\odot$ as an upper limit on the black hole mass. This upper limit is compatible with our adopted value of $M_{\rm BH}=1.1\times10^9\ M_{\odot}$ within the uncertainty.

This value also agrees with that inferred from the empirical relation between the black hole mass and stellar velocity dispersion ($M_{\rm BH}$–$\sigma_\star$ relation; \citealt{kormendy13}). The stellar velocity dispersion measured in NGC~1275 is $\rm 265 \pm 26~km~s^{-1}$, corresponding to a black hole mass of $M_{\rm BH} = 1.1^{+0.9}_{-0.5} \times 10^{9}\ M_\odot$ \citep{riffel20}.
The consistency among these independent estimates justifies adopting a black hole mass of $M_{\rm BH}=1.1\times10^{9}\ M_\odot$ for NGC~1275\footnote{Black hole masses derived from the single-epoch method, using broad emission lines, typically fall below $1\times10^8\ M_{\odot}$ for NGC~1275 \citep[e.g.,][]{son12, onori17, sani18}. However, these estimates are likely underestimated because the standard virial factors used in these methods were calibrated for AGN with higher Eddington ratios (e.g., \citealt{woo15}). For sources with low Eddington ratios like NGC~1275 ($\lesssim0.1\%$), the virial factor is expected to be larger \citep[e.g.,][]{marconi08, marconi09}}.

The Bondi radius reported by \citet{fujita16} is 8.6~pc, assuming a black hole mass of $M_{\rm BH}=8\times10^{8}\ M_\odot$. Since the Bondi radius scales linearly with black hole mass, rescaling this Bondi radius to our adopted value of $M_{\rm BH}=1.1\times10^{9}\ M_\odot$ yields $r_{\rm B}=11.8$~pc, which we adopt in this work.

To estimate the bolometric luminosity, we use two independent indicators: the X-ray luminosity and the broad emission lines, which offer complementary constraints on the bolometric luminosity of the AGN. Several indications from X-ray observations, including spectral properties, correlations with the radio light curve, and X-ray polarization, suggest that the parsec-scale jet may partially contribute to the observed X-ray emission in NGC~1275 \citep{fabian15, rani18, reynolds21,imazato21, sinitsyna25, liodakis25}. To minimize possible contamination from the jet, we adopt the 2–10~keV X-ray luminosity measured with \textit{XMM-Newton} in 2001 January \citep{panessa06, ho09}, prior to the onset of the current episode of jet activity that began around 2003 \citep{nagai10, suzuki12}. Rescaling this measurement to a distance of 73.2~Mpc yields $L_{\rm 2–10,keV}=7.9\times10^{42}$~erg~s$^{-1}$. This epoch corresponds to a period when both the X-ray and radio flux densities were near their minima, just before the emergence of the parsec-scale jet, thereby reducing the potential jet contribution to the X-ray emission. Applying a bolometric correction factor of 7.7–10 appropriate for low-luminosity AGN \citep{ho09, vasudevan09, duras20}, we obtain a bolometric luminosity of $L_{\rm Bol}\simeq(6–8)\times10^{43}$~erg~s$^{-1}$.

This estimate is consistent with an independent approach based on the broad H$\beta$ line. Using the empirical correlation between the broad H$\beta$ luminosity and the 5100\AA\ continuum luminosity \citep{greene05}, and adopting a bolometric correction factor of 10 \citep{woo02}, the broad H$\beta$ luminosities observed from the 1980s to 2010s \citep[Table~1 in][]{punsly18} correspond to $L_{\rm Bol}=(6-12)\times10^{43}$ erg s$^{-1}$. 

Given the consistency between the two independent estimates, we conservatively adopt $L_{\rm Bol}=(6 – 12)\times10^{43}$~erg~s$^{-1}$ as the representative bolometric luminosity of NGC~1275. Combined with the black hole mass of $M_{\rm BH}=1.1\times10^9\ M_{\odot}$, this yields an Eddington ratio $L_{\rm Bol}/L_{\rm Edd}=0.04-0.09\%$. 

We note that our estimated Eddington ratio is approximately 4--10 times smaller than the commonly quoted value of $L_{\rm bol}/L_{\rm Edd}=0.4\%$ \citep{plambeck14}, which was derived from the the International Ultraviolet Explorer (IUE) spectrum of NGC~1275 \citep{kinney91}. When this estimate is corrected for our adopted distance and black hole mass, the corresponding IUE-based Eddington ratio decreases by more than a factor of two. More importantly, the original calculation assumed a typical spectral energy distribution (SED) of bright quasars \citep{levinson95}, whereas low-luminosity AGN like NGC~1275 are known to have different SEDs from bright quasars, making this assumption inappropriate \citep{ho99}. This mismatch likely leads to a systematic overestimation of the commonly quoted Eddington ratio for NGC~1275. For these reasons, we adopt $L_{\rm Bol}/L_{\rm Edd}=0.04-0.09\%$ as the representative value for NGC~1275.

\subsubsection{Radiatively inefficient accretion flow?}\label{sec:riaf?}

The RIAF has often been considered as a possible origin of the RM in 3C~84. Accretion onto black holes is broadly categorized into hot accretion flows and cold accretion disks, depending on their radiative efficiency. The standard cold accretion disk is geometrically thin and optically thick, as expected when cooling is efficient \citep[e.g.,][]{shakura73, lynden74}. When the accretion rate is either very low or very high, cooling becomes inefficient, resulting in hotter and lower-density plasma compared to a standard accretion disk \citep{narayan05}. This leads to the formation of a hot accretion flow, often referred to as a RIAF, which is geometrically thick but optically thin \citetext{e.g., advection-dominated accretion flow; \citealp{narayan94, narayan95_outflow, abramowicz95}}. 

In this scenario, the accretion flow is expected to develop a two-zone structure, consisting of an inner RIAF surrounded by an outer, truncated cold thin disk \citep[e.g.,][]{narayan96}. As the Eddington ratio decreases, radiative cooling becomes less efficient, causing the cold thin disk to be truncated at progressively larger radii and the transition radius between the thin disk and the inner hot flow to move outward (e.g., \citealp{esin97, yuan04, narayan08, you23}; see \citealp{yn14} for a review). Consequently, the innermost regions become increasingly dominated by hot, geometrically thick gas, and the radial extent of the RIAF grows as the Eddington ratio decreases. Due to its geometrically thick structure, which could intersect the line of sight to the jet and/or black hole, the RIAF has therefore been discussed as a primary contributor to the observed RM in AGN with very low Eddington ratios ($L_{\rm Bol}/L_{\rm Edd}\ll 0.1\%$), such as M87 \citep[e.g.,][]{kuo14, park19_rm, yuan22} and Sgr~A* \citep[e.g.,][]{quataert00, bower03, bower18, marrone06, marrone07}.

Theoretical studies and numerical simulations suggest that the transition between hot and cold accretion states occur at an Eddington ratio of $L_{\rm bol}/L_{\rm Edd}=0.1-1\%$ \citep[e.g.,][]{xie12, inayoshi20}. \citealt{plambeck14} first suggested RIAF models as the origin of the RM toward 3C~84, as the Eddington ratio in 3C~84 is in this transition regime. Our updated estimate of $L_{\rm bol}/L_{\rm Edd}\sim 0.04-0.09\%$, nearly an order of magnitude lower than the value commonly adopted in earlier studies, further implying the presence of a RIAF around the black hole. Due to its geometrically thick structure, which could intersect the line of sight to the jet, such an inner RIAF has been discussed as a potential contributor to the observed RM \citep[e.g.,][]{plambeck14, li16, kim19}. 

However, the RIAF in 3C~84 is unlikely to extend radially to parsec-scale distances, as the surrounding environment is significantly cooler than expected for RIAF models. To illustrate this, we examine the temperature structure of the circumnuclear region from tens of parsecs down to sub-parsec scales.

Gemini/NIFS observations revealed a molecular disk traced by the H$_{2}$ line within $\sim$50~pc of the black hole \citep{scharwachter13, riffel20}, with temperatures of a few 1000~K \citep{wilman05, scharwachter13}. At similar distances, ALMA detected a molecular disk in CO(2-1), HCN(3-2), and HCO$^{+}$(3-2) lines, with temperatures below 100~K \citep{nagai19}. 
On smaller scales of 2--3~pc, VLBA observations show that the northern counter-jet exhibits an inverted spectrum due to free-free absorption (FFA), unlike the southern approaching jet \citep{walker94, wrb94, walker00, fujita17, wajima20}. This indicates the presence of an ionized gas structure along the line of sight to the counter-jet, likely associated with the inner region of the molecular disk found on larger scales. Photoionization modeling suggests that the surface layers of this FFA disk have temperatures of $\rm T\sim1.5\times10^4$~K, while the mid-plane would be cooler as it is shielded by the surface layers \citep{rybicki79, levinson95}. The detection of the Fe-K$\alpha$ emission line at 6.4~keV further supports the presence of the FFA disk, as it indicates neutral or weakly ionized iron \citep{hitomi18, reynolds21}. The measured line width of FWHM 500--1600 km s$^{-1}$ is narrower than that of the broad H$\alpha$ and H$\beta$ emission lines with the FWHM of $\sim$5000 km s$^{-1}$ \citep[e.g.,][]{punsly18}, suggesting that the Fe-K$\alpha$ emitting region lies at a larger radius of 1.4--14~pc of the black hole. This region is likely associated with the FFA disk.

The innermost region of the FFA disk appears to be delineated by the broad-line region (BLR). Broad emission lines have been consistently detected in 3C~84, indicating the presence of partially ionized, line-emitting gas orbiting the central black hole \citep[e.g.,][]{ho97, sosa01, veron06, son12, koss17, punsly18}. The BLR is generally inferred to have a flattened, disk-like geometry \citep[e.g.,][]{gaskell09, baskin18, gravity18} and to be photoionized by radiation from the accretion disk \citep[e.g.,][]{kaspi05, bentz13}. In such gas, temperatures stabilize near $T\sim 10^4$~K, where photoionization heating is balanced by radiative cooling \citep{osterbrock89, netzar90, sutherland93, netzar01, kaastra08}. The outer boundary of the BLR is set by the dust sublimation radius, beyond which it connects smoothly to the dusty or molecular disk \citep{czerny11, czerny17, baskin18, Shablovinskaya20}, likely corresponding to the FFA disk in 3C~84. 

Given these temperatures and spatial scales, it is highly unlikely that a RIAF in 3C~84 extends to parsec-scale distances. If a RIAF were to expand radially to such scales, both the BLR and the FFA disk would be embedded in gas with temperatures exceeding $10^8$~K \citep[e.g.,][]{yn14, li16}. Under these conditions, the gas would be completely ionized and incapable of producing either broad emission lines or FFA, which is incompatible with the observed properties of 3C~84. Therefore, the RIAF in 3C~84 cannot dominate the circumnuclear environment on parsec scales or serve as the primary contributor to the observed RM. Instead, it would be confined to a region smaller than the BLR radius, which may occupy the innermost region of the FFA disk. 

It is noteworthy that the BLR in Seyfert galaxies tends to disappear when the Eddington ratio falls below $L_{\rm Bol}/L_{\rm Edd}\sim0.1-1\%$, where the transition between hot and cold accretion is expected to occur \citep{czerny04, kollmeier06, elitzur09, trump09}. Similarly, hidden broad line regions (HBLR)\footnote{Early AGN unification models suggested that Seyfert 2 galaxies also possess a BLR that is obscured from direct view by a dusty torus or molecular disk when the source is viewed nearly edge-on \citep[see][for reviews]{antonucci93, urry95, netzar15}. This picture is supported by the detection of broad emission lines in the polarized spectra of Seyfert 2 galaxies, which reach us via scattering \citep[e.g.,][]{antonucci85, tran92}. Such obscured BLRs are commonly referred to as hidden broad line regions (HBLR).}, observed in polarized spectra of Seyfert 2 galaxies \citep[e.g.,][]{antonucci85, tran92, heisler97}, also vanish below $L_{\rm Bol}/L_{\rm Edd}\sim0.1-1\%$ \citep[e.g.,][]{gu02, nicastro03, wu11, marinucci12}. The simultaneous disappearance of both the BLR and HBLR is unlikely to be caused solely by obscuration. Instead, it may reflects a physical disappearance of the BLR as the RIAF expands outward at low accretion rates, suppressing or eliminating the cold structures required to sustain both the BLR and dusty torus \citep{elitzur09, cao10, trump11, wu11, marinucci12, liu22}. 

The Eddington ratio of 3C~84 ($L_{\rm Bol}/L_{\rm Edd}=0.04$–$0.09\%$) lies just below the transition between cold and hot accretion. At such Eddington ratios, the radial extent of a RIAF is expected to remain limited, since a RIAF expands to larger radii as the Eddington ratio decreases further. Thus, while a RIAF is likely present in 3C~84, it is expected to be radially compact. The continued detection of broad emission lines and a FFA disk requires the presence of cold, partially ionized gas at $T\sim10^4$~K. This coexistence implies that the RIAF must be confined to radii smaller than the BLR, with its outer boundary lying inside the BLR radius. Consequently, while a RIAF likely exists in the immediate vicinity of the black hole, it does not dominate the parsec-scale circumnuclear environment. 

\subsubsection{Jet boundary layer?}

The jet boundary layer or sheath has also been proposed as a potential origin of the RM in 3C~84 \citep{plambeck14, kim19, paraschos24_eht}. In this scenario, the jet is transversely stratified, consisting of a faster inner spine and a slower outer boundary layer\footnote{Recent studies of the NGC~315 jet suggest that the jet structure may deviate from the classical faster-spine and slower-sheath picture \citep{park24_ngc315}.}. The outer boundary layer is magnetized and can act as a Faraday screen (see Fig.~14 in \citealt{kim19}). 

In several nearby AGN jets, the observed RM values exhibiting systematic transverse gradients \citep[e.g.,][]{asada02, gabuzda04, osullivan09, nikonov23} or temporal variations including sign changes \citep[e.g.,][]{asada08, lico17, park18, toscano25} have been interpreted as evidence that the Faraday rotation originates in a medium associated with the jet, such as a magnetized sheath surrounding the emitting spine. In these sources, the typical RM values reported at $\lesssim43$~GHz are of order $10^2$ to $\rm 10^3~rad\ m^{-2}$. 

In case of 3C 84, however, RM values exceeding $10^4$--$\rm 10^5\ rad\ m^{-2}$ are already measured at 43~GHz, a frequency range in which other nearby AGN jets typically show only $10^2$ to $\rm 10^3~rad\ m^{-2}$. In addition, the RM measured in 3C~84 remains consistently positive and does not show sign reversals either over time or across different jet components. This contrasts with the temporal RM variability and sign changes reported in several nearby AGN jets, suggesting that the dominant Faraday screen in 3C~84 is relatively stable in time and not directly tied to dynamical jet structures.

Another important point is the location of the polarized emission. These large RM values are observed toward the jet termination region, including components C3 and K. In these regions, the polarized emission originates at the jet head where the jet interacts with the ambient medium \citep{laing80, meisenheimer89}. In a simplified geometric picture of a cylindrical or conical jet, and adopting the viewing angle of $\sim45^\circ$, radiation from the jet head can reach the observer without passing through a substantial column of boundary-layer plasma. This geometry therefore reduces the expected contribution from a jet sheath to the observed RM.

Moreover, the RM values measured across different jet components follow a common radial trend with distance from the center (Figure~\ref{fig:rm_dist}). For example, even when the RM is measured toward the side of the jet (e.g., the C4 component), the values remain $\rm \gtrsim 10^5\ rad\ m^{-2}$ and follow the same overall radial trend observed for the jet termination region and the southern lobe. This suggests the presence of a single large-scale Faraday screen affecting all components, rather than a localized Faraday screen associated with the jet boundary layer.

Therefore, these results disfavor the jet boundary layer as the dominant origin of the observed RM. Although a sheath may contribute to specific components (e.g., C4) at some level, its contribution would be negligible compared to the dominant Faraday screen and cannot account for the common radial trend observed across all jet components. 

\begin{figure*}[t!]
\centering 
\includegraphics[width=0.55\textwidth]{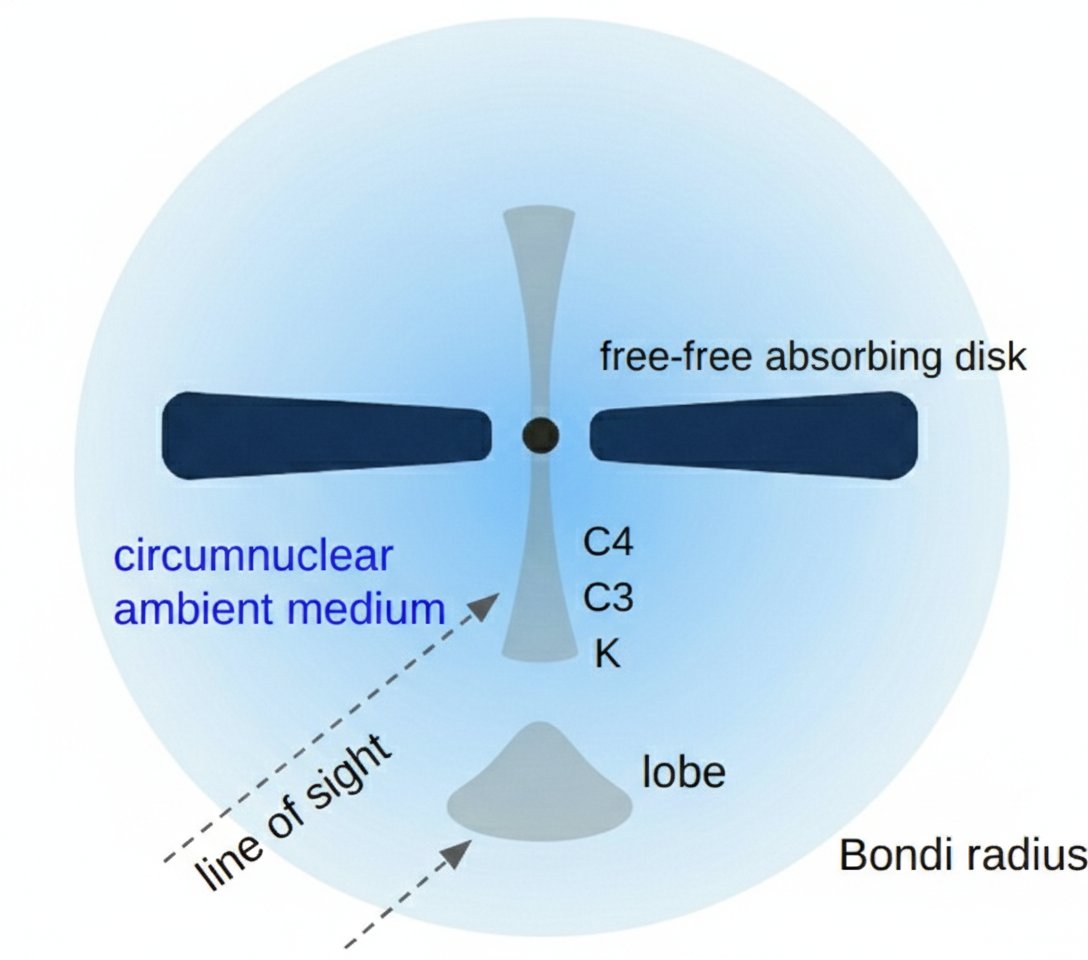}
\caption{Schematic illustration of the central region of NGC~1275. The components C4, C3, and K are located along the inner jet. The southern lobe lies between the inner jet and the Bondi radius. Distances and relative scales are not to scale.
\label{fig:schematic_pic}}
\end{figure*}

\subsubsection{A gas cloud?}
At the end of 2015, both the total and polarized flux densities at C3 began to increase, forming a bright hotspot \citep{nagai17, kino18}. This enhancement enabled the measurement of the RM of $\rm (3-6)\times10^5~rad~m^{-2}$ at C3. If the flux increase resulted from a collision between the jet and a small gas cloud, the cloud could account for the observed RM \citep{nagai17}.

However, our first KVN observations in 2018 February were conducted after the hotspot and associated flaring activity had disappeared in late 2017 \citep{kino21}. Nevertheless, polarized emission and RM values exceeding $\rm 10^5\ rad\ m^{-2}$ were still clearly detected at C3. In addition, similarly large RM values were consistently detected not only at C3 but also across multiple positions along the inner jet. Such persistent and spatially extended Faraday rotation cannot be explained by a single, localized gas cloud. Furthermore, the RM measured toward both the inner jet components and the southern lobe follow the same power-law relation, indicating that the Faraday screen extends to the southern lobe at distances of 7--8~pc from the center. This makes a compact gas cloud an unlikely origin for the observed RM in 3C~84. 

\subsubsection{The most likely origin: circumnuclear ambient medium}\label{sec:themostlikelyorigin}

Finally, we consider the scenario in which the circumnuclear ambient medium filling the central region of NGC~1275, where both the inner jet and the southern lobe are embedded, serves as the origin of the RM \citep{trippe12}. The observed RM profile (Figure~\ref{fig:rm_time} and \ref{fig:rm_dist}) provides three key constraints on the Faraday screen: (i) the RM is relatively stable over time while decreasing systematically with distance from the black hole, following a single power-law relation, (ii) the Faraday screen extends over at least 1--8~pc from the black hole, and (iii) the consistent sign of all detected RMs indicates that the magnetic field along the line of sight within the Faraday screen has a unidirectional structure. These observations are naturally explained if the entire jet of 3C~84, including both the inner jet and the southern lobe, is embedded in a common circumnuclear ambient medium (Figure~\ref{fig:schematic_pic}), which acts as the Faraday screen responsible for all the observed RM. 

The outer boundary of the Faraday screen can be reasonably associated with the Bondi radius of $r_{\rm B}=11.8$~pc. All RM measurements along the jet, including the outermost points at the southern lobe located at 7--8~pc from the nucleus, lie within the Bondi radius and follow the same RM–distance relation. This continuity implies that the electron density and magnetic-field strength decrease smoothly with radius throughout this region, as expected for gas governed by the gravitational potential of the central black hole. This supports the presence of a common, large-scale Faraday screen within the Bondi radius, which therefore provides a natural outer boundary for the Faraday screen. 

These results demonstrate that all the observed RM are naturally explained by a common circumnuclear ambient medium within the Bondi radius of NGC~1275. In this framework, the observed RM–distance relation arises from the smooth radial decline of density and magnetic-field strength in the ambient medium. This scenario provides a coherent physical explanation for the Faraday rotation in 3C~84 and resolves the long-standing question regarding the origin of its RM.

\subsection{Implication for the RM toward the core}

One direct implication of the RM-distance relation is that, if the relation continues inward toward the black hole, the RM toward the core should be much larger than those measured toward downstream components. As the electron density, magnetic-field strength, and line-of-sight path length are all expected to increase toward smaller radii, the core RM is expected to reach or exceed $\sim10^{7}$–$10^{8}\ {\rm rad\ m^{-2}}$, although this estimate is sensitive to the distance at which the RM profile begins to flatten and the extent of this flattening near the core (see Section~\ref{sec:both_profiles}). A detailed analysis of the plausible range of the core RM, constrained by inward extrapolation of the RM-distance relation and by depolarization mechanisms, will be presented in a separate study (Kam et al. in preparation). 

Several previous studies have attempted to estimate the RM toward the core of 3C~84. For example, \citealt{kim19} reported a value of $\rm 2\times10^{5}\ rad\ m^{-2}$, while \citealt{paraschos24_eht} suggested a larger value of $\rm 6\times10^{6}\ rad\ m^{-2}$. However, these estimates partly rely on EVPA measurements with single-dish telescopes and interferometric arrays (e.g., ALMA), which do not spatially resolve the parsec-scale jet. As a result, the detected polarization was assumed to originate from the core, whereas in reality it may predominantly arise from unresolved downstream jet components.

Our VLBI monitoring from 8 to 141~GHz shows that polarization has never been reliably detected at the core. Instead, it has been consistently dominated by downstream components. Furthermore, both the location and strength of polarized emission vary systematically with frequency due to Faraday depolarization, with polarized emission detected progressively farther upstream at higher frequencies (e.g., Figure~\ref{fig:nov2024_01cd}). These findings show that unresolved measurements at different frequencies probe different spatial regions, thereby preventing a reliable RM measurement from single-dish or other spatially-unresolved observations.

\subsection{Electron density and magnetic-field strength}\label{sec:both_profiles}

The observed RM profile provides a direct constraint on the radial distributions of the electron density and magnetic-field strength in the circumnuclear ambient medium. Within the Bondi radius, which we adopt as the outer boundary of the circumnuclear medium (see Section~\ref{sec:themostlikelyorigin}), the gas dynamics are expected to be governed primarily by the gravitational potential of the central black hole. This naturally leads to smooth radial gradients in both the electron density and magnetic-field strength. Consistent with this physical expectation, all RM measurements along the jet follow a single power-law relation with distance, indicating that any angular variation in density or magnetic-field strength is intrinsically smaller than the radial variation and/or remains below the RM measurement uncertainty. We therefore approximate the circumnuclear ambient medium as spherically symmetric within the region probed by the jet, and describe the electron density and magnetic-field strength as functions of radius only. 

Given this consistency between the physical picture and the observed RM--distance relation, we model the Faraday screen using one-dimensional radial scaling relations. The polarized jet emission probes this medium along the line of sight toward the jet, sampling a polar region set by the jet viewing angle of 45$^\circ$.

\begin{figure}[t!]
\centering 
\includegraphics[width=\columnwidth]{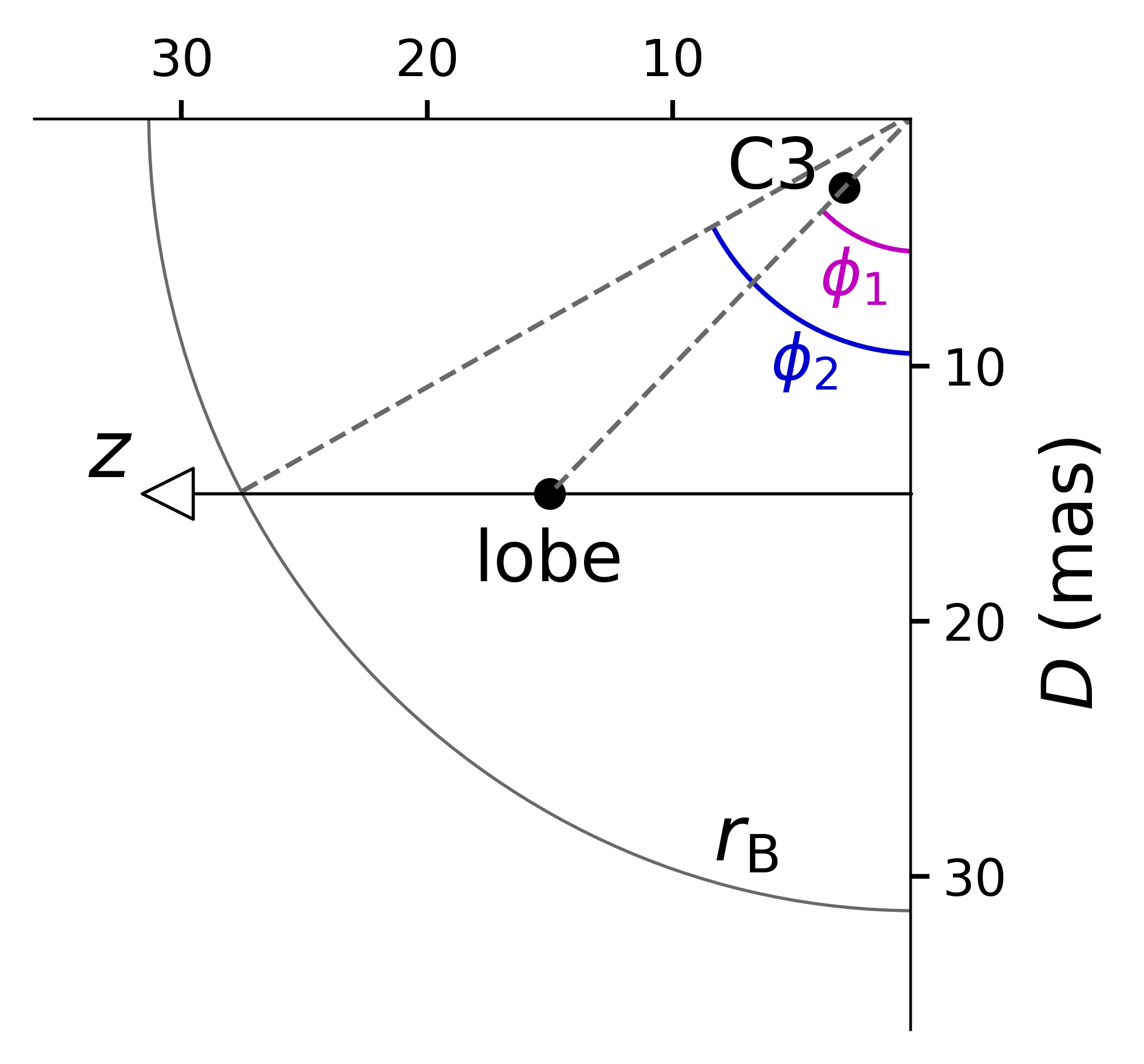}
\caption{Schematic diagram illustrating the geometry of the jet and the conventions adopted for the line-of-sight integration. The projected distances of $\sim$3~mas and $\sim$15~mas, corresponding to the approximate locations of the C3 component and the southern lobe, represent the inner and outer boundaries of the RM measurement region. The angle $\phi_1$ is defined between the jet axis and the plane of the sky, while $\phi_2$ is defined between the plane of the sky and the line connecting the center (0,0) to a point along the line of sight. The angle $\phi_1$ is fixed at $45^{\circ}$, and the Bondi radius $r_{\rm B}$ marks the outer boundary of the circumnuclear ambient medium.
\label{fig:schematic}}
\end{figure}

The RM is defined as
\begin{equation}\label{eq:rm_formula}
    {\rm RM} = 812 \int n_e(r)\, B_\parallel(r)\, dr \quad [{\rm rad\ m^{-2}}],
\end{equation}

\noindent where $n_e(r)$ is in units of cm$^{-3}$, $B_\parallel(r)$ is the line-of-sight magnetic field in $\mu$G, and $r$ is in parsecs. Assuming both the electron density and the magnetic field follow power-law profiles within the Bondi radius, we write
\begin{align}
    n_e(r) &= n_0 \left(\frac{r}{r_0}\right)^{-\alpha}, \\
    B_\parallel(r) &= B_0 \left(\frac{r}{r_0}\right)^{-\beta},
\end{align}

\noindent where $n_0$ and $B_0$ are the electron density and magnetic field at a reference radius $r_0$, and $\alpha$ and $\beta$ are the power-law indices. Equation~\ref{eq:rm_formula} can be expressed in terms of the projected distance $D$ (in the sky plane) and the line-of-sight coordinate $z$:
\begin{equation}\label{eq:rm_r}
    {\rm RM} \propto \int_{z_1}^{z_2} r^{-(\alpha+\beta)}\,dz ,
\end{equation}  

\noindent where $r^2=D^2+z^2$ (Figure~\ref{fig:schematic}). We introduce the angle $\phi$, defined as the angle between the line of sight and the radius vector, so that
\begin{gather}
    r = D \sec\phi, \label{eq:r}\\
    z = D\tan\phi, \quad dz = d(D\tan\phi). \label{eq:dz}
\end{gather}  

\noindent Substituting Equations~\ref{eq:r} and \ref{eq:dz} into Equation~\ref{eq:rm_r}, the RM can be written as
\begin{equation}\label{eq:rm_full}
    {\rm RM} \propto D^{-(\alpha+\beta)+1}\int_{\phi_1}^{\phi_2} (\cos\phi)^{\alpha+\beta}\, d(\tan\phi),
\end{equation}  

\noindent where $\phi_1$ and $\phi_2$ represent the inner and outer edges of the integration: $\phi_1$ is the angle at the location where polarization is detected, which is assumed to be nearly constant at 45$^\circ$ \citep{asada06, fujita17}, and $\phi_2$ corresponds to the outer boundary of the circumnuclear ambient medium, taken as the Bondi radius $r_{\rm B}=11.8~\text{pc}$. Although $\phi_2$ formally varies along the jet (e.g., between C3 and the southern lobe), the integral is dominated by the innermost regions because of the steep decline of $n_e$ and $B$. Therefore, the change in the integral is negligible (see Appendix~\ref{app:rm_integral}), allowing us to approximate
\begin{equation}\label{eq:rm_final}
    {\rm RM} \propto D^{-(\alpha+\beta)+1}.
\end{equation}  

\noindent Given the observed RM profile index is $-2.7\pm0.2$, this relation implies $\alpha+\beta \simeq 3.7$ and allows us to constrain the power-law indices of the electron density and magnetic-field strength profiles.

We note that the RM-distance relation derived above is observationally constrained only out to the southern lobe at 7–8 pc. Adopting the Bondi radius of 11.8~pc as the outer boundary provides a physically motivated reference radius, since the gravitational potential of the central black hole is expected to dominate within this scale and the Bondi radius lies close to the location of the outermost RM measurements. However, it remains unclear whether the observed RM–distance relation extends beyond the Bondi radius, and it may flatten at larger radii, as discussed later in this section.

To proceed, we adopt $\alpha=1.5$ for the electron density profile inside the Bondi radius, assuming mass conservation and a free-fall velocity $v\propto r^{-0.5}$ expected in spherical Bondi accretion \citep{bondi52, fujita16}. With this choice, the observed RM implies a magnetic field profile with $\beta=2.2\pm0.2$. Notably, this value is consistent with $\beta=2$ as expected for a magnetic field dominated by the radial component. 

However, the gas distribution could deviate from pure free-fall due to additional physical factors. To account for this, we also consider steeper electron density profiles. From the continuity equation $n v r^2 = \text{constant}$, the radial velocity scales as $v \propto r^{\alpha-2}$. Thus, $\alpha = 1.5$ corresponds to Bondi-like free-fall with $v \propto r^{-0.5}$, values $1.5 < \alpha < 2$ describe partially supported inflows where the infall velocity grows more slowly than in free-fall, and $\alpha = 2$ corresponds to a steady flow that sets an upper limit on the density slope. 

Comparing these cases with the observed RM yields $\beta\simeq2.2$ for $\alpha=1.5$, $\beta\simeq1.7$ for $\alpha=2$, and intermediate $\beta$ values for $1.5<\alpha<2$. These results constrain the possible ranges of $\alpha$ and $\beta$ inside the Bondi radius. 

A density profile shallower than $\alpha=1.5$ is unlikely for the circumnuclear ambient medium of NGC~1275. Although such shallow slopes can arise either inside a RIAF at very low Eddington ratios or in radiatively driven disk winds at higher accretion rates, NGC~1275 lies close to the transition between these regimes, where neither scenario is expected to dominate the parsec-scale ambient medium (Appendix~\ref{app:wind}).

\begin{figure*}[t!]
\centering 
\includegraphics[width=\textwidth]{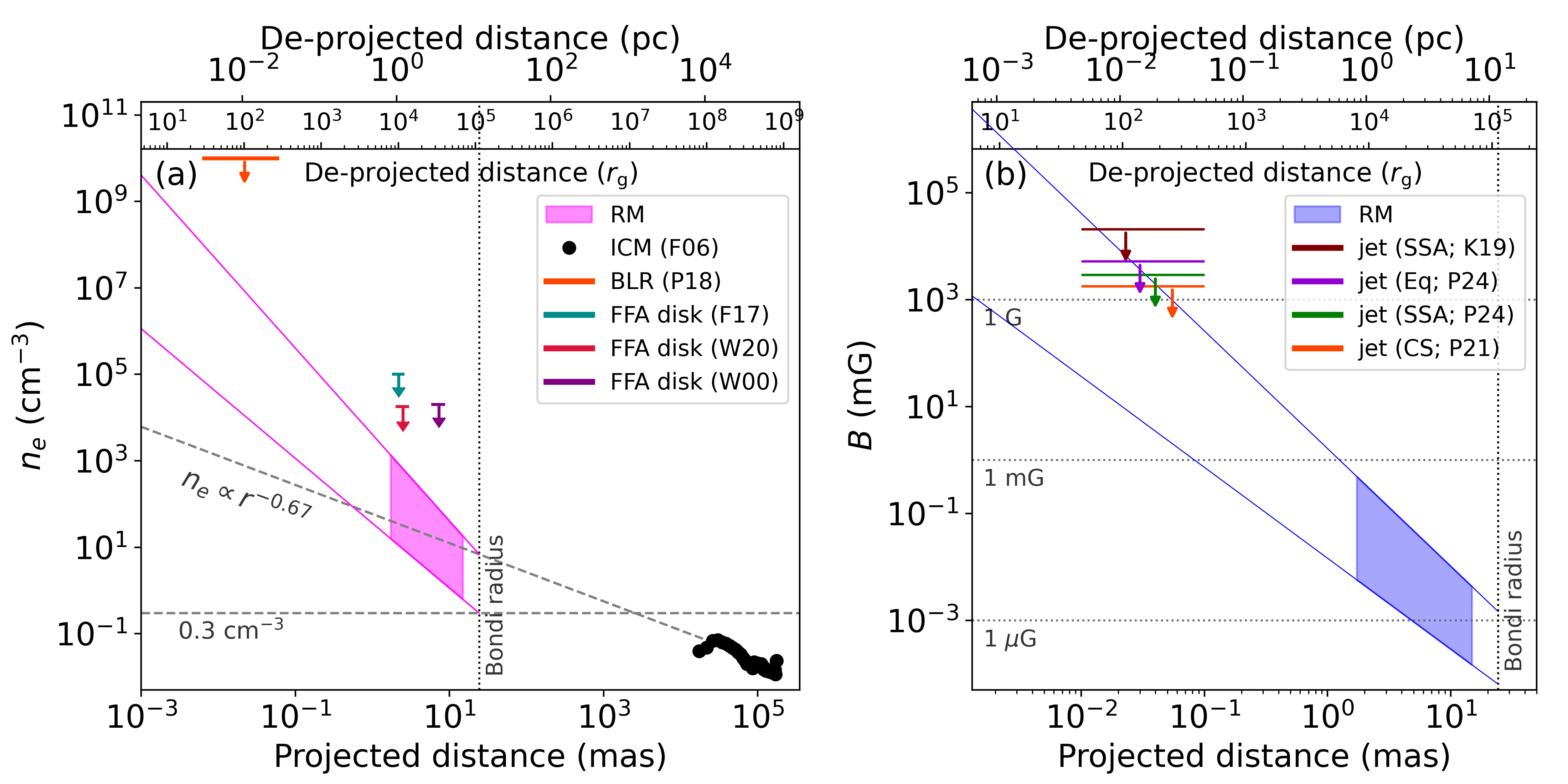}
\caption{The electron density (left) and magnetic field strength (right) as a function of distance from the black hole. \textit{Left}: Black dots show the electron density of the ICM measured at 10--80~kpc from X-ray observations \citep{fabian06}. The two gray dashed lines represent (i) the best-fit to these measurements $n_e\propto r^{-0.67}$, and (ii) an adopted lower-limit density of $n_{\rm e}=0.3\ \text{cm}^{-3}$ within 2~kpc \citep{taylor06}. The magenta shaded region marks the density range directly constrained by the RM measurements, and the unshaded region between the upper and lower bounding profiles represents the extrapolated range inside the Bondi radius. The orange line shows the electron density of the BLR \citep{punsly18, oh22}, and the purple, green, and red bars show the electron density of the FFA disk at specific radii \citep{walker00, fujita17, wajima20}, providing upper limits on the ambient density at those distances. 
\textit{Right}: The blue shaded region shows the magnetic-field strength directly constrained by RM measurements, while the unshaded region between the lower and upper bounds represents the corresponding extrapolation. Brown, purple, green, and orange lines indicate the magnetic-field strength at the jet core estimated from SSA (K19; \citealp{kim19}), equipartition (Eq) (P24; \citealp{paraschos24_eht}), SSA (P24), and core-shift (CS) (P21; \citealp{paraschos21}), respectively. These values provide upper limits on the magnetic-field strength in the ambient medium.
\label{fig:both_profiles}}
\end{figure*}

Figure~\ref{fig:both_profiles} shows the possible ranges for the electron density and magnetic field profiles. On kiloparsec scales, the electron density is constrained by X-ray observations \citep{fabian06}. The best-fit profile, $n_e \propto r^{-0.6}$, is much shallower than expected within the Bondi radius, where the gravitational influence of the black hole is expected to steepen the density gradient. Extrapolating this density profile inward to the Bondi radius provides the outer boundary condition for the circumnuclear density. Inside the Bondi radius, we adopt slopes between $\alpha=1.5$ and $\alpha=2$, where $\alpha=1.5$ corresponds to Bondi-like inflow and $\alpha=2$ represents the steepest case. Thus, the combination of the inward extrapolation and an inner slope of $\alpha=2$ defines the upper envelope of the possible density distribution.  

Alternatively, the density may remain nearly flat from kiloparsec scales down to the Bondi radius due to the inflating radio bubbles \citep[e.g.,][]{boehringer93, fabian03, fabian06, fabian11} or the presence of a multiphase medium, in which cold gas sinks toward the center, enhancing mixing and flattening the overall density profile \citep{fabian06, gaspari12, gaspari13, oosterloo24, olivares25}. Adopting an average density of $n_e = 0.3$~cm$^{-3}$ within 2~kpc \citep{taylor06} as the outer boundary condition and applying an inner slope of $\alpha=1.5$ then sets the lower envelope. The region between these two limiting cases represents the possible range of electron-density profiles inside the Bondi radius, shown as magenta shaded area in Figure~\ref{fig:both_profiles}a.

These results are consistent with independent constraints provided by the BLR and the FFA disk. The BLR is often regarded as tracing the inner edge of the dusty torus or disk \citep[e.g.,][]{czerny11, czerny17, baskin18,  Shablovinskaya20}, and the FFA disk in 3C~84 may represent a related circumnuclear disk structure. Both the BLR and the FFA disk are much denser than the diffuse ambient medium at comparable radii. Their densities therefore provide upper limits on the density of the ambient medium at those distances. 
The density of the BLR in 3C~84 is estimated to be $(0.6-3)\times10^{10}$~cm$^{-3}$ \citep{punsly18}. Its size derived from photoionization modeling with CLOUDY \citep{ferland17} and from empirical correlations between BLR size, continuum luminosity \citep{bentz13}, and line flux \citep{greene05} yields $10^{16}-10^{17}$~cm \citep{punsly18, oh22}, corresponding to 0.003--0.03~pc. For the FFA disk, the electron density has been measured to lie in the range $1.8\times10^4\ \text{cm}^{-3} < n_{\rm e} < 10^6\ \text{cm}^{-3}$ \citep{wajima20}, with independent estimates suggesting $n_{\rm e} > 10^5\ \text{cm}^{-3}$ \citep{fujita16} at 2--2.5~mas from the core, and $n_{\rm e} > 2\times10^4$~$\rm cm^{-3}$ at $\sim$8~mas \citep{walker00}. Importantly, these upper limits lie well above the density range derived from both the RM profile and the jet–ambient pressure balance, providing a consistent and mutually supporting picture.

At the smallest radii near the black hole, however, the electron density profile may deviate from a simple power-law trend and become flatter. In the vicinity of the black hole, a larger fraction of the ambient medium within a given radius can be affected by interactions with the jet. This effect is likely enhanced by the rapid and irregular changes in the jet direction close to the core \citep[e.g.,][]{dominik21, park24_3c84, foschi25}, while the inner jet as a whole maintains a stable orientation on larger scales \citep{kam24}. Such behavior can increase jet-ambient interactions and promote efficient mixing of the surrounding medium, which may flatten the density at small radii and, in turn, lead to a corresponding flattening of the RM profile near the black hole. 

The magnetic field slope $\beta$ inside the Bondi radius is constrained by the observed RM profile together with the allowed range of electron density distributions. In particular, the two limiting cases of $\alpha=1.5$ and $\alpha=2.0$ bracket the possible range of $\beta$. For $\alpha=1.5$, corresponding to a Bondi-like inflow, we find $\beta \simeq 2.2$, while for the upper-limit case of $\alpha=2.0$, we obtain $\beta \simeq 1.7$. Intermediate values of $\alpha$ between these limits yield intermediate slopes, $1.7 < \beta < 2.2$, thereby defining the allowed range of magnetic-field profiles within the Bondi radius.

This range of $\beta$ is close to that expected for a magnetic field dominated by its radial component, for which $B_r \propto r^{-2}$. This implies that the magnetic field in the circumnuclear ambient medium of NGC~1275 is predominantly radial within the Bondi radius. Such a steep radial decline is clearly different from the magnetic-field structure generally inferred for AGN jets, where theoretical models \citep{bz77, bp82} and polarimetric observations and simulations \citep[e.g.,][]{asada02, asada08, gabuzda04, gabuzda15, osullivan09, broderick10} indicate the presence of helical magnetic fields with a substantial toroidal component. In idealized jet models, the toroidal component ($B_\phi \propto r^{-1}$) is expected to dominate over the more rapidly decaying radial component ($B_r \propto r^{-2}$), resulting in a comparatively flatter magnetic-field profile \citep[e.g.,][]{jorstad07, ro23}. Although recent work suggests that non-ideal MHD effects such as magnetic dissipation could reduce toroidal-field dominance in jets \citep{park26}, the range $1.7 \le \beta \le 2.2$ derived here remains steeper than typically inferred for jets. This favors a poloidal-dominated magnetic field in the circumnuclear ambient medium of NGC~1275 and highlights a physical distinction between the ambient medium and the jet itself.


At the Bondi radius, the electron density spans $n_e = 0.3$–$6.8$~cm$^{-3}$. Using these density limits together with the best-fit to the observed RM and adopting the Bondi radius as the effective outer boundary of the Faraday screen (beyond which the RM contribution is negligible), we derive upper and lower bounds on the magnetic field strength of $1.5~\mu$G and $0.1~\mu$G, respectively. The higher slope $\beta=2.2$ anchored at the upper limit sets the maximum field strength at smaller radii, while the lower slope $\beta=1.7$, anchored at the lower limit sets the minimum. This constrains both the magnitude and radial dependence of the magnetic field, providing the possible range within the Bondi radius. This provides the first spatially resolved measurement of the ambient magnetic-field strength within the parasec-scale Bondi radius of an elliptical galaxy.

Figure~\ref{fig:both_profiles}b shows the possible range of the magnetic field strength. When extrapolated inward, the ambient magnetic-field profile approaches the values independently measured for the VLBI core at sub-milliarcsecond distances from the black hole. To place these core magnetic-field measurements at the appropriate distances, we briefly summarize current constraints on the location of the VLBI core. Core-shift measurements between 15, 43, and 86~GHz place the 86~GHz core at a projected distance of $\sim$0.083~mas from the black hole \citep{paraschos21}. However, independent observational constraints suggest that this value likely represents an upper limit. RadioAstron space-VLBI observations indicate that the 22~GHz core lies within 0.03~mas of the black hole \citep{giovannini18}, and more recent studies further suggest that the true core shift may be smaller by up to an order of magnitude than inferred from the 15–86~GHz measurements \citep{savolainen23, park24_3c84}. To account for these uncertainties, we conservatively adopt a projected distance range of 0.01–0.1~mas for the location of the 86~GHz core.

At these distances, magnetic-field strengths measured for the VLBI core include $21\pm14$~G from synchrotron self-absorption (SSA) \citep{kim19}, $2.9\pm1.6$~G from SSA \citep{paraschos24_eht}, $5.2\pm1.6$~G from an equipartition condition, and $1.8$--$4.0$~G from the core-shift \citep{paraschos21}. These values should be regarded as upper limits on the ambient magnetic field at similar radii, since relativistic jets are expected to be significantly more strongly magnetized than their surroundings \citep[e.g.,][]{nakamura18}. This implies that the steep magnetic-field profile with $\beta \sim 2$ cannot extend to the innermost regions near the black hole without exceeding these upper limits. Instead, it would flatten at some inner radii so that the extrapolated field strength remains below or at most comparable to the measured field strength at the jet core. 

On larger scales, the magnetic-field strength at the Bondi radius of 11.8~pc in NGC~1275 is $0.1$–$1.5~\mu$G, lies well within the range of the galaxy-averaged field strength in elliptical galaxies, which typically span $0.1-10~\mu$G \citep[e.g.,][]{moss96, shah21, seta21}. This implies that the ambient magnetic field increases rapidly inside the Bondi radius, from $\mu$G levels at parsec scales toward the mG–G field strengths at subparsec distances. 

It is interesting to note that the magnetic-field strength at the Bondi radius is also comparable to the magnetic-field strength of $24~\mu$G measured in the kiloparsec-scale filamentary structures of NGC~1275 \citep{fabian08}, as well as to field strengths typically inferred in cool-core galaxy clusters such as Perseus (see \citealp{carilli02, govoni04} for reviews). These filaments are thought to have been dragged outward from the galaxy center by rising radio-emitting bubbles \citep{conselice01, fabian03_drag, fabian08, hatch06, fabian12, vigneron24}. In this picture, the similarity in magnetic-field strength across these scales is naturally explained by efficient mixing driven by AGN feedback, implying that, like the electron density, the magnetic-field profile flattens outside the Bondi radius due to the influence of inflating radio bubbles and mixing associated with the multiphase medium.

We note that this work represents the first spatially resolved measurement of the ambient magnetic-field strength on parsec-scales of an elliptical galaxy, and the first determination of its radial structure within the Bondi radius. Measuring magnetic fields in elliptical galaxies has long been challenging, primarily because they lack the ongoing star formation that supplies the relativistic electrons required for synchrotron emission and for tracing magnetic fields through polarization \citep{widrow02, beck15}. For this reason, previous magnetic-field estimates have relied either on theoretical estimates from fluctuation-dynamo theory \citep{moss96, seta21} or RM measurements averaged over galaxy scales on hundreds of parsecs to kiloparsecs \citep{shah21}. 

Our RM profile directly traces the radial variation of the magnetic-field strength within the Bondi radius, providing information that was inaccessible in earlier studies. Moreover, the fact that all measured RM values are positive indicates that the line-of-sight magnetic field maintains a coherent orientation throughout the circumnuclear region. 

\section{Conclusions} \label{sec:Conclusions}
We conducted multi-frequency polarimetric observations of 3C~84 using the KVN, VLBA, and HSA from 2015 to 2024. BU data obtained between 2015 December and 2017 May show that linear polarization was consistently detected from component C3, yielding an average RM of $(3.9\pm0.6)\times10^5~{\rm rad~m^{-2}}$ within the observing bandwidth. Subsequent KVN monitoring at 43–141~GHz, beginning in 2018 February, also consistently detected polarized emission from C3, which persisted throughout the monitoring period until 2024 December. In addition to C3, polarized emission was detected from other components along the inner jet, including C4 and K, distributed between 3 and 7~mas from the black hole. Notably, this represents the first detection of polarization from components C4 and K. We also detected polarized emission from the southern lobe at a distance of $\sim15$~mas using the HSA. 

We measured the RM toward these components, finding values that span from $\rm (3\pm6)\times10^3\ rad\ m^{-2}$ to $\rm (6.1\pm1.1)\times10^5\ rad\ m^{-2}$. All RM values toward the inner jet and the southern lobe decrease systematically with distance from the black hole, following a single power-law relation, $\mathrm{RM}\propto r^{-k}$ with $k=2.7\pm0.2$. RM values measured with CARMA, SMA, and IRAM PdBI between 2010 and 2014 also follow this RM–distance trend when attributed to component C3. These results show that all RM measurements over multiple decades and across a wide range of spatial scales follow a single relation, providing strong evidence that they originate in a common external Faraday screen surrounding both the inner jet and the southern lobe.

The observed RM properties are naturally explained if the Faraday screen corresponds to the circumnuclear ambient medium filling the central region of the host galaxy NGC~1275. In this framework, the Bondi radius of 11.8~pc defines the outer boundary of the Faraday screen, as all RM sites lie within this radius and exhibit a continuous radial dependence. The observed RM–distance relation implies that both the electron density and magnetic-field strength decrease smoothly with radius, consistent with gas gravitationally bound to the central black hole. 

For possible density profiles within the Bondi radius, characterized by $n_e \propto r^{-\alpha}$ with $\alpha=1.5$–2.0, the derived densities are consistent with independent constraints from the BLR and the FFA disk, whose much higher densities at comparable radii provide robust upper limits on the circumnuclear ambient medium. 

Given these density profiles, the RM-distance relation yields magnetic-field profiles of $B \propto r^{-\beta}$ with $\beta \sim 1.7$–2.2. This indicates that the magnetic-field strength in the circumnuclear ambient medium ranges from $0.1$ to $1.5~\mu$G at the Bondi radius and increases rapidly toward the black hole, approaching the magnetic-field strength measured in the core, 1--20~G. This provides the first spatially resolved constraints on the ambient magnetic-field strength within the parsec-scale Bondi radius of an elliptical galaxy.

In summary, these results provide a unified interpretation of the wide range of RM measurements in 3C~84 and suggest that they arise from a common circumnuclear ambient medium. By linking the electron density and magnetic-field strength across a broad range of spatial scales, this study provides a spatially resolved and physically consistent view of the environment of NGC~1275/3C~84.

\section*{Acknowledgments}
M.Kam acknowledges S.Jorstad and A.Marscher for useful discussions regarding the BU data. M.Kam also acknowledges A.Marconi for helpful discussions on black hole mass estimates based on the single-epoch method.
This research was supported by Basic Science Research Program through the National Research Foundation of Korea (NRF) funded by the Ministry of Education (RS-2024-00412117). 
This work was supported by National Research Foundation of Korea (NRF) grants funded by the Korea government (MSIT; RS-2024-00449206 and RS2025-02214038). 
This research has been supported by the POSCO Science Fellowship of POSCO TJ Park Foundation.
This research was supported by the Global-Learning \& Academic research institution for Master’s, PhD students, and Postdocs (G-LAMP) Program of the National Research Foundation of Korea (NRF) grant funded by the Ministry of Education (RS-2025-25442355). 
This work was supported by the BK21 FOUR program through the National Research Foundation of Korea (NRF) under the Ministry of Education (Kyung Hee University, Human Education Team for the Next Generation of Space Exploration).
This research was supported by the Korea Astronomy and Space Science Institute under the R\&D program(Project No. 2025-9-844-00) supervised by the Korea AeroSpace Administration.
This work was supported by National Research Foundation of Korea (NRF) grants funded by the Korea government (RS-2025-16067786). 
K.A. acknowledges financial support from AS-CDA-110-M05, NSTC 114-2124-M-001-015, and NSTC 114-2112-M-001-016. 
M.Kino is supported by the JSPS KAKENHI Grant number JP22H00157 and JP21H01137. We are grateful to the staff of the KVN who helped to operate the array and to correlate the data. 
LCH was supported by the China Manned Space Program (CMS-CSST-2025-A09) and the National Science Foundation of China (12233001).
J.A.H. acknowledges the support of the National Research Foundation of Korea (NRF) (NRF-2021R1C1C1009973) and that this work was supported by the National Research Foundation of Korea (NRF) grant funded by the Korea government(MSIT) RS-2025-16302968.
The KVN and a high-performance computing cluster are facilities operated by the KASI (Korea Astronomy and Space Science Institute). The KVN observations and correlations are supported through the high-speed network connections among the KVN sites provided by the KREONET (Korea Research Environment Open NETwork), which is managed and operated by the KISTI (Korea Institute of Science and Technology Information). This study makes use of VLBA data from the VLBA-BU Blazar Monitoring Program (BEAM-ME and VLBA-BU-BLAZAR; http://www.bu.edu/blazars/BEAM-ME.html), funded by NASA through the Fermi Guest Investigator Program. The VLBA is an instrument of the National Radio Astronomy Observatory. The National Radio Astronomy Observatory is a facility of the National Science Foundation operated by Associated Universities, Inc.

%

\vspace{5mm}

\bibliography{ref}{}
\bibliographystyle{aasjournal}

\appendix

\section{KVN D-terms}\label{app:dterm}

\begin{table*}[hb!]
    \centering 
    \caption{KVN D-terms [\%] in 2018--2023 \label{tab:dterm1}}
    \begin{tabular}{cc cc cc cc}
        \hline
    \multicolumn{2}{c}{Frequency} & \multicolumn{2}{c}{TN} & \multicolumn{2}{c}{US} & \multicolumn{2}{c}{YS}  \\
        \cline{3-8}
        & & LCP & RCP & LCP & RCP & LCP & RCP  \\ 
        \cline{1-8}
        \multirow{2}{*}{Q-high} & 43 & $3.7\pm0.2$ & $3.7\pm0.2$ & $3.5\pm0.2$ & $3.8\pm0.2$ & $4.8\pm0.8$ & $6.5\pm0.3$  \\
        & 46 & $3.6\pm2.2$ & $6.9\pm2.2$ & $2.0\pm0.5$ & $3.4\pm1.2$ & $6.5\pm0.8$ & $5.3\pm0.4$  \\
        \hline
        \multirow{5}{*}{W-low} & 86 & $9.4\pm1.4$ & $9.4\pm2.3$ & $11.6\pm1.3$ & $12.8\pm2.4$ & $7.8\pm3.6$ & $4.8\pm2.8$  \\
        & 88 & 5.5 & 6.3 & 11.5 & 11.6 & 10.3 & 5.9  \\
        & 90 & 3.8 & 11.8 & 7.9 & 10.4 & 8.6 & 5.1  \\
        & 92 & $10.2\pm1.3$ & $5.1\pm0.6$ & $8.3\pm0.8$ & $8.5\pm1.3$ & $9.1\pm2.2$ & $11.9\pm1.6$  \\
        & 94 & $9.7\pm0.3$ & $12.5\pm1.6$ & $9.1\pm1.1$ & $10.6\pm0.5$ & $2.4\pm1.2$ & $4.0\pm0.7$  \\
        \hline
        \multirow{3}{*}{D} & 129 & $3.7\pm1.6$ & $6.8\pm3.2$ & $2.6\pm1.9$ & $10.1\pm3.2$ & $9.3\pm1.9$ & $10.0\pm1.5$  \\
        & 139 & $5.4\pm0.6$ & $6.4\pm0.6$ & $3.0\pm0.8$ & $3.3\pm0.9$ & $1.5\pm0.4$ & $4.8\pm0.6$  \\
        & 141 & 4.2 & 8.2 & 2.5 & 2.9 & 3.5 & 5.4  \\
        \hline
    \end{tabular}
\tablecomments{KVN D-terms measured from the following epochs between 2018 and 2023: p18st01b, n18jp01a, n18jp01b, p18st01o, p18st01p, p19st01g, p18st01h, p21st02e, p22st02b, p22stf02, and t23jp01a.}
\end{table*}

\begin{table*}[ht!]
    \centering 
    \caption{KVN D-terms [\%] in 2022--2024 \label{tab:dterm2}}
    \begin{tabular}{cc cc cc cc cc}
        \hline
        \multicolumn{2}{c}{Frequency} & \multicolumn{2}{c}{PC} & \multicolumn{2}{c}{TN} & \multicolumn{2}{c}{US} & \multicolumn{2}{c}{YS}  \\
        \cline{3-10}
        & & LCP & RCP & LCP & RCP & LCP & RCP & LCP & RCP \\ 
        \cline{1-10}
        \multirow{4}{*}{Q-low} & 35.5 & 7.4 & 9.0 & 5.6 & 4.2 & 7.0 & 1.8 & 4.6 & 10.0 \\
        & 36.0 & 7.5 & 8.6 & 5.8 & 5.3 & 5.5 & 3.5 & 4.9 & 9.3 \\
        & 37.0 & 5.5 & 6.0 & 4.7 & 4.5 & 2.4 & 1.8 & 4.4 & 7.1 \\
        & 38.0 & 6.4 & 6.5 & 5.6 & 4.4 & 3.2 & 3.6 & 4.9 & 5.2 \\
        \hline
        \multirow{4}{*}{Q-high} & 43.1 & $2.9\pm1.3$ & $3.9\pm1.4$ & $3.4\pm0.8$ & $3.6\pm0.8$ & $3.3\pm0.9$ & $3.6\pm0.9$ & $4.7\pm0.8$ & $5.7\pm0.8$ \\
        & 44.0 & $3.8\pm0.3$& $3.0\pm0.2$& $2.9\pm0.2$& $3.8\pm0.2$& $2.8\pm0.4$& $3.7\pm0.4$& $5.0\pm0.3$& $5.2\pm0.5$ \\
        & 45.5 & $6.2\pm1.1$& $7.6\pm1.3$& $4.6\pm0.9$& $4.2\pm0.9$& $2.6\pm0.9$& $10.0\pm1.6$& $6.0\pm2.5$& $7.3\pm0.9$\\
        & 46.5 & $9.7\pm1.9$& $10.5\pm2.0$& $3.7\pm1.0$& $6.6\pm1.3$& $2.0\pm1.1$& $3.5\pm1.9$& $6.2\pm1.2$& $6.1\pm1.0$\\
        \hline
        \multirow{4}{*}{W-low} & 86.2 & $7.1\pm0.8$& $5.3\pm1.1$& $11.3\pm0.6$& $7.4\pm1.7$& $13.6\pm0.6$& $11.0\pm1.3$& $11.8\pm0.9$& $7.1\pm0.8$\\
        & 88.0 & $8.3\pm0.6$& $5.5\pm0.5$& $3.9\pm0.6$& $6.0\pm0.6$& $10.4\pm0.7$& $10.1\pm0.8$& $10.2\pm0.6$& $4.8\pm0.9$\\
        & 91.0 & $7.0\pm0.3$& $4.3\pm0.6$& $7.7\pm1.2$& $7.8\pm1.0$& $8.7\pm1.2$& $9.7\pm0.6$& $10.6\pm0.7$& $10.4\pm0.5$\\
        & 93.0 & $5.7\pm0.3$& $3.7\pm0.4$& $3.5\pm1.7$& $5.5\pm0.8$& $7.7\pm3.0$& $7.5\pm1.9$& $7.4\pm2.2$& $2.7\pm1.2$\\
        \hline
        \multirow{4}{*}{W-high}  & 106.5 & 5.4 & 2.3 & 5.6 & 6.6 & 5.8 & 7.0 & 8.3 & 10.9 \\
        & 108.0 & 4.2 & 1.6 & 9.9 & 7.6 & 6.4 & 5.1 & 9.8 & 11.8 \\
        & 111.0 & 3.8 & 1.7 & 7.1 & 8.9 & 4.7 & 4.9 & 8.3 & 9.2 \\
        & 114.0 & 2.8 & 3.5 & 15.4 & 13.1 & 4.5 & 14.0 & 5.8 & 15.9 \\
        \hline
    \end{tabular}
\tablecomments{KVN D-terms measured from the following epochs between 2022 and 2024, using the same frequency setup within each frequency band: p22st02d, p22st02e, n24jp01a, n24jp01c, n24jp01d, and n24jp01e.}
\end{table*}

\begin{figure*}[t!]
\centering 
\includegraphics[width=0.9\textwidth]{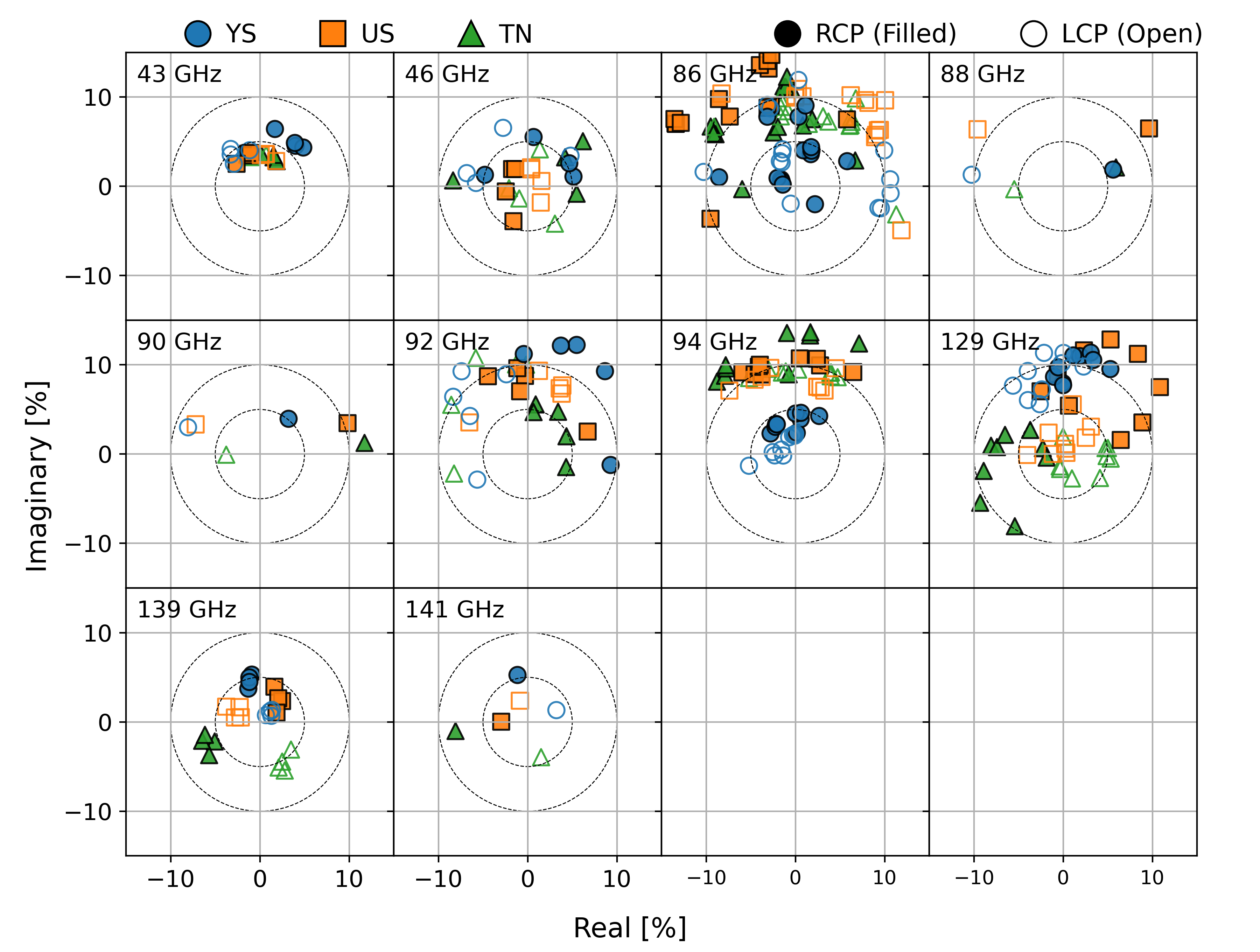}
\caption{KVN D-terms in Table~\ref{tab:dterm1}.
\label{fig:app_dterm1}}
\end{figure*}

\begin{figure*}[t!]
\centering 
\includegraphics[width=0.9\textwidth]{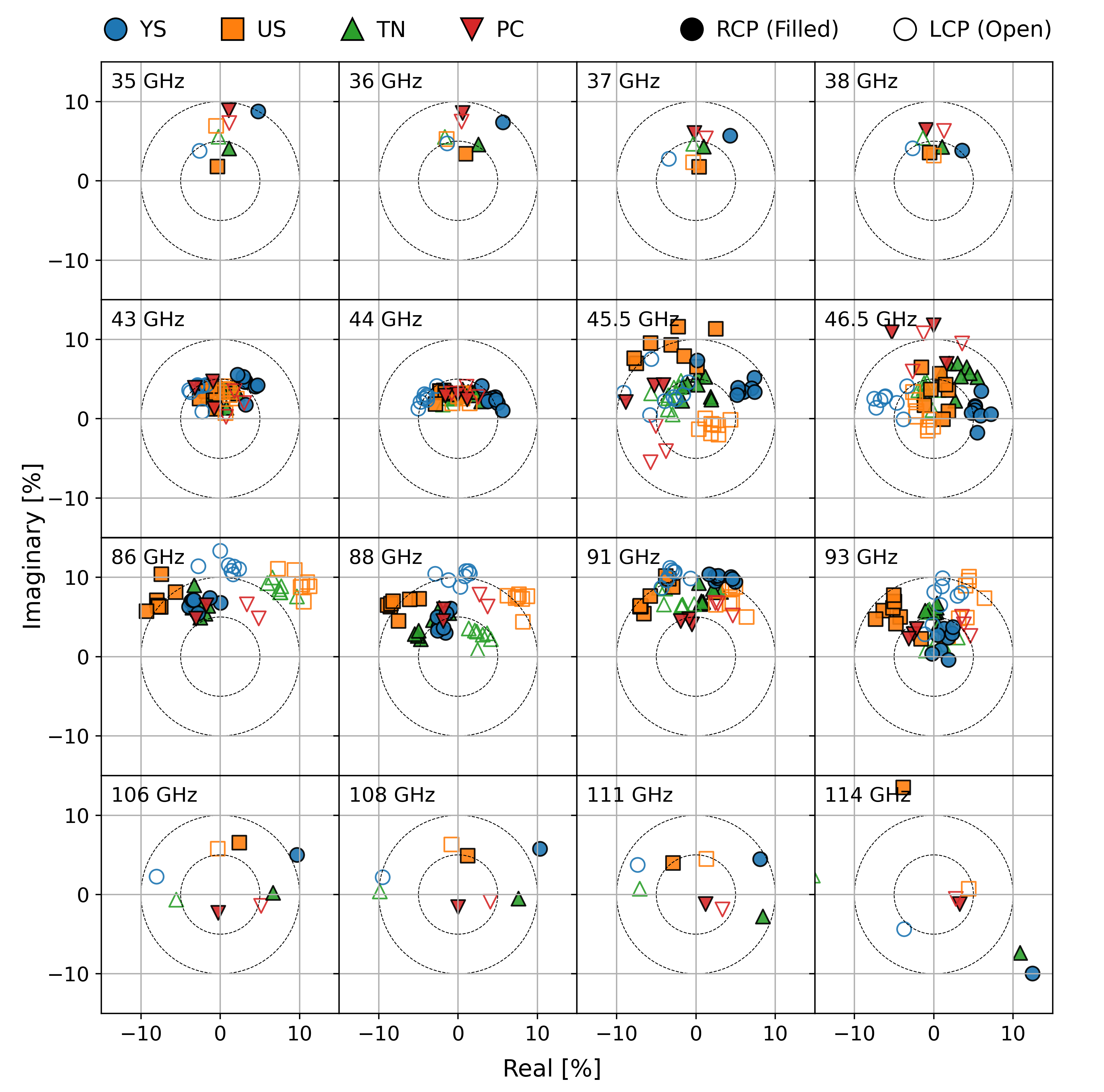}
\caption{KVN D-terms in Table~\ref{tab:dterm2}.
\label{fig:app_dterm2}}
\end{figure*}

\noindent We summarize the KVN D-terms in Tables~\ref{tab:dterm1} and \ref{tab:dterm2}. Table~\ref{tab:dterm1} lists D-terms measured from observations conducted between 2018 February and 2023 February, which used different frequencies and bandwidths setups. These values are shown in Figure~\ref{fig:app_dterm1}. In contrast, Table~\ref{tab:dterm2} presents D-terms obtained at identical frequencies between 2022 November and 2024 December, allowing us to evaluate their temporal stability. These values are displayed in Figure~\ref{fig:app_dterm2}. 

In the Q-low (35--38~GHz) and W-high (106--114~GHz) bands, observations were conducted only once (Exp. code: p24st01d), so stability cannot be evaluated. In contrast, multiple observations in the Q-high (43–46.5~GHz) and W-low (86–93~GHz) bands between 2022 November and 2024 December enable us to examine their stability. In the Q-high band, the D-terms are stable at 43 and 44~GHz, but show larger scatter and higher amplitudes at 45.5 and 46.5~GHz, indicating larger measurement uncertainty or weaker at these frequencies. In the W-low band, the D-terms remain stable across all frequencies within the measurement uncertainties, reported as the standard deviation in Table~\ref{tab:dterm2}. They tend to decrease at higher frequencies, opposite to the trend seen in the Q-high band. 

We note that the observations in 2024 November and December (Exp. code: p24st01a, p24st01c, p24st01e), which included all four telescopes, yield D-terms consistent with those obtained without the Pyeongchang station in 2022 November and December (Exp. code: p22st02d, p22st02e). The consistency over a two-year interval and across different station configurations demonstrates the stability of the KVN D-terms within the measurement uncertainties.

\section{Verification of KVN EVPA Uncertainty Estimates} \label{app:evpa}

We verified the EVPA uncertainties using two representative datasets. For the Q band, we used p22st02b (43.1, 43.228, 46.0, and 46.128~GHz; 128~MHz each), and for the W band, p22stf02 (86.2, 86.328, 92.0, and 92.128~GHz), comparing EVPAs between the first two and last two frequencies in each band (Table~\ref{tab:pol_error}). 
For both datasets, all polarization calibrations from D-term correction to absolute EVPA calibration were performed independently for each frequency. 

The EVPAs between those frequency pairs are mostly consistent within their measurement uncertainties, except for component K in dataset p22stf02 at 46.0 and 46.1~GHz. Even in this case, the values remain consistent within twice the uncertainty. This minor deviation is likely due to relatively unstable D-term behavior at 45–46~GHz compared to other frequencies, which may not be fully captured in the residual maps center.

\begin{table*}[ht]
    \centering 
    \caption{EVPA measurement uncertainty [$^\circ$] in the Q and W bands \label{tab:pol_error}}
    \begin{tabular}{cc cccc cccc}
        \hline
    Exp. code & Comp. & \multicolumn{4}{c}{Frequency (Q-band)} & \multicolumn{4}{c}{Frequency (W-band)} \\
        \cline{3-10}
        & & 43.1 & 43.228 & 46.0 & 46.1 & 86.2 & 86.328 & 92.0 & 92.1 \\ 
        \cline{1-10}
        \multirow{2}{*}{p22st02b} & C3 & $172.3\pm5.2$ & $173.4\pm5.4$ & $108.7\pm5.0$ & $112.2\pm5.5$ & - & - & - & - \\
        & K & $64.4\pm2.9$& $63.4\pm2.6$& $31.5\pm3.4$& $37.2\pm4.6$& - & - & - & - \\
        \hline
        \multirow{2}{*}{p22stf02} & C3 & - & - & - & - & $14.8\pm1.5$& $13.9\pm1.5$& $5.7\pm1.3$& $6.3\pm1.1$\\
        & K & $58.2\pm1.8$& $57.8\pm2.2$& $34.6\pm4.3$& $25.7\pm4.6$& -& -& -& -\\
        \hline
    \end{tabular}
\end{table*}

\section{Comparison of KVN and BU data}\label{app:comparison}

\begin{figure*}[t!]
\centering 
\includegraphics[width=\textwidth]{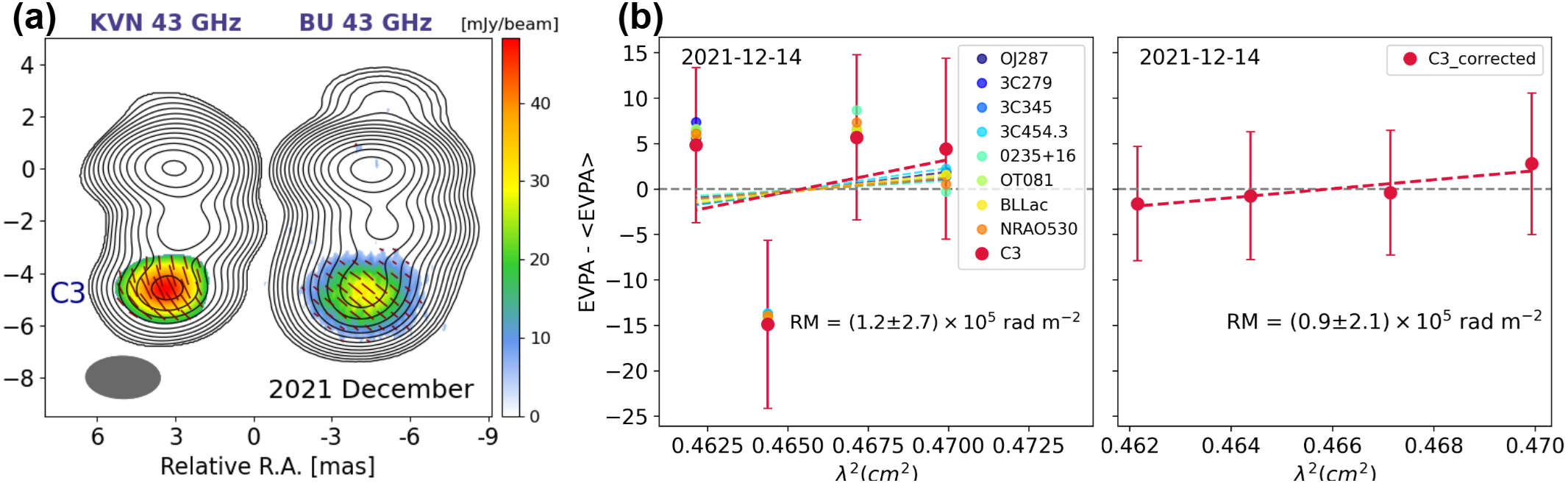}
\caption{(a): KVN and BU images of 3C~84 at 43 GHz obtained on 2021 December 1 and 2021 December 14, respectively. The VLBA image is convolved with the KVN beam size. (b): Same as Figure~\ref{fig:evpa_correction} but for the BU data obtained on 2021 December 14. The in-band RM at C3 is $\rm (0.9\pm2.1)\times10^5\ rad\ m^{-2}$. 
\label{fig:evpa_correction1}}
\end{figure*}

After the consistent detection of the polarization at C3 between 2015 December and 2017 May (Figure~\ref{fig:vlba_img1}), polarization was not reliably detected at C3 with the BU data for the next few years. While polarized features occasionally appear in the C3 region in some epochs, they are not consistently detected when applying a conservative threshold of ${\rm SNR} > 7$. However, it became consistently detected again from mid-2020 onward, after the bandwidth was expanded from 256~MHz to 512~MHz. In particular, the BU data obtained on 2021 December 14 provides a useful opportunity for comparison with our KVN 43 GHz data obtained on 2021 December 1.

To compare the two datasets, we convolved the BU image with the KVN beam of $2.84\times1.61$~mas at $\rm P.A.=89^\circ$. In both images, a single polarized feature is detected in the jet termination region. This morphology differs from the KVN 86–129~GHz images obtained on 2021 December 1 (Figure~\ref{fig:19jan21dec}), which resolve two polarized components located at C3 and K. This can be understood because the KVN angular resolution at 43~GHz is insufficient to separate these components. The single polarized feature detected at 43~GHz therefore likely represents blended emission from both C3 and K. 

The EVPA values at the jet termination region are not exactly identical between the two images, likely due to time variability over the two-week interval between the KVN and BU observations (see Figure~\ref{fig:evpa_correction1}a). However, the RM in 3C~84 is expected to be stable over time and can therefore be used to check the consistency between the KVN and BU data. We derived an in-band RM from the BU data obtained on 2021 December 14 using the full 512~MHz bandwidth. This bandwidth is two times wider than that used in earlier BU observations between 2015 December and 2017 May, from which we derived an average RM of $\rm (3.9\pm0.6)\times10^5\ rad\ m^{-2}$ at C3 (Table~\ref{tab:rm_bu}). With this wider bandwidth, the RM of similar magnitude would be expected to produce a more clearly detectable in-band EVPA rotation.

However, we did not detect a significant RM in this epoch, deriving an in-band RM of $\rm (0.9\pm2.1)\times10^5\ rad\ m^{-2}$ at the jet termination region (Figure~\ref{fig:evpa_correction1}b). Although the uncertainty is large, this value is smaller than the average in-band RM of $\rm (3.9\pm0.6)\times10^5\ rad\ m^{-2}$ measured at C3 between 2015 December and 2017 May. This decrease is consistent with the radial RM trend (Figure~\ref{fig:rm_dist}), as the C3 component continues to propagate outward from the black hole after 2017 May and therefore probes regions with lower RM. This value is also consistent with the contemporaneous KVN measurements obtained on 2021 December 1, yielding $\rm (1.0\pm0.1)\times10^5\ rad\ m^{-2}$ at C3 and $\rm (0.5\pm0.2)\times10^5\ rad\ m^{-2}$ at K (Table~\ref{tab:rm_kvn}). Because the C3 and K components cannot be resolved at the KVN 43 GHz resolution, the in-band RM derived from the BU data likely reflects blended contributions from these two components. This naturally explains the overall consistency between the RM values derived from the BU and KVN data.

Overall, the BU and KVN 43~GHz images in 2021 December show consistent polarization morphology, and the in-band RM derived from the BU data is consistent with the results from the multi-frequency KVN results. This comparison supports the robustness of the KVN polarization and RM measurements.

\begin{figure*}[t!]
\centering 
    \includegraphics[width=0.91\textwidth]{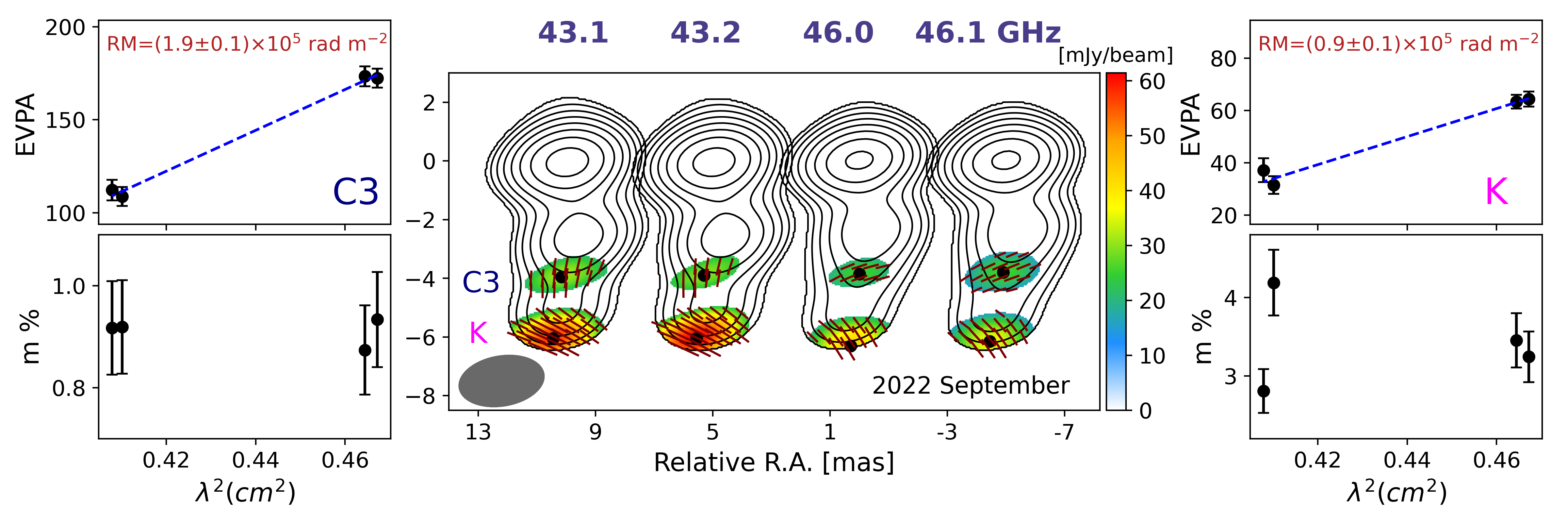}
    \caption{KVN images of 3C~84 at 43.1, 43.2, 46.0, and 46.1~GHz obtained on 2022 September 15. Black dots in the images show the positions of the polarized intensity peaks. Blue dashed lines show the best fit to the EVPA values.  
\label{fig:sep2022}}
\end{figure*}

\begin{figure*}[t!]
\centering 
    \includegraphics[width=0.71\textwidth]{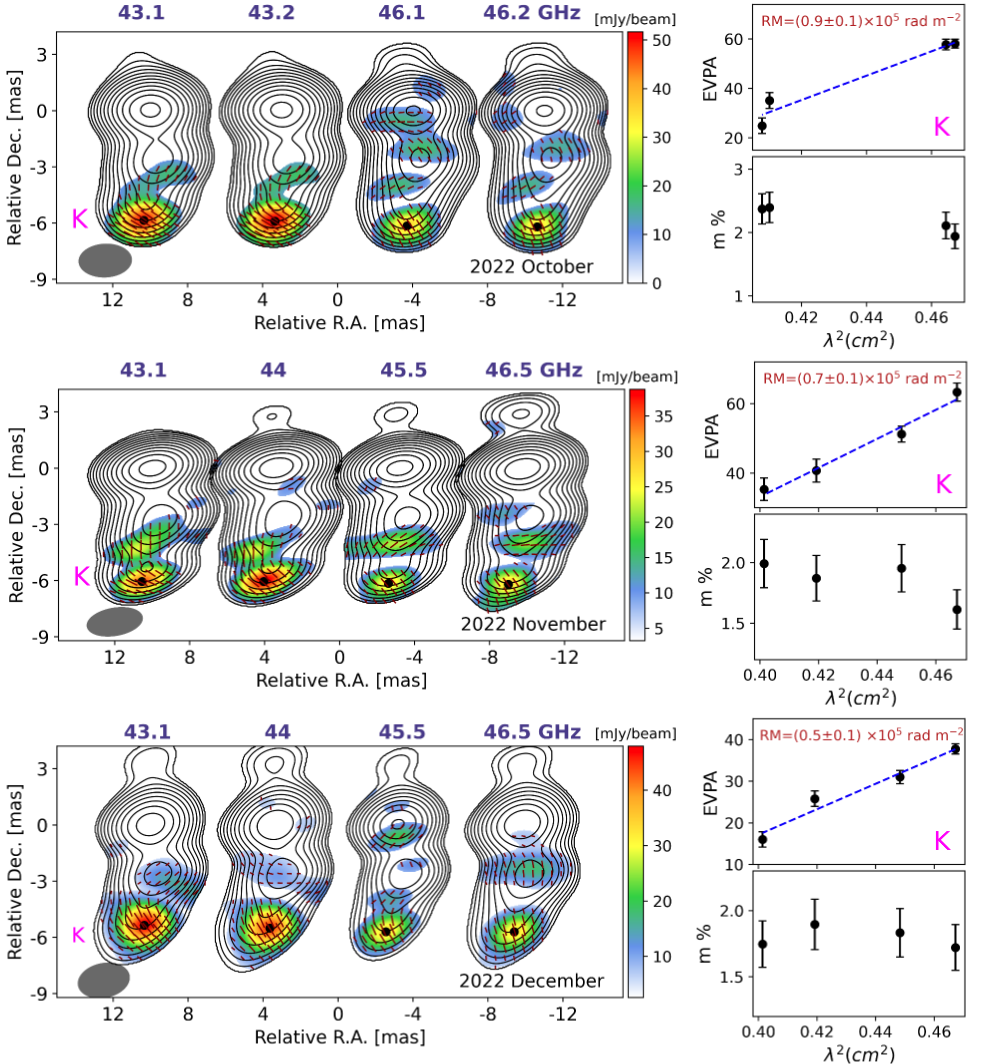}
    \caption{KVN images of 3C~84 obtained on 2022 October 20, November 5, and December 27. The frequencies used for each epoch are 43.1, 43.2, 46.0, and 46.1~GHz on 2022 October 20 and 43.1, 44.0, 45.5, and 46.5~GHz on November 5 and December 27. The right panels show the EVPA, RM, and fractional polarization measured at the polarized intensity peak of K.
\label{fig:22octdec_q}}
\end{figure*}

\begin{figure*}[t!]
\centering 
    \includegraphics[width=0.75\textwidth]{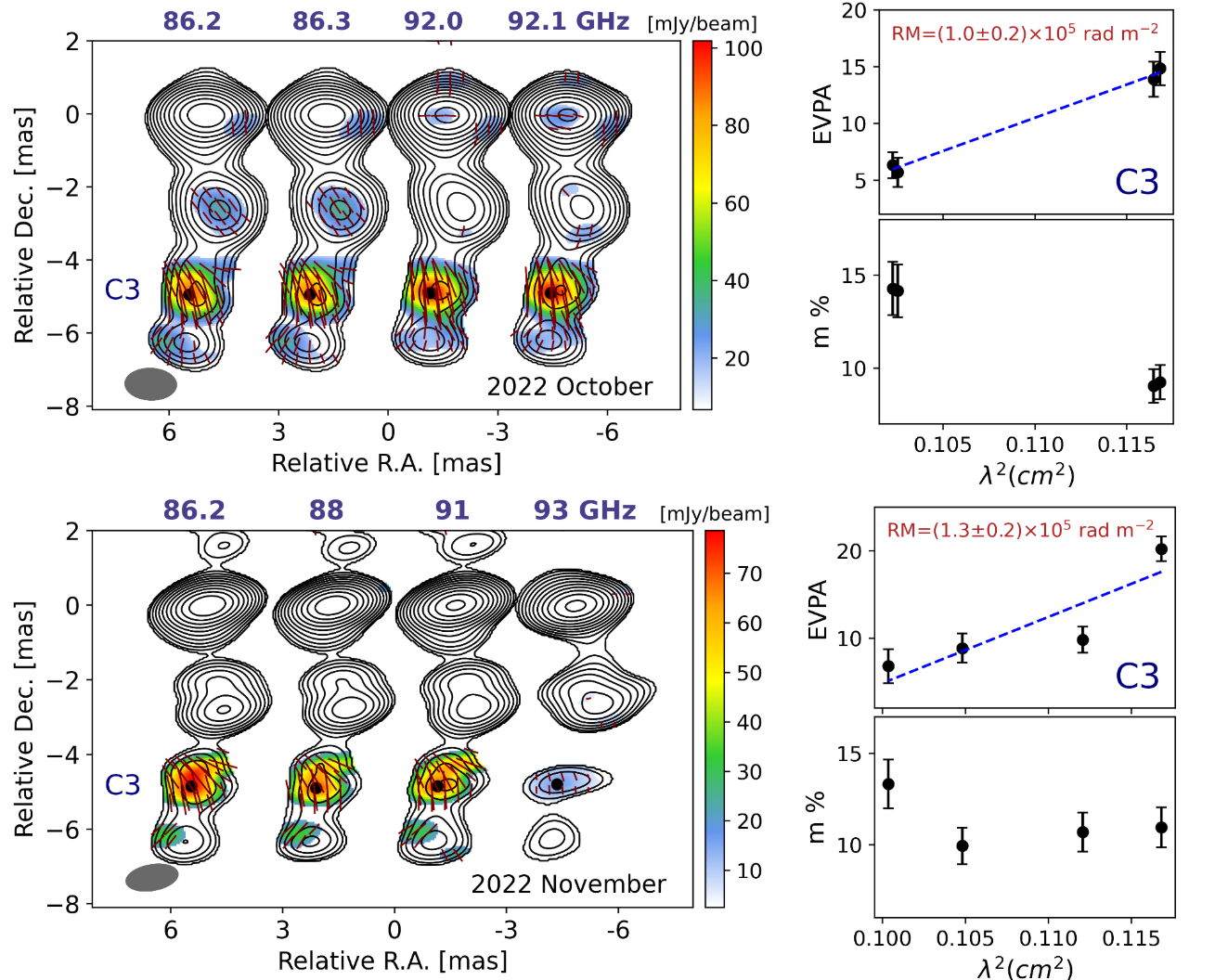}
    \caption{KVN W-band images of 3C~84 obtained on 2022 October 20 and November 5. The EVPA, RM, and fractional polarization measured at the polarized intensity peak of C3 are shown in the right panels.  
\label{fig:22octnov_w}}
\end{figure*}

\section{Polarization and RM (2022 September -- 2023 February)} \label{app:kvn_images_2022}

Figure~\ref{fig:sep2022} presents the KVN images at 43.1, 43.2, 46.0, and 46.1~GHz. As observed in 2021 December (Figure~\ref{fig:19jan21dec}), the RM at component K remains smaller than that at C3. Specifically, the RM at C3 is $\rm (1.9\pm0.1)\times10^5\ rad\ m^{-2}$, while the RM at K is $\rm (0.9\pm0.1)\times10^5\ rad\ m^{-2}$. 

In the Q band observations conducted between 2022 October and December, polarization was reliably detected only at component K and not at C3 (Figure~\ref{fig:22octdec_q}). The observing frequencies were 43.1, 43.2, 46.0, and 46.1~GHz in October, and 43.1, 44.0, 45.5, and 46.5~GHz in November and December. Polarization was consistently detected at K across all these frequencies and epochs, whereas at C3 it was either undetected or exhibits large positional scatter. This is probably due to stronger depolarization at C3 than at K, which prevented reliable polarization measurements at C3. Consequently, we derived the RM only at K, with values of $\rm (0.9\pm0.1)\times10^5\ rad\ m^{-2}$ in October, $\rm (0.7\pm0.1)\times10^5\ rad\ m^{-2}$ in November, and $\rm (0.5\pm0.2)\times10^5\ rad\ m^{-2}$ in December.

\begin{figure*}
\centering 
    \includegraphics[width=\textwidth]{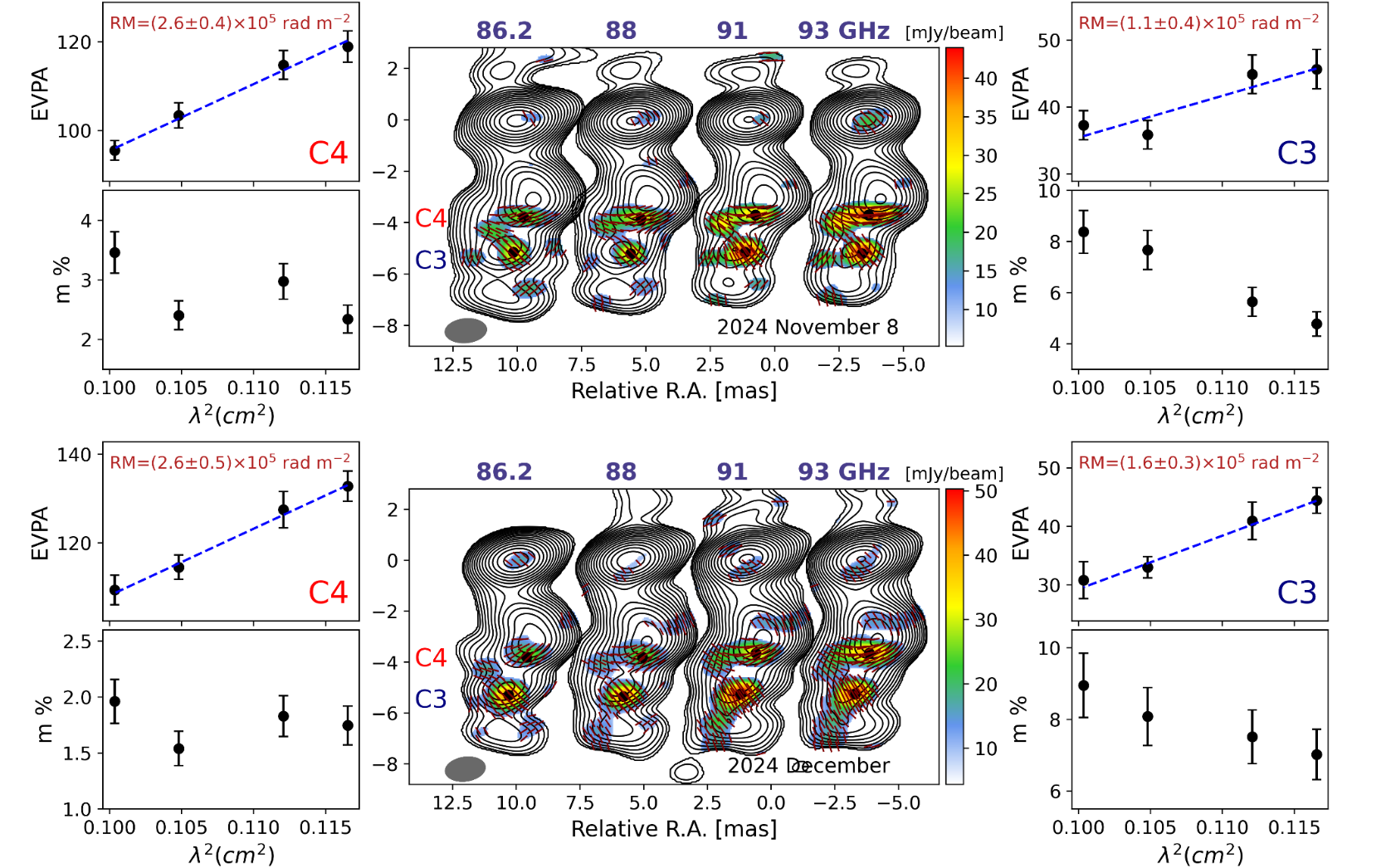}
    \caption{KVN W-band images of 3C~84 obtained at 86.2, 88, 91, and 93~GHz on 2024 November 8 (top) and December 27 (bottom). The EVPA, RM, and fractional polarization measured at the polarized intensity peak of C4 are shown in the left panels, and those of C3 are shown on the right panels. 
\label{fig:novdec2024_w}}
\end{figure*}

\begin{figure*}
\centering 
    \includegraphics[width=\textwidth]{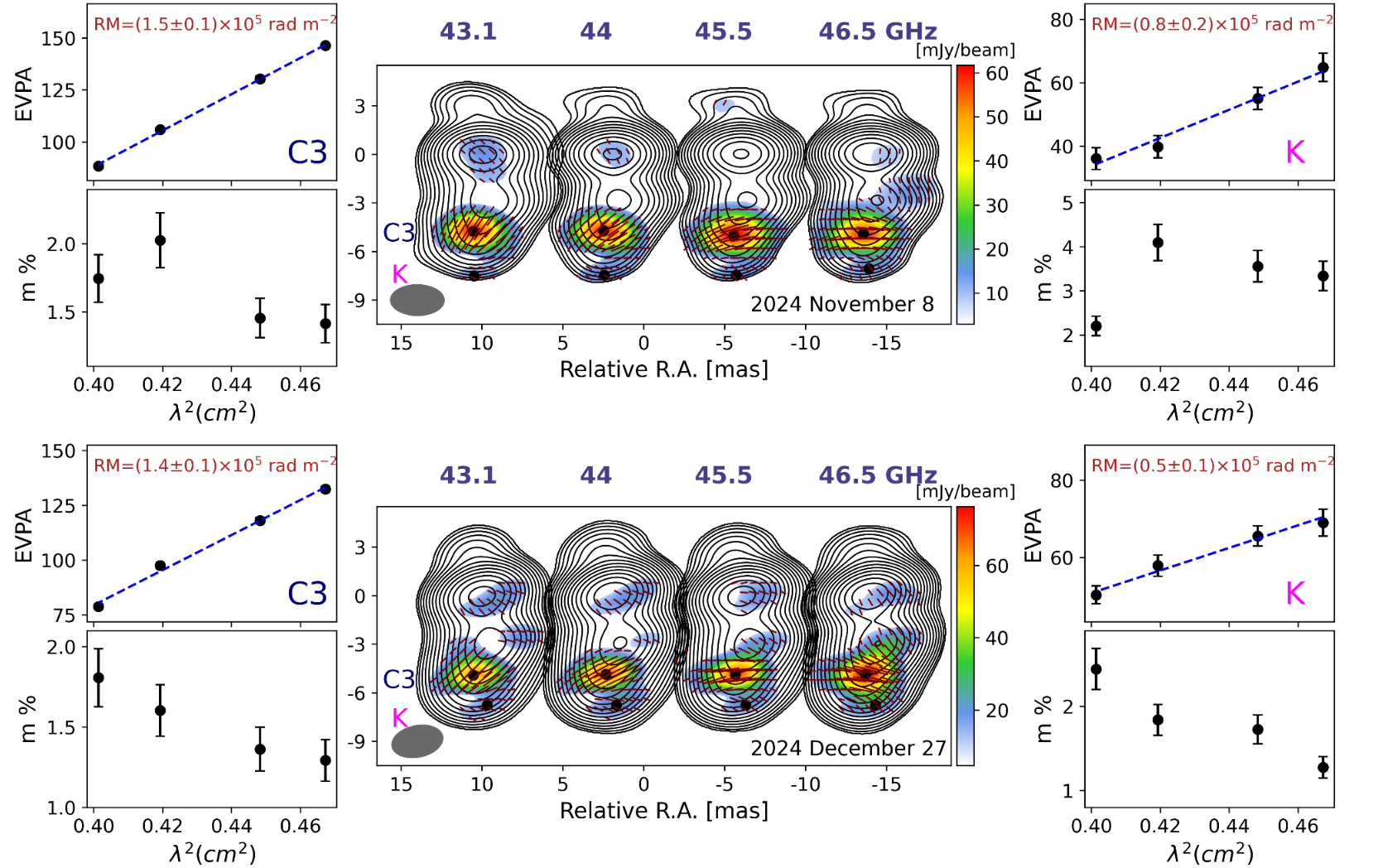}
    \caption{KVN Q-band images of 3C~84 obtained at 43.1, 44.0, 45.5, and 46.5~GHz on 2024 November 8 (top) and December 27 (bottom). The EVPA, RM, and fractional polarization measured at the polarized intensity peak of C3 are shown in the left panels, and those of K are shown on the right panels. 
\label{fig:novdec2024_q}}
\end{figure*}

In contrast, in the W band data from the same observing epochs, polarization becomes clearly detectable at C3 but not reliably detected at K (Figure~\ref{fig:22octnov_w}). 
Polarization features appear to be present at K in October and November, but their peak positions shift inconsistently with frequency: in October, the peak lies on the eastern side at 86.2 and 86.3~GHz but moves to the western side at 92.0 and 92.1~GHz; in November, it appears on the eastern side at 86.2, 86.3, and 91.0~GHz but is absent at 93.0~GHz. Although the RM values derived from these detections, $\rm (1.1\pm1.2)\times10^5\ rad\ m^{-2}$ in October and $\rm (1.2\pm1.6)\times10^5\ rad\ m^{-2}$ in November, are broadly consistent with the Q-band estimates, the shifting polarized-peak locations and inconsistent detectability across frequencies make these results unreliable. We therefore do not report these RM values at K from the W-band.

This reversal relative to the Q band, C3 becoming polarized at high frequency while K fades, can be explained by two competing effects. At K, the polarized flux density decreases with increasing frequency in this optically thin region, and this decline outweighs the increase in fractional polarization from reduced Faraday depolarization. At C3, which has intrinsically stronger polarized emission and larger RM, the reduced depolarization at W-band frequencies significantly enhances the observed polarized flux. This leads to reliable detection of polarization at C3 that was not observed in the Q band.

The RM at C3 measured from 86.2, 86.3, 92.0, and 92.1~GHz in October is $\rm (1.0\pm0.2)\times10^5\ rad\ m^{-2}$, and that measured from 86.2, 88, 91, and 93~GHz in November and December are $\rm (1.3\pm0.2)\times10^5\ rad\ m^{-2}$ in November (Figure~\ref{fig:22octnov_w}), and  $\rm (1.3\pm0.1)\times10^5\ rad\ m^{-2}$ in December (Figure~\ref{fig:22dec23feb_w}). These RM values at C3 are generally larger than those at K from the same observing epochs. Furthermore, the RM at C3 measured from 86.2, 88, 90, and 92~GHz in 2023 February is $\rm (1.0\pm0.2)\times10^5\ rad\ m^{-2}$ (Figure~\ref{fig:22dec23feb_w}), also larger than the RM at K measured in the Q band data from 2022 October to December. These results consistently show that the RM at C3 is larger than that at K. 

The RM at C3 measured from 86.2, 86.3, 92.0, and 92.1~GHz in October is $\rm (1.0\pm0.2)\times10^5\ rad\ m^{-2}$, and from 86.2, 88, 91, and 93~GHz in November and December are $\rm (1.3\pm0.2)\times10^5\ rad\ m^{-2}$ in November and $\rm (1.3\pm0.1)\times10^5\ rad\ m^{-2}$ in December (Figures~\ref{fig:22octnov_w} and \ref{fig:22dec23feb_w}). In February 2023, the RM at C3 measured from 86.2, 88, 90, and 92~GHz is $\rm (1.0\pm0.2)\times10^5\ rad\ m^{-2}$ (Figure~\ref{fig:22dec23feb_w}). In all cases, these RM values at C3 are larger than those at K during this period, consistently showing that C3 exhibits higher RM than K.

\section{Polarization and RM (2024 November -- December)}\label{app:kvn_images_2024}

Figure~\ref{fig:novdec2024_w} shows the W-band images from two of these epochs, obtained on 2024 November 8 and December 27.  Polarization is still consistently detected at C3 and C4, although the polarized intensity at these positions has decreased compared to the previous epochs in 2022 December and 2023 February (Figure~\ref{fig:22dec23feb_w}). The EVPA at both positions continue to show Faraday rotation, corresponding to the RM of $\rm (2.6\pm0.4)\times10^5\ rad\ m^{-2}$ at C4 and $\rm (1.1\pm0.4)\times10^5\ rad\ m^{-2}$ at C3 on November 8, and $\rm (2.6\pm0.5)\times10^5\ rad\ m^{-2}$ at C4 and $\rm (1.6\pm0.3)\times10^5\ rad\ m^{-2}$ at C3 on December 27. In both epochs, the RM is again larger at C4 than at C3. 

In the Q-band images from the same two epochs, polarization is not detected at C4, whereas it is reliably detected at C3 and K (Figure~\ref{fig:novdec2024_q}). This is naturally explained by stronger depolarization at C4, which exhibits a larger RM than C3 and K. The RM at C3 measured across the Q-band is $\rm (1.5\pm0.1)\times10^5\ rad\ m^{-2}$ on November 8 and $\rm (1.4\pm0.1)\times10^5\ rad\ m^{-2}$ on December 27. Both values are consistent with those obtained in the W-band within the measurement uncertainties. The fractional polarization at C3 is at the level of 1--2\% in the Q-band, systematically lower than the 4--9\% observed in the W-band, which is also consistent with stronger Faraday depolarization at lower frequencies.

The RM at K measured in the Q-band is $\rm (0.8\pm0.1)\times10^5\ rad\ m^{-2}$ on November 8 and $\rm (0.5\pm0.1)\times10^5\ rad\ m^{-2}$ on December 27. Again, the RM at K is smaller than C3 and C4.

\section{Effect of the line-of-sight Integral on RM Scaling}\label{app:rm_integral}

As discussed in Section~\ref{sec:both_profiles}, the observed RM as a function of projected distance $D$ can be expressed as Equation~\ref{eq:rm_full}, which consists of a power-law dependence on $D$ and a line-of-sight (LOS) integral that depends on the viewing geometry. In this appendix, we examine whether the the LOS integral in Equation~\ref{eq:rm_full} affects the resulting RM--distance relation $\textrm{RM}\propto r^{-(\alpha+\beta)+1}$. 

Equation~\ref{eq:rm_full} can be written as
\begin{equation}
\mathrm{RM}(D) \propto D^{-(\alpha+\beta)+1} I(D),
\end{equation}
where the LOS integral is
\begin{equation}
I(D) = \int_{\phi_1}^{\phi_2(D)} (\cos\phi)^{\alpha+\beta} \, d(\tan\phi).
\end{equation}

\noindent Here, $\phi_1$ and $\phi_2$ denote the lower and upper limits of the line-of-sight integration, respectively, as defined in Section~\ref{sec:both_profiles}. The lower limit $\phi_1$ is fixed at $45^\circ$, while $\phi_2(D)$ is determined by the outer boundary of the circumnuclear ambient medium. First, we examine how $\phi_2(D)$ varies between projected distances of $D=3$~mas and $D=15$~mas. Its maximum angle along the LOS is given by

\begin{equation}
\phi_2(D) = \arctan\frac{\sqrt{r_{\rm B}^2-D^2}}{D}.
\end{equation}

\noindent where $r_{\rm B}$ is the Bondi radius, corresponding to $r_{\rm B}=34.4$~mas. Evaluating this yields
\begin{align*}
\phi_2(3~{\rm mas}) = 85.0^\circ, \\
\phi_2(15~{\rm mas}) = 64.2^\circ,
\end{align*}

\noindent corresponding to a difference of $\Delta\phi_2 \approx 20.8^\circ$. Using $d(\tan\phi) = d\phi / \cos^2\phi$, the LOS integral can be written as
\begin{equation}
I(D) = \int_{\phi_1}^{\phi_2(D)} (\cos\phi)^{\alpha+\beta - 2} \, d\phi.
\end{equation}

\noindent For the observed RM slope, $-(\alpha+\beta)+1 = -2.7$, this simplifies to
\begin{equation}
I(D) = \int_{\phi_1}^{\phi_2(D)} (\cos\phi)^{1.7} \, d\phi.
\end{equation}

\noindent Calculating the integral $I(D)$ for the two limiting projected distances yields
\begin{align*}
    I(3~{\rm mas}) = 0.174, \\ 
    I(15~{\rm mas}) = 0.133,
\end{align*}

\noindent corresponding to a change of $\sim 24\%$. While this change appears to be noticeable in absolute terms, it is still significantly smaller than the variation of the power-law factor $D^{-(\alpha+\beta)+1}$, which changes by nearly two orders of magnitude over the same distance range (Figure~\ref{fig:rm_dist}). 

Taking the logarithm of the RM expression yields:
\[
\log \mathrm{RM}(D) = [-(\alpha+\beta)+1]\log D + \log I(D) + \text{constant}.
\]

Differentiating with respect to $\log D$ gives the slope in log-log space:
\[
\frac{d \log \mathrm{RM}}{d \log D} = -(\alpha+\beta)+1 + \frac{d \log I(D)}{d \log D}.
\]

The first term $-(\alpha+\beta)+1$ corresponds to the power-law slope, while the second term $d \log I(D)/d \log D$ accounts for the variation of the integral with $D$. Using the numerical values above, the variation term changes by
\[
\frac{d \log I(D)}{d \log D} = -0.17
\]
\noindent between $D=3$~mas and $D=15$~mas. This indicates that the LOS integral introduces only a small systematic shift in the RM slope, with a maximum magnitude of 0.17. As the RM slope directly constrains $\alpha+\beta$, this corresponds to at most a 0.17 change in the combined index $\alpha+\beta$, and thus no more than a 0.17 shift in either $\alpha$ or $\beta$ individually. Since this possible bias is smaller than the observational uncertainty of 0.2, the relation $-(\alpha+\beta)+1=-2.7\pm0.2$ remains valid. Notably, decreasing $D$ further to $0.3$, $0.03$, or $0.003$~mas (i.e., approaching the black hole) still yields $I(D)=0.174$, identical to the value at $D=3$~mas. Accordingly, the variation term remains $d\log I(D)/d\log D=-0.17$, indicating that the relation between the observed RM slope and $-(\alpha+\beta)+1$ holds even down to the vicinity of the black hole. This justifies treating $I(D)$ as effectively constant when estimating the RM scaling.

\section{Physical constraints on the circumnuclear density profile}\label{app:wind}

Here we discuss the physical conditions under which a shallow density profile $n\propto r^{-\alpha}$ with $\alpha<1.5$ may arise and explain why such scenarios are unlikely to characterize the parsec-scale circumnuclear medium responsible for the observed Faraday rotation in NGC~1275. Throughout this section, we generally use the term \emph{wind} to refer to outflows launched from a radiatively efficient (cold) accretion disk, and the term \emph{outflow} to describe mass loss associated with radiatively inefficient (hot) accretion flows.

\subsection{Winds from cold accretion disk}

In Seyfert galaxies, disk winds can be driven by radiation pressure and/or magnetic processes, and are often observed in sources accreting at relatively high Eddington ratios (see \citealp{tombesi13} and references therein). Such winds are typically traced by warm absorbers, ionized gas producing blueshifted absorption features in the ultraviolet (UV) and X-ray spectra (e.g., \citealp{halpern84, reynolds97, blustin05}). These features arise because disk winds launched from the accretion disk are photoionized by the AGN ionizing continuum, leading to absorption features in the UV and X-ray spectra (see \citealp{crenshaw03} for a review). In some cases, the resulting wind density profile can be relatively shallow, depending on the wind geometry, acceleration profile, and mass-loading (e.g., $1.2<\alpha<1.5$; \citealt{tombesi13, laha16}). However, disk winds do not uniquely produce shallow density distributions in the ambient medium and may also exhibit steeper profiles with $\alpha>1.5$ \citep{wang22}. 

In the case of NGC~1275, there is no evidence for warm-absorber features in the X-ray spectra, implying that any disk wind would carry only a negligible amount of mass \citep{reynolds21}. This is consistent with accretion-mode studies \citep[e.g.,][]{giustini19}, which show that sources with Eddington ratios of $\sim$1\% produce only weak winds, while at lower ratios ($\lesssim 0.1\%$) winds become ineffective or disappear. Given the low Eddington ratio of NGC~1275 (0.04--0.09\%), disk winds are therefore expected to be negligible or absent.

On larger spatial scales, kinematically distinct outflows with velocities exceeding a few hundred km~s$^{-1}$ are nevertheless observed in NGC~1275, on scales of several tens of parsec \citep{nagai19} and a few hundred parsecs \citep{riffel20}. 
Given the lack of warm absorbers and the low Eddington ratio, these large-scale outflows are unlikely to be driven by the accretion disk. Instead, they are more plausibly associated with previous episodes of AGN activity, as evidenced by rising radio bubbles and X-ray cavities, along with associated shocks and ripples \citep[e.g.,][]{conselice01, fabian06, hatch06, dunn06, falceta10, vigneron24, rhea25}. Consequently, these outflows occur well outside the parsec-scale Bondi radius and their contribution to the observed Faraday rotation is expected to be negligible, since both the gas density and magnetic-field strength at these distances are much lower than those within the Bondi radius.

\subsection{Outflows from hot accretion flow}

At lower Eddington ratios, shallow density profiles with $\alpha<1.5$ may instead arise in hot accretion flows, where the net mass accretion rate decreases toward smaller radii. If the mass accretion rate were constant with distance from the black hole, a spherical inflow would produce a density profile with $\alpha=1.5$ \citep[e.g.,][]{narayan94}. However, theoretical studies suggest that the mass accretion rate in RIAFs decreases with decreasing distance from the black hole \citep[e.g.,][]{blandford99, narayan00, stone01, hawley01}. This leads to flatter density profiles, $\rho \propto r^{-\alpha}$ with $\alpha<1.5$, and numerical simulations often find $0.5\le\alpha<1$ within such RIAFs \citep[e.g.,][]{stone99, igumenshchev00, machida01, pang11, yuan12_1}. 

However, such shallow density profiles are expected only within the hot accretion flow itself. As discussed in Section~\ref{sec:riaf?}, a compact RIAF is likely present in NGC~1275, but it would be confined to radii smaller than the BLR\footnote{While we allow for the presence of a compact RIAF in 3C~84, our interpretation of the observed RM does not rely on its presence. Given its highly limited radial extent, any such RIAF would contribute negligibly to the observed Faraday rotation. If RIAF does not exist, the interpretation is even more straightforward. In either case, the RM measurements primarily probe the parsec-scale circumnuclear ambient medium.}. As a result, the density flattening associated with the hot accretion flow is not expected to extend to parsec scales, and therefore does not affect the circumnuclear medium probed by the RM measurements. 

In some cases, such as the adiabatic inflow-outflow solution (ADIOS; \citealp{blandford99}), outflows may be launched from RIAFs. However, even if such a RIAF-driven outflow is present in NGC~1275, an outflow launched from such a compact region is unlikely to have a significant impact on the parsec-scale environment. This is because the small radial extent of the RIAF in NGC~1275 limits the cumulative mass loss associated with hot accretion. Moreover, once the outflow propagates beyond the hot accretion region, its velocity is expected to be determined by energy conservation, yielding an initial scaling $v(r)\propto r^{-0.5}$ \citep[e.g.,][]{cao10}. If the mass of the outflow is conserved  ($\dot M_{\rm out} = 4\pi r^{2} \rho v$), this velocity scaling corresponds to a density profile with $\alpha=1.5$. As the outflow propagates outward and becomes unbound, the velocity approaches a nearly constant value at large radii, because for such an unbound outflow, energy conservation gives

\begin{equation}
    v^2(r) = v_\infty^2 + \frac{2GM}{r}.
\end{equation}

\noindent This implies that the velocity approaches a constant value at large radii. In this case, mass conservation requires the outflow density profile to steepen toward $\alpha\sim2$, rather than $\alpha<1.5$. 

Notably, $\alpha\sim2$ also corresponds to the upper limit expected for a steady-flow circumnuclear medium, as discussed in Section~\ref{sec:both_profiles}. Once a RIAF-driven outflow propagates beyond the hot accretion region, its velocity approaches a nearly constant value, causing the outflow density profile to asymptotically steepen toward this limit. Even if such an outflow is partially decelerated by the surrounding medium, it would act to provide pressure support against gravitational infall, thereby reducing the infall velocity and leading the ambient medium toward a quasi-steady state, corresponding to the upper limit $\alpha=2$. Accordingly, whether freely propagating or decelerated by the ambient medium, a potential RIAF-driven outflow would tend to drive the circumnuclear density profile toward $\alpha=2$, rather than producing a shallow profile with $\alpha<1.5$ on parsec scales. 

In summary, while shallow density profiles with $\alpha<1.5$ may arise either in disk winds at relatively high Eddington ratio or within the RIAF at very low Eddington ratios, NGC~1275 lies near the transition between these regimes, where neither scenario is expected to dominate the parsec-scale circumnuclear environment. Instead, the physical and observational constraints favor a circumnuclear ambient medium with $1.5\lesssim\alpha\lesssim2$.

\end{document}